\documentclass{aa}  

\usepackage{graphicx}
\usepackage{txfonts}
\usepackage{xcolor}
\usepackage{lipsum}
\usepackage{natbib}
\usepackage{ulem}
\usepackage{multirow}
\definecolor{mygray}{gray}{0.9}
\begin{document}

   \title{Horizontal shear instabilities in rotating stellar radiation zones}

   \subtitle{II. Effects of the full Coriolis acceleration}

   \author{J. Park
   	\inst{1,2}
	\and
          V. Prat
          \inst{1}
          \and
          S. Mathis
          \inst{1}
          \and
          L. Bugnet
          \inst{1}
          }

   \institute{
             \inst{1} AIM, CEA, CNRS, Universit\'e Paris-Saclay, Universit\'e Paris Diderot, Sorbonne Paris Cit\'e, F-91191 Gif-sur-Yvette, France\\
   		\inst{2} Fluid and Complex Systems Research Centre, Coventry University, Coventry CV1 5FB, UK\\
              \email{junho.park@cea.fr}
                           }

   \date{}

 
  \abstract
   {Stellar interiors are the seat of efficient transport of angular momentum all along their evolution. 
   In this context, understanding the dependence of the turbulent transport triggered by the instabilities of the vertical and horizontal shears of the differential rotation in stellar radiation zones as a function of their rotation, stratification, and thermal diffusivity is mandatory. 
   Indeed, it constitutes one of the cornerstones of the rotational transport and mixing theory which is implemented in stellar evolution codes to predict the rotational and chemical evolutions of stars.}
   {We investigate horizontal shear instabilities in rotating stellar radiation zones by considering the full Coriolis acceleration with both the dimensionless horizontal Coriolis component $\tilde{f}$ and the vertical component $f$. 
   }
   {We performed a linear stability analysis using linearized equations derived from the Navier-Stokes and heat transport equations in the rotating non-traditional $f$-plane. 
   We considered a horizontal shear flow with a hyperbolic tangent profile as the base flow. 
   The linear stability was analyzed numerically in wide ranges of parameters, and we performed an asymptotic analysis for large vertical wavenumbers using the Wentzel-Kramers-Brillouin-Jeffreys (WKBJ) approximation for non-diffusive and highly diffusive fluids. 
}
   {As in the traditional $f$-plane approximation, we identify two types of instabilities: the inflectional and inertial instabilities.
   The inflectional instability is destabilized as $\tilde{f}$ increases and its maximum growth rate increases significantly, while the thermal diffusivity stabilizes the inflectional instability similarly to the traditional case. 
   The inertial instability is also strongly affected; for instance, the inertially unstable regime is also extended in the non-diffusive limit as $0<f<1+\tilde{f}^{2}/N^{2}$, where $N$ is the dimensionless Brunt-V\"ais\"al\"a frequency. 
   More strikingly, in the high-thermal-diffusivity limit, it is always inertially unstable at any colatitude $\theta$ except at the poles (i.e., $0^{\circ}<\theta<180^{\circ}$).
   We also derived the critical Reynolds numbers for the inertial instability using the asymptotic dispersion relations obtained from the WKBJ analysis. 
   Using the asymptotic and numerical results, we propose a prescription for the effective turbulent viscosities induced by the inertial and inflectional instabilities that can be possibly used in stellar evolution models. 
   The characteristic time of this turbulence is short enough so that it is efficient to redistribute angular momentum and mix chemicals in stellar radiation zones.}
   {}

   \keywords{hydrodynamics --
   		turbulence --
                stars: rotation --
                stars: evolution
               }

   \maketitle
%

\section{Introduction}

Stellar rotation is one of the key physical processes to build a modern picture of stellar evolution \citep[e.g.][and references therein]{Maeder2009}. 
Indeed, it triggers transport of angular momentum and of chemicals, which drives the rotational and chemical evolution of stars, respectively \citep[e.g.][]{Zahn1992,MaederZahn1998,MathisZahn2004}. 
This has major impact on the late stages of their evolution \citep[e.g.][]{Hirschietal2004}, their magnetism \citep[e.g.][]{BrunBrowning2017}, their winds and mass losses \citep[e.g.][]{Ud-Doulaetal2009,Mattetal2015}, and the interactions with their planetary and galactic environment \citep[e.g.][]{Galletetal2017,Strugareketal2017}.

A robust ab-initio evaluation of the strength of each (magneto-)hydrodynamical mechanism that transports momentum and chemicals is thus mandatory to understand astrophysical observations. 
In particular, our knowledge of the internal rotation of stars and its evolution has been revolutionized thanks to space-based asteroseismology with the {\it Kepler} space mission (NASA) \citep[][and references therein]{Aertsetal2019}. 
It has proven that stars are mostly hosting weak differential rotation in the whole Hertzsprung-Russell diagram, like our Sun. 
Stellar interiors are thus the seat of efficient mechanisms that transport angular momentum all along their evolution.
These mechanisms have not yet been identified even if several candidates have been proposed, such as stable/unstable magnetic fields \citep[e.g.]{Moss1992,Charbonneau1993,Spruit1999,Spruit2002,Fuller2019}, stochastically-excited internal gravity waves \citep[e.g.][]{TalonCharbonnel2005,Rogers2015,Pinconetal2017}, and mixed gravito-acoustic modes \citep[][]{Belkacemetal2015a,Belkacemetal2015b}. 
In this framework, improving our knowledge of the hydrodynamical turbulent transport induced by the instabilities of the stellar differential rotation is mandatory since it has been proposed recently as another potential efficient mechanism to transport angular momentum \citep[][]{Barkeretal2020,Garaud2020} while it constitutes one of the cornerstones of the theory of the rotational transport and mixing along the evolution of stars \citep{Zahn1992}.
   
Since stellar radiation zones are stably stratified, rotating regions, it is expected that the turbulent transport triggered there by the instabilities of the vertical and horizontal shear of the differential rotation should be anisotropic \citep{Zahn1992}. 
As pointed out in \cite{Mathis2018} and \cite{PPM2020}, it is the vertical shear instabilities that have received important attention in the literature, in particular with taking into account the impact of the high thermal diffusion in stellar radiation zones \citep[][]{Zahn1983,Lignieres1999} and the interactions with the horizontal turbulence induced by the horizontal shear \citep{TalonZahn1997}. 
The obtained prescriptions, mainly derived using phenomenological modelings, have been broadly implemented in state-of-the-art evolution models of rotating stars \citep[e.g.][]{Ekstrometal2012,Marquesetal2013,Amardetal2019}. 
They are now tested using direct numerical simulations devoted to the stellar regime \citep{Pratetal2013,Pratetal2014,Garaudetal2017,Pratetal2016,GagnierGaraud2018,KG2018}. 
The horizontal turbulence induced by horizontal gradients of the differential rotation has only been examined in the stellar regime in few works using again phenomenological arguments \citep{Zahn1992,Maeder2003}, results from lab experiments studying differentially rotating flows \citep{RichardZahn1999,Mathis2004}, and first devoted numerical simulations \citep{Copeetal2019}. 
The systematic study of the combined effect of stable stratification, rotation, and thermal diffusion in the stellar context has only been recently undertaken.            
   
In \cite{PPM2020}, we thus examined the behavior of shear instabilities sustained by a horizontal shear as a function of stratification, rotation, and thermal diffusion.
However, in this first work, we neglected the horizontal projection of the rotation vector and the corresponding terms of the Coriolis acceleration, following the traditional approximation of rotation (TAR) which is often used to describe geophysical and astrophysical flows in stably stratified, rotating regions \citep{Eckart1960}. 
However, recent works have disputed this approximation as it can fail in some cases. 
For instance, the dynamics of near-inertial waves is strongly influenced by the non-traditional effects in such a way that properties of the wave reflection and associated mixing are largely modified \citep[]{Gerkema2005,Gerkema2008}. 
Moreover, couplings between gravito-inertial waves, inertial waves, and wave-induced turbulence in stars can only be properly treated when the full Coriolis acceleration is considered \citep[]{Mathis2014}. 
Finally, \citet{Zeitlin2018} demonstrated analytically that the instability of a linear shear flow can be significantly modified if the full Coriolis acceleration is considered. 
Non-traditional effects can be particularly important for stellar structure configurations where the Coriolis acceleration can compete with the Archimedean force in the direction of both entropy and chemical stratification, for instance during the formation of the radiative core of pre-main-sequence, low-mass stars or in the radiative envelope of rapidly-rotating upper-main-sequence stars. 
These regimes should be treated properly to build robust one- or two-dimensional (1D or 2D) secular models of the evolution of rotating stars \citep[e.g.][]{Ekstrometal2012,Amardetal2019,Gagnieretal2019}.

In this paper, we, therefore, continue our previous work that examined the effect of thermal diffusion on horizontal shear instabilities in stably stratified rotating stellar radiative zones \citep[]{PPM2020}, but we now consider the full Coriolis acceleration. 
In particular, we investigate how the full Coriolis acceleration modifies the dynamics of two types of shear instabilities: the inflectional and inertial instabilities. 
In Sect.~\ref{sec:problem}, we formulate linear stability equations derived from the Navier-Stokes equations when the fluid is stably stratified and thermally diffusive in the rotating non-traditional $f$-plane where both vertical and horizontal components of the rotation vector are taken into account. 
We consider a base shear flow in a hyperbolic tangent form as a canonical shear profile used in previous studies \citep[]{Schmid2001,Deloncle2007,Griffiths2008,Arobone2012}. 
In Sect.~\ref{sect:general}, we provide general numerical results on the inflectional and inertial instabilities. 
In Sect.~\ref{sect:WKBJ}, we use the Wentzel-Kramers-Brillouing-Jeffreys (WKBJ) approximation in the asymptotic limit of large vertical wavenumbers to investigate how the inertial instability is modified by non-traditional effects in the asymptotic non-diffusive and high-diffusivity cases. 
In Sect.~\ref{sect:parametric}, we analyze in detail numerical results in wide ranges of parameters for both the inflectional and inertial instabilities. 
In Sects.~\ref{sect:turbulent_viscosity} and \ref{sect:discussion}, we propose expressions for the horizontal turbulent viscosity and the characteristic time of the turbulent transport, which can be possibly applied to 1D and 2D secular evolution models of rotating stars. 
Finally, in Sect.~\ref{sect:conclusion}, we provide our conclusions and discussions about the non-traditional effects on the horizontal shear instabilities, and we propose the perspectives of this work for astrophysical and geophysical flows. 

\section{Problem formulation}
\label{sec:problem}
\subsection{Navier-Stokes equations and base steady state}
%
   \begin{figure}
   \centering
   \includegraphics[height=5cm]{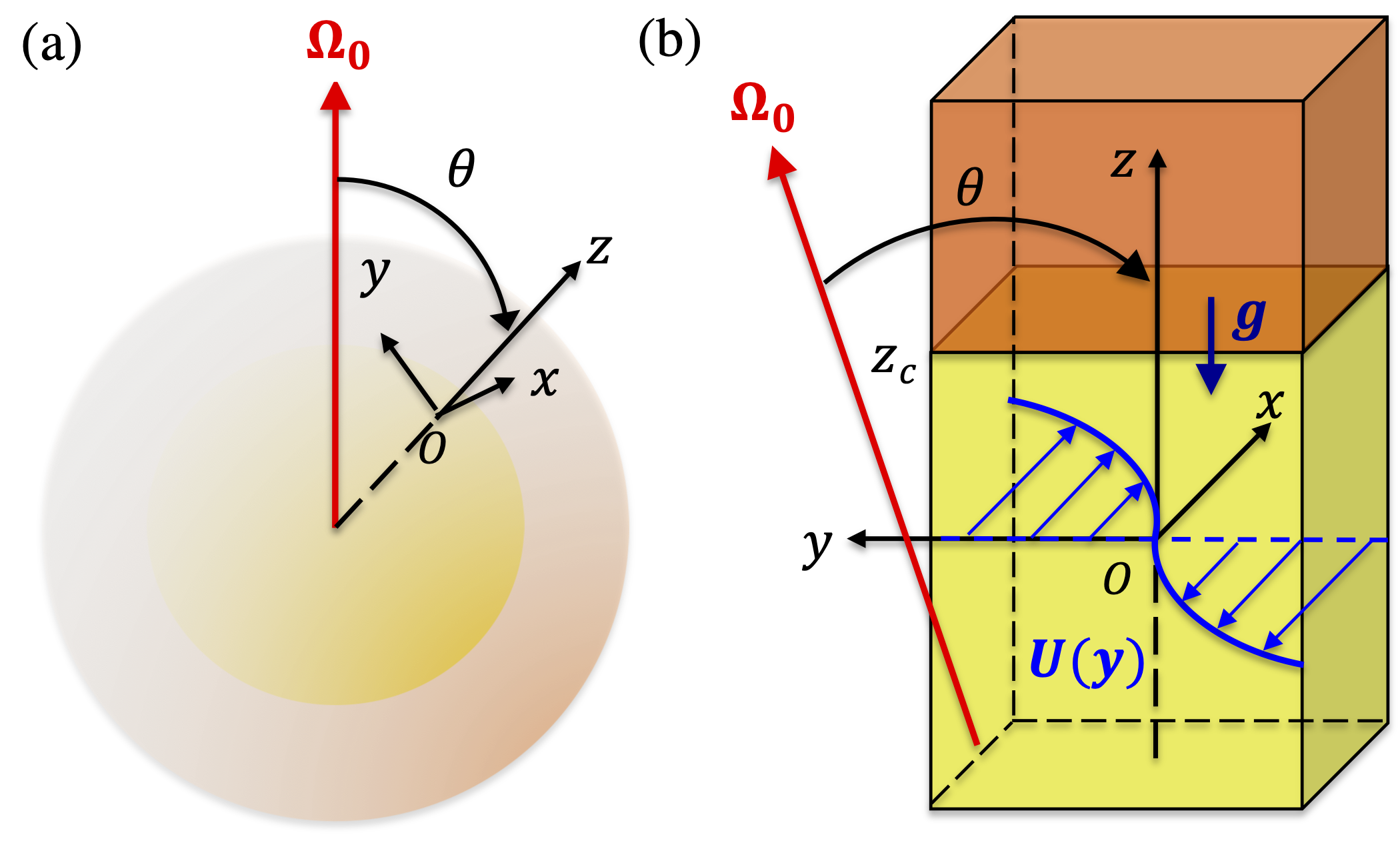}
      \caption{(a) Schematic of the radiative and convective zones, colored as yellow and orange, respectively, for the configuration of low-mass stars rotating with angular speed $\Omega_{0}$. 
      (b) Horizontal shear flow $U(y)$ in a local non-traditional $f$-plane at a colatitude $\theta$ in the radiative zone; $z_{c}$ denotes the transition altitude between the radiative and convective zones.
      }
         \label{Fig_cartoon}
   \end{figure}
We consider the Navier-Stokes {and heat transport} equations under the Boussinesq approximation in a rotating frame.
We define the local Cartesian coordinates $(x,y,z)$ with $x$ the longitudinal coordinate, $y$ the latitudinal coordinate, and $z$ the vertical coordinate:
\begin{equation}
\label{eq:total_continuity}
	\nabla\cdot\vec{u}=0,
\end{equation}
\begin{equation}
\label{eq:total_momentum}
	\frac{\partial\vec{u}}{\partial t}+\left(\vec{u}\cdot\nabla\right)\vec{u}+\vec{f}\times\vec{u}=-\frac{1}{\rho_{0}}\nabla p-\alpha_{\rm{T}}\Theta\vec{g}+\nu_{0}\nabla^{2}\vec{u},
\end{equation}
\begin{equation}
\label{eq:total_diffusion}
	\frac{\partial \Theta}{\partial t}+\vec{u}\cdot\nabla \Theta=\kappa_{0}\nabla^{2}\Theta,
\end{equation}
where $\vec{u}=\left(u,v,w\right)$ is the velocity, $p$ is the pressure, $\Theta$ is the temperature deviation from the reference temperature $T_{0}$, $\vec{f}=\left(0,{f}_{\rm{h},0},{f}_{\rm{v},0}\right)$ is the Coriolis vector with the horizontal (latitudinal) Coriolis component $f_{\rm{h},0}=2\Omega_{0}\sin\theta$ and the vertical Coriolis component $f_{\rm{v},0}=2\Omega_{0}\cos\theta$ where $\Omega_{0}=\left(f_{\rm{h},0}^{2}+f_{\rm{v},0}^{2}\right)^{1/2}/2$ is the stellar rotation rate, $\theta$ is the colatitude, $\rho_{0}$ is the reference density, $\vec{g}=(0,0,-g)$ is the gravity, $\nu_{0}$ is the reference viscosity, $\kappa_{0}$ is the reference thermal diffusivity, $\alpha_{\rm{T}}$ is the thermal expansion coefficient that assumes a linear relation between the density deviation and the temperature deviation $\Theta$, and $\nabla^{2}=\partial^{2}/\partial x^{2}+\partial^{2}/\partial y^{2}+\partial^{2}/\partial z^{2}$ denotes the Laplacian operator. 
Figure~\ref{Fig_cartoon} illustrates the local coordinate system of the horizontal shear flow in the radiative zone when the non-traditional $f$-plane is considered at a given colatitude $\theta$.
For the base state, we consider a canonical example of the steady horizontal shear flow $\vec{U}=\left(U(y),0,0\right)$ in a hyperbolic tangent form:
\begin{equation}
\label{eq:base_shear}
	U(y)=U_{0}\tanh\left(\frac{y}{L_{0}}\right),
\end{equation}
where $U_{0}$ and $L_{0}$ are the reference velocity and length scales, respectively. 
{This hyperbolic tangent profile has an inflection point at $y=0$. Therefore, it has been considered in many previous stability studies \citep[see e.g.,][]{Schmid2001,Deloncle2007,Arobone2012} to investigate the inflectional instability. 
We also adopt this profile to compare with other studies and extend our understanding of the inflectional instability for a horizontal shear when the full Coriolis acceleration is taken into account.}
In the presence of the horizontal Coriolis parameter $f_{\rm{h},0}$, such a base flow is steady if the thermal wind balance between $U(y)$ and the base temperature $\bar{\Theta}(y,z)$ is satisfied as follows:
\begin{equation}
\label{eq:base_thermal_wind_balance}
	\alpha_{\rm{T}}g\frac{\partial\bar{\Theta}}{\partial y}=-f_{\rm{h},0}\frac{\partial U}{\partial y}.
\end{equation}
Such a thermal-wind balance has been adopted for cases where the rotation vector and the base shear are not aligned in stratified fluids; for instance, ageostrophic instability in vertical shear flows in stratified fluids in the traditional $f$-plane \citep[]{Wang2014}.
In addition to the thermal-wind balance, the base temperature $\bar{\Theta}(y,z)$ is considered to be stably stratified with a linearly increasing profile in $z$; therefore, it has the form
\begin{equation}
\label{eq:base_temperature}
	\bar{\Theta}(y,z)=\frac{\Delta\bar{\Theta}_{0}}{\Delta z}z-\frac{f_{\rm{h},0}}{\alpha_{\rm{T}}g}U(y),
\end{equation}
where $\Delta\bar{\Theta}_{0}$ is the difference in base temperature along the vertical distance $\Delta z$ at a given $y$.

\subsection{Linearized equations}
We consider the velocity perturbation $\check{\vec{u}}=\vec{u}-\vec{U}=\left(\check{u},\check{v},\check{w}\right)$, the pressure perturbation $\check{p}=p-P$ where $P$ is the base pressure, and the temperature perturbation $\check{T}=\Theta-\bar{\Theta}$. 
We nondimensionalize variables in Eqs.~(\ref{eq:total_continuity}-\ref{eq:total_diffusion}) by considering the length scale $L_{0}$, the velocity scale $U_{0}$, the time scale as $t_{0}=L_{0}/U_{0}$, the pressure scale as $\rho_{0}U_{0}^{2}$, and the temperature scale as $L_{0}\Delta\bar{\Theta}_{0}/\Delta z$. 
For infinitesimally small perturbations, we have the following non-dimensional linearized equations:
\begin{equation}
\label{eq:ptb_continuity}
	\frac{\partial \check{u}}{\partial x}+\frac{\partial \check{v}}{\partial y}+\frac{\partial \check{w}}{\partial z}=0,
\end{equation}
\begin{equation}
\label{eq:ptb_x_mom}
	\frac{\partial \check{u}}{\partial t}+U\frac{\partial \check{u}}{\partial x}+\left(U'-f\right)\check{v}+\tilde{f}\check{w}=-\frac{\partial \check{p}}{\partial x}+\frac{1}{Re}\nabla^{2}\check{u},
\end{equation}
\begin{equation}
\label{eq:ptb_y_mom}
	\frac{\partial \check{v}}{\partial t}+U\frac{\partial \check{v}}{\partial x}+f\check{u}=-\frac{\partial \check{p}}{\partial y}+\frac{1}{Re}\nabla^{2}\check{v},
\end{equation}
\begin{equation}
\label{eq:ptb_z_mom}
	\frac{\partial \check{w}}{\partial t}+U\frac{\partial \check{w}}{\partial x}-\tilde{f}\check{u}=-\frac{\partial \check{p}}{\partial z}+N^{2}\check{T}+\frac{1}{Re}\nabla^{2}\check{w},
\end{equation}
\begin{equation}
\label{eq:ptb_diffusion}
	\frac{\partial \check{T}}{\partial t}+U\frac{\partial \check{T}}{\partial x}-\frac{\tilde{f}U'}{N^{2}}\check{v}+\check{w}=\frac{1}{Pe}\nabla^{2}\check{T},
\end{equation}
where the prime symbol ($'$) denotes the derivative with respect to $y$, $f=2\Omega\cos\theta$ and $\tilde{f}=2\Omega\sin\theta$ are the dimensionless vertical and horizontal Coriolis parameters, respectively, $\Omega$ is the dimensionless stellar rotation rate, $Re$ is the Reynolds number
\begin{equation}
\label{eq:Reynolds}
Re=\frac{U_{0}L_{0}}{\nu_{0}}, 
\end{equation}
$N$ is the dimensionless Brunt-V\"ais\"al\"a frequency\footnote{{The non-dimensional parameter $N^{2}$ is equivalent to the Richardson number $Ri$ with our choice of normalization.}}
\begin{equation}
\label{eq:Brunt_Vaisala}
N=\sqrt{\frac{\alpha_{\rm{T}}gL_{0}^{2}}{U_{0}^{2}}\frac{\Delta\bar{\Theta}_{0}}{\Delta z}}, 
\end{equation}
and $Pe$ is the P\'eclet number 
\begin{equation}
\label{eq:Peclet}
Pe=\frac{U_{0}L_{0}}{\kappa_{0}}. 
\end{equation}
In Eq.~(\ref{eq:ptb_diffusion}), the term $-\tilde{f}U'\hat{v}/N^{2}$ on the left-hand side appears due to the thermal-wind balance (\ref{eq:base_thermal_wind_balance}).

To derive linear stability equations, we apply a normal mode expansion to perturbations as
\begin{equation}
\label{eq:ptb_normal_mode}
	\left[\check{\vec{u}},\check{p},\check{T}\right]=\left[\hat{\vec{u}}(y),\hat{p}(y),\hat{T}(y)\right]\exp\left[\mathrm{i}\left(k_{\rm{x}} x+k_{\rm{z}} z\right)+\sigma t\right]+c.c.,
\end{equation}
where $\hat{\vec{u}}=\left(\hat{u},\hat{v},\hat{w}\right)$, $\hat{p}$ and $\hat{T}$ are the mode shapes of velocity, pressure, and temperature perturbations, respectively, $\mathrm{i}^{2}=-1$, $k_{\rm{x}}$ is the streamwise (longitudinal) wavenumber, $k_{\rm{z}}$ is the vertical wavenumber, $\sigma=\sigma_{r}+\mathrm{i}\sigma_{i}$ is the complex growth rate where the real part $\sigma_{r}$ is the growth rate and the imaginary part $\sigma_{i}$ is the temporal frequency, and $c.c.$ denotes the complex conjugate. 
Using the normal mode expansion, we obtain the following linear stability equations
\begin{equation}
\label{eq:lse_continuity}
	\mathrm{i}k_{\rm{x}}\hat{u}+\frac{\mathrm{d}\hat{v}}{\mathrm{d}y}+\mathrm{i}k_{\rm{z}}\hat{w}=0,
\end{equation}
\begin{equation}
\label{eq:lse_x_mom}
	\left(\sigma+\mathrm{i}k_{\rm{x}} U\right)\hat{u}+\left(U'-f\right)\hat{v}+\tilde{f}\hat{w}=-\mathrm{i}k_{\rm{x}}\hat{p}+\frac{1}{Re}\hat{\nabla}^{2}\hat{u},
\end{equation}
\begin{equation}
\label{eq:lse_y_mom}
	\left(\sigma+\mathrm{i}k_{\rm{x}} U\right)\hat{v}+f\hat{u}=-\frac{\mathrm{d}\hat{p}}{\mathrm{d}y}+\frac{1}{Re}\hat{\nabla}^{2}\hat{v},
\end{equation}
\begin{equation}
\label{eq:lse_z_mom}
	\left(\sigma+\mathrm{i}k_{\rm{x}} U\right)\hat{w}-\tilde{f}\hat{u}=-\mathrm{i}k_{\rm{z}}\hat{p}+N^{2}\hat{T}+\frac{1}{Re}\hat{\nabla}^{2}\hat{w},
\end{equation}
\begin{equation}
\label{eq:lse_diffusion}
	\left(\sigma+\mathrm{i}k_{\rm{x}} U\right)\hat{T}-\frac{\tilde{f}U'}{N^{2}}\hat{v}+\hat{w}=\frac{1}{Pe}\hat{\nabla}^{2}\hat{T},
\end{equation}
where $\hat{\nabla}^{2}=\frac{\rm{d}^{2}}{\rm{d}y^{2}}-k^{2}$ with $k^{2}=k_{\rm{x}}^{2}+k_{\rm{z}}^{2}$. 
By eliminating $\hat{p}$ using the continuity equation (\ref{eq:lse_continuity}), Eqs.~(\ref{eq:lse_continuity}-\ref{eq:lse_diffusion}) can be gathered into a matrix form as an eigenvalue problem:
\begin{equation}
\label{eq:lse_matrix}
	\sigma\mathcal{B}
	\left(
	\begin{array}{c}
	\hat{u}\\
	\hat{v}\\
	\hat{w}\\
	\hat{T}
	\end{array}
	\right)=
	\mathcal{A}
	\left(
	\begin{array}{c}
	\hat{u}\\
	\hat{v}\\
	\hat{w}\\
	\hat{T}
	\end{array}
	\right),
\end{equation}
where $\mathcal{A}$ and $\mathcal{B}$ are the operator matrices expressed as
\begin{equation}
\label{eq:appendix_matrixA}
	\mathcal{A}=
	\left[
	\begin{array}{cccc}
	\mathcal{A}_{11} & \mathcal{A}_{12} & \mathcal{A}_{13} & \mathcal{A}_{14}\\
	\mathcal{A}_{21} & \mathcal{A}_{22} & \mathcal{A}_{23} & \mathcal{A}_{24}\\
	\mathcal{A}_{31} & \mathcal{A}_{32} & \mathcal{A}_{33} & \mathcal{A}_{34}\\
	0 & \mathcal{A}_{42} & \mathcal{A}_{43} & \mathcal{A}_{44}
	\end{array}
	\right],
\end{equation}
\begin{equation}
\label{eq:appendix_matrixB}
	\mathcal{B}=
	\left[
	\begin{array}{cccc}
	-k^{2} & \mathrm{i}k_{\rm{x}}\frac{\rm{d}}{\rm{d}y} & 0 & 0\\
	0 & \hat{\nabla}^{2} & 0 & 0\\
	0 & \mathrm{i}k_{\rm{z}}\frac{\rm{d}}{\rm{d}y} & -k^{2} & 0\\
	0 & 0 & 0 & 1
	\end{array}
	\right],
\end{equation}
where
\begin{equation}
\label{eq:Appendix1_A1}
\begin{aligned}
&\mathcal{A}_{11}=k_{\rm{x}}\left(\mathrm{i}k^{2}U+k_{\rm{z}}\tilde{f}\right)-\frac{k^{2}}{Re}\hat{\nabla}^{2},\\
&\mathcal{A}_{12}=k_{\rm{z}}^{2}(U'-f)+k_{\rm{x}}^{2}U\frac{\rm{d}}{\rm{d}y}+\frac{\mathrm{i}k_{\rm{x}}}{Re}\hat{\nabla}^{2}\frac{\rm{d}}{\rm{d}y},\\
&\mathcal{A}_{13}=k_{\rm{z}}^{2}\tilde{f},~~
\mathcal{A}_{14}=k_{\rm{x}}k_{\rm{z}} N^{2},
\end{aligned}
\end{equation}
\begin{equation}
\label{eq:Appendix1_A2}
\begin{aligned}
&\mathcal{A}_{21}=k^{2}f-\mathrm{i}k_{\rm{z}}\tilde{f}\frac{\rm{d}}{\rm{d}y},\\
&\mathcal{A}_{22}=\mathrm{i}k_{\rm{x}}\left(k^{2}U+U''-f\frac{\rm{d}}{\rm{d}y}-U\frac{\rm{d}^{2}}{\rm{d}y^{2}}\right)+\frac{1}{Re}\hat{\nabla}^{4},\\
&\mathcal{A}_{23}=\mathrm{i}k_{\rm{x}}\tilde{f}\frac{\rm{d}}{\rm{d}y},~~
\mathcal{A}_{24}=-\mathrm{i}k_{\rm{z}} N^{2}\frac{\rm{d}}{\rm{d}y},
\end{aligned}
\end{equation}
\begin{equation}
\label{eq:Appendix1_A3}
\begin{aligned}
&\mathcal{A}_{31}=-k_{\rm{x}}^{2}\tilde{f},\\
&\mathcal{A}_{32}=k_{\rm{x}}k_{\rm{z}}\left(U\frac{\rm{d}}{\rm{d}y}+f-U'\right)+\frac{\mathrm{i}k_{\rm{z}}}{Re}\hat{\nabla}^{2},\\
&\mathcal{A}_{33}=k_{\rm{x}}\left(\mathrm{i}k^{2}U-k_{\rm{z}}\tilde{f}\right)-\frac{k^{2}}{Re}\hat{\nabla}^{2},~~
\mathcal{A}_{34}=-k_{\rm{x}}^{2}N^{2},
\end{aligned}
\end{equation}
\begin{equation}
\label{eq:Appendix1_A4}
\mathcal{A}_{42}=\frac{\tilde{f}U'}{N^{2}},~~
\mathcal{A}_{43}=-1,~~
\mathcal{A}_{44}=-\mathrm{i}k_{\rm{x}} U+\frac{1}{Pe}\hat{\nabla}^{2}.
\end{equation}
We discretize numerically the operators $\mathcal{A}$ and $\mathcal{B}$ in the $y$-direction using the rational Chebyshev function that maps the Chebyshev domain $y_{\rm{cheb}}\in(-1,1)$ onto the physical space $y\in (-\infty,\infty)$ via the mapping $y/{L}_{\rm{map}}=y_{\rm{cheb}}/\sqrt{1+y_{\rm{cheb}}^{2}}$ where $L_{\rm{map}}$ is the mapping factor \citep[]{Deloncle2007,PPM2020}. 
{To distinguish physical and numerically-converged modes from spurious modes, we use a convergence criterion based on the residual of coefficients on the Chebyshev functions proposed by \citet{Fabre2004}. 
This technique allows us to separate highly-oscillatory spurious modes from physical modes that decays smoothly at boundaries as $y\rightarrow\pm\infty$. 
Furthermore, we chose the number of collocation points in the $y$-direction from 100 to 200, which was sufficient in our study to confirm the convergence of physical modes. 
}
We impose vanishing boundary conditions as $y\rightarrow\pm\infty$ by suppressing terms in the first and last rows of the operator matrices \citep[]{Antkowiak2005,Park2012}. 
Numerical results of the horizontal shear instability are compared and validated with results of \citet{Deloncle2007,Arobone2012,PPM2020} in stratified and rotating fluids for the traditional case when $\tilde{f}=0$. 

While the stability of horizontal shear flows in stratified-rotating fluids in the traditional $f$-plane has one symmetry condition on $\sigma$ with $\pm k_{\rm{x}}$ and $\pm k_{\rm{z}}$ \citep[]{Deloncle2007,PPM2020}, an analogous symmetry condition does not exist due to the horizontal Coriolis parameter $\tilde{f}>0$. 
Instead, there exist two separate symmetry conditions on $\sigma$ in terms of $k_{\rm{x}}$ and $k_{\rm{z}}$ as follows:
\begin{equation}
\sigma(k_{\rm{x}},k_{\rm{z}})=\sigma^{*}(-k_{\rm{x}},-k_{\rm{z}}),
\end{equation}
and
\begin{equation}
\sigma(k_{\rm{x}},-k_{\rm{z}})=\sigma^{*}(-k_{\rm{x}},k_{\rm{z}}),
\end{equation}
where the asterisk (*) denotes the complex conjugate.
Therefore, in this paper, we will investigate both negative and positive $k_{\rm{z}}$ for $k_{\rm{x}}\geq0$.

\subsection{Simplified equations for $\hat{v}$}
Equations~(\ref{eq:lse_continuity}-\ref{eq:lse_diffusion}) can be simplified into ordinary differential equations (ODEs) for $\hat{v}$ in particular cases. 
For instance, in the inviscid and non-diffusive limits ($Re\rightarrow\infty$ and $Pe\rightarrow\infty$), we can derive the single 2nd-order ODE for $\hat{v}$ as follows:
\begin{equation}
\label{eq:2ODE_v}
	\frac{\rm{d}^{2}\hat{v}}{\rm{d}y^{2}}+V_{1}\frac{\rm{d}\hat{v}}{\rm{d}y}+V_{0}\hat{v}=0,
\end{equation}
where
\begin{equation}
	V_{1}=\frac{2sk_{\rm{x}} U'-2\mathrm{i}k_{\rm{z}} f\tilde{f}}{s^{2}-N^{2}-\tilde{f}^{2}}-\frac{Q'}{Q},
\end{equation}
$s=-\mathrm{i}\sigma+k_{\rm{x}} U$ is the Doppler-shifted frequency, $Q=k^{2}s^{2}-k_{\rm{x}}^{2}N^{2}$,
\begin{equation}
\begin{aligned}
	V_{0}=&-k_{\rm{z}}^{2}\frac{\Gamma}{s^{2}-N^{2}-\tilde{f}^{2}}-k_{\rm{x}}^{2}-\frac{k_{\rm{x}}}{s}\left(U''-\frac{Q'}{Q}U'\right)-\frac{k_{\rm{x}} f}{s}\frac{Q'}{Q}\\
	&-\frac{2k_{\rm{x}}^{2}U'(U'-f)}{s^{2}-N^{2}-\tilde{f}^{2}}+\frac{\tilde{f}}{s^{2}-
	N^{2}-\tilde{f}^{2}}\left[\frac{fQ'}{Q}\left(\mathrm{i}k_{\rm{z}}-\frac{k_{\rm{x}}\tilde{f}}{s}\right)-\tilde{f}k_{\rm{x}}^{2}\right],
\end{aligned}
\end{equation}
and 
\begin{equation}
\label{eq:Gamma}
\Gamma=s^{2}+f\left(U'-f\right). 
\end{equation}

For finite $Pe$ in the inviscid limit, we can find the single 4th-order ODE when $k_{\rm{x}}=0$ as follows:
\begin{equation}
\label{eq:4ODE_v}
	\frac{\rm{d}^{4}\hat{v}}{\rm{d}y^{4}}+\mathcal{V}_{3}\frac{\rm{d}^{3}\hat{v}}{\rm{d}y^{3}}+\mathcal{V}_{2}\frac{\rm{d}^{2}\hat{v}}{\rm{d}y^{2}}+\mathcal{V}_{1}\frac{\rm{d}\hat{v}}{\rm{d}y}+\mathcal{V}_{0}\hat{v}=0,
\end{equation}
where
\begin{equation}
\mathcal{V}_{3}=-\frac{\mathrm{i}k_{\rm{z}}\tilde{f}(U'-2f)}{\sigma^{2}+\tilde{f}^{2}},
\end{equation}
\begin{equation}
\mathcal{V}_{2}=k_{\rm{z}}^{2}\left(\frac{\Gamma}{\sigma^{2}+\tilde{f}^{2}}-1\right)-\frac{3\mathrm{i}k_{\rm{z}}\tilde{f}U''}{\sigma^{2}+\tilde{f}^{2}}-Pe\frac{\sigma(\sigma^{2}+\tilde{f}^{2}+N^{2})}{\sigma^{2}+\tilde{f}^{2}},
\end{equation}
\begin{equation}
\mathcal{V}_{1}=k_{\rm{z}}^{3}\frac{\mathrm{i}\tilde{f}(U'-2f)}{\sigma^{2}+\tilde{f}^{2}}+\frac{2k_{\rm{z}}^{2}fU''}{\sigma^{2}+\tilde{f}^{2}}-\frac{3k_{\rm{z}}\mathrm{i}\tilde{f}U'''}{\sigma^{2}+\tilde{f}^{2}}-Pe\frac{2\mathrm{i}k_{\rm{z}}\tilde{f}f\sigma}{\sigma^{2}+\tilde{f}^{2}},
\end{equation}
\begin{equation}
\begin{aligned}
\mathcal{V}_{0}=-\frac{k_{\rm{z}}^{4}\Gamma}{\sigma^{2}+\tilde{f}^{2}}+\frac{k_{\rm{z}}^{3}\mathrm{i}\tilde{f}U''}{\sigma^{2}+\tilde{f}^{2}}+\frac{k_{\rm{z}}^{2}fU''}{\sigma^{2}+\tilde{f}^{2}}-\frac{k_{\rm{z}}\mathrm{i}\tilde{f}U''''}{\sigma^{2}+\tilde{f}^{2}}-Pe\frac{\sigma k_{\rm{z}}^{2}\Gamma}{\sigma^{2}+\tilde{f}^{2}}.
\end{aligned}
\end{equation}
In Sect.~\ref{sect:WKBJ}, the above single ODEs (\ref{eq:2ODE_v}) and (\ref{eq:4ODE_v}) will be used for asymptotic analyses with the WKBJ approximation for large $k_{\rm{z}}$ to derive asymptotic dispersion relations for the inertial instability. 
    
 \section{General stability results}
 \label{sect:general}
  \begin{figure*}
   \centering
   \includegraphics[height=5.15cm]{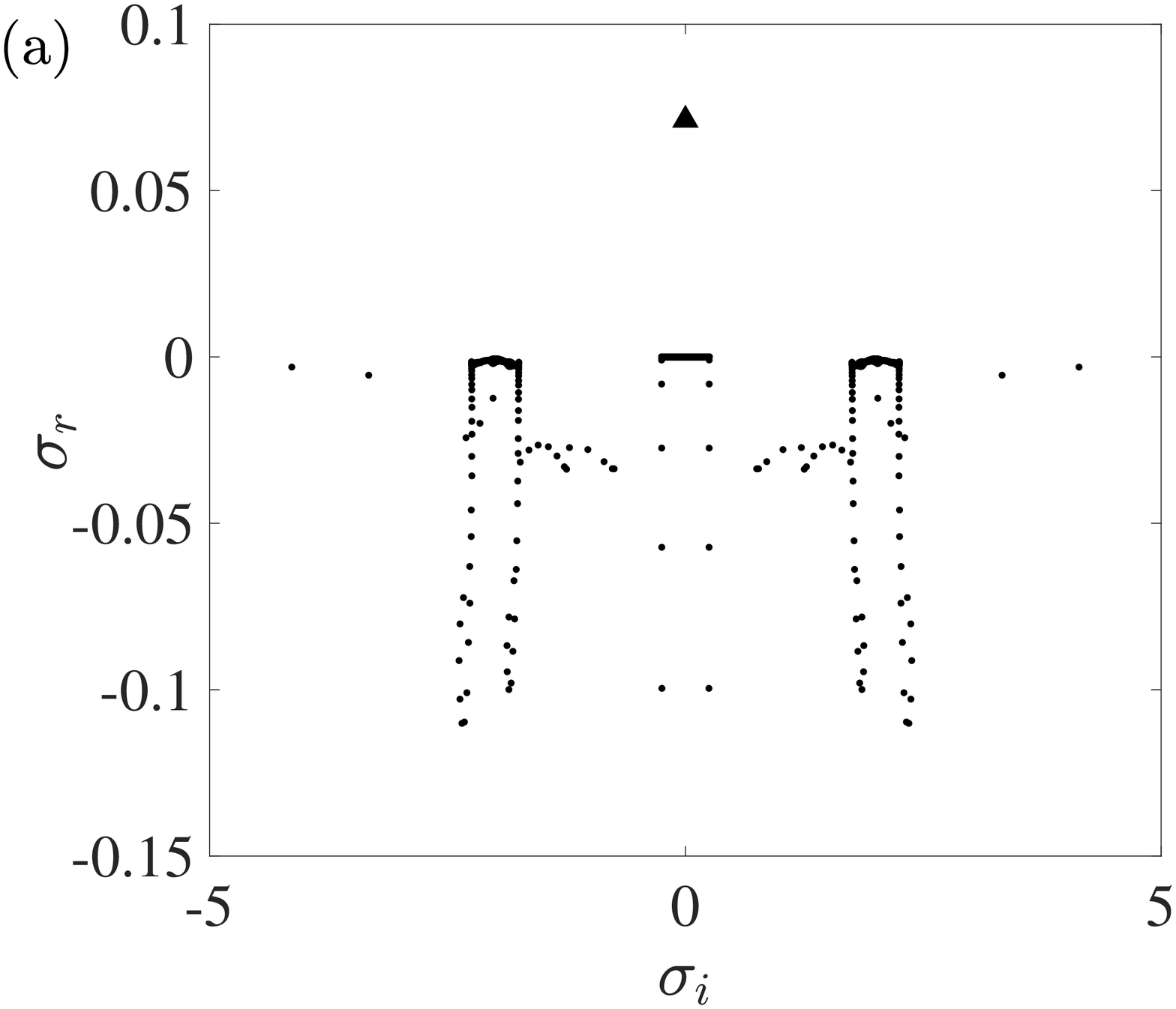}
   \includegraphics[height=5.15cm]{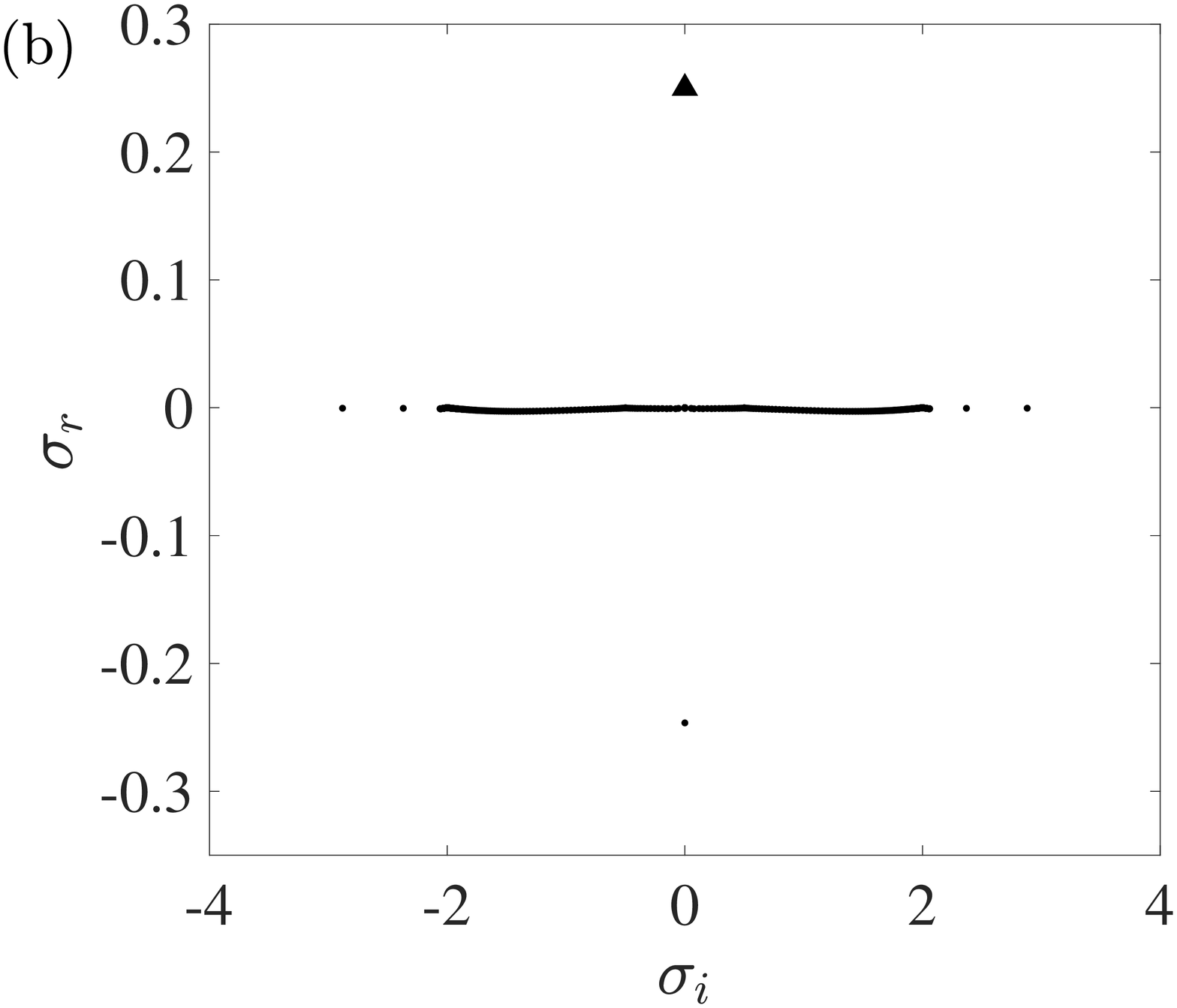}
   \includegraphics[height=5.15cm]{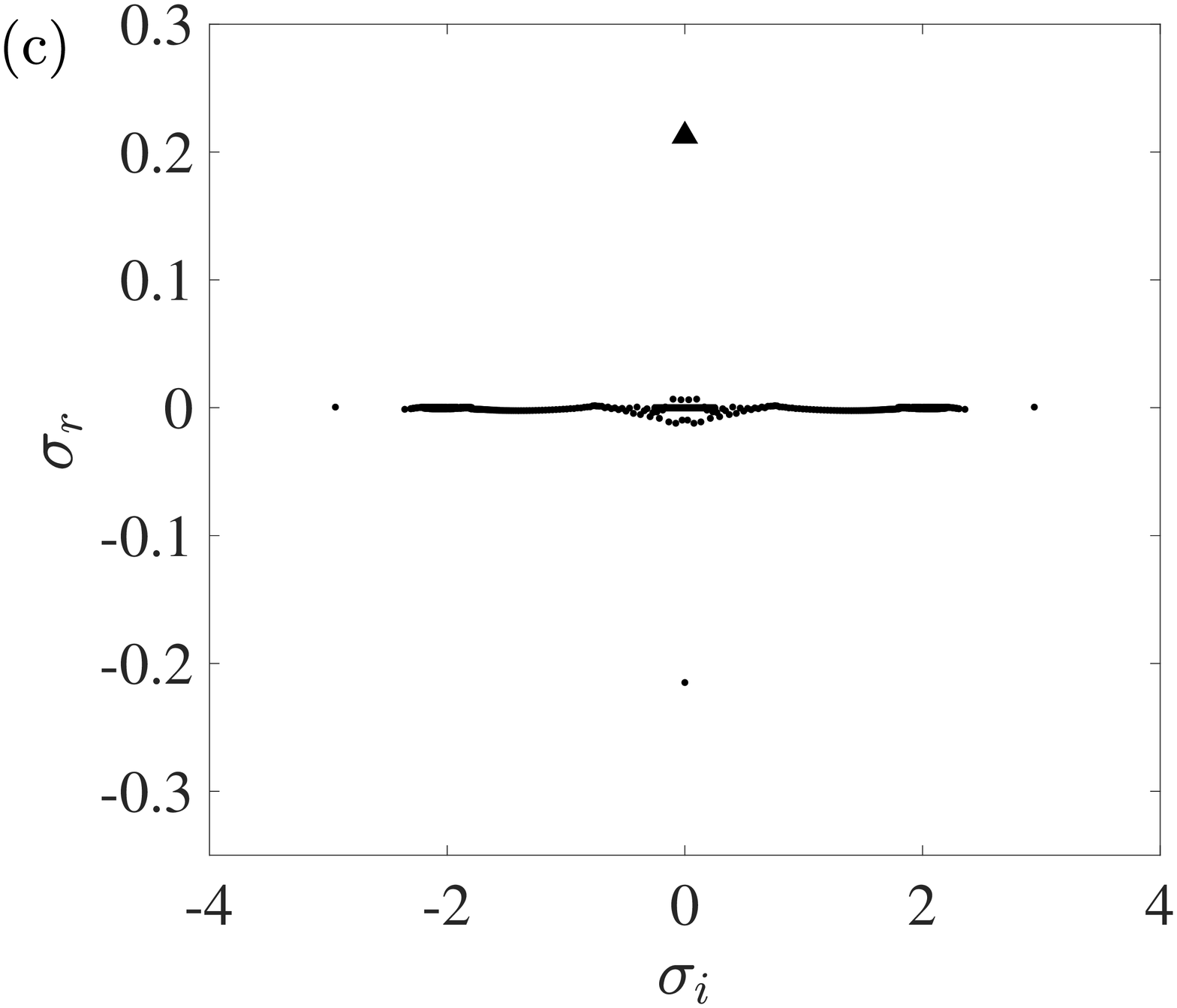}
   \caption{
Eigenvalue spectra in the space $(\sigma_{i},\sigma_{r})$ at $Re=\infty,$ $N=1$, $Pe=1$, $f=0.5$, and $\tilde{f}=2${ (i.e., $(\Omega,\theta)=(1.031,76^{\circ}$))} for (a) $(k_{\rm{x}},k_{\rm{z}})=(0.25,0)$, (b) $(k_{\rm{x}},k_{\rm{z}})=(0,6)$, and (c) $(k_{\rm{x}},k_{\rm{z}})=(0.25,6)$. 
Triangles denote the maximum growth rates. 
   }
              \label{Fig_spectrum}%
    \end{figure*}
  \begin{figure*}
   \centering
   \includegraphics[height=3.7cm]{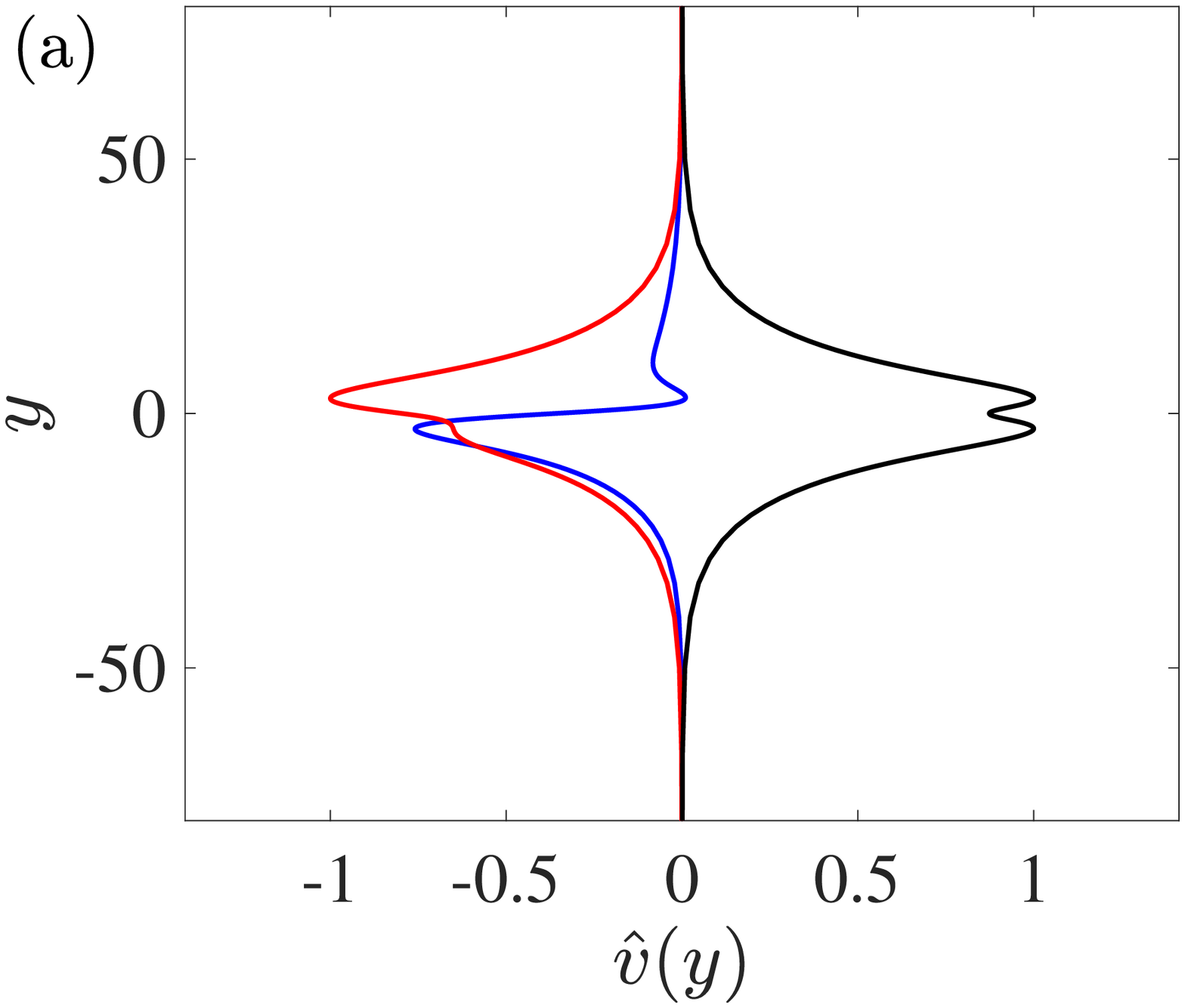}
   \includegraphics[height=3.7cm]{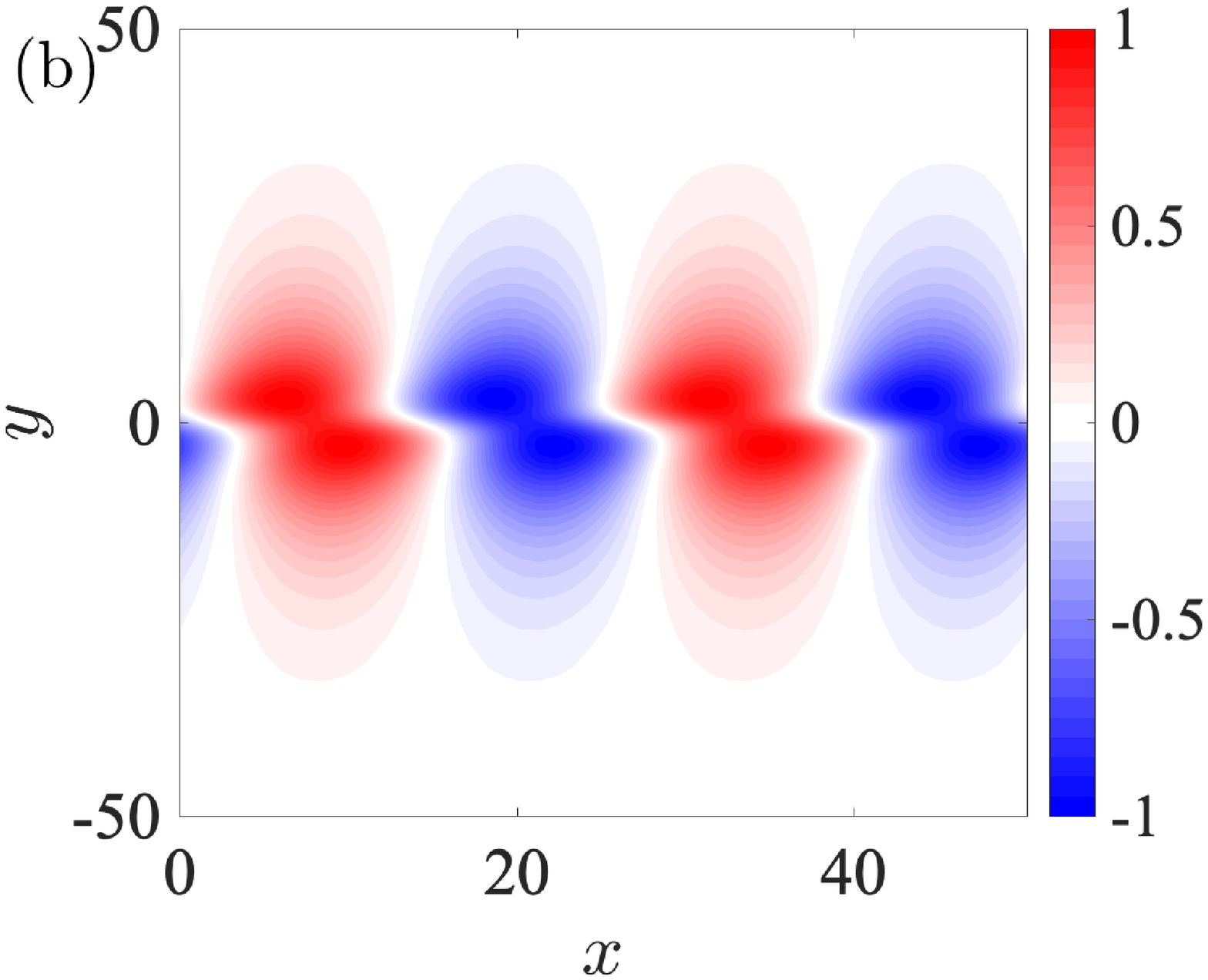}
   \includegraphics[height=3.7cm]{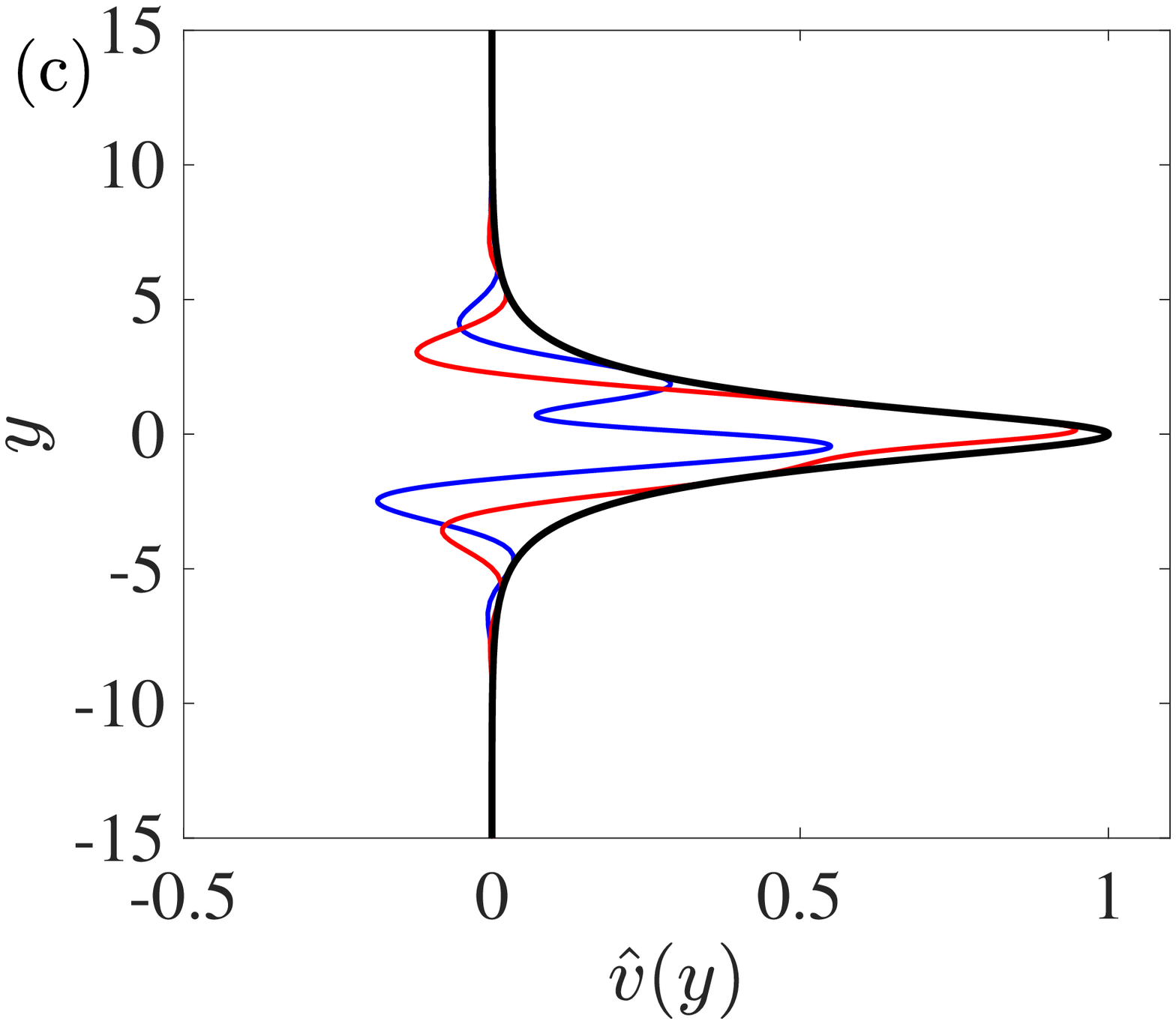}
   \includegraphics[height=3.7cm]{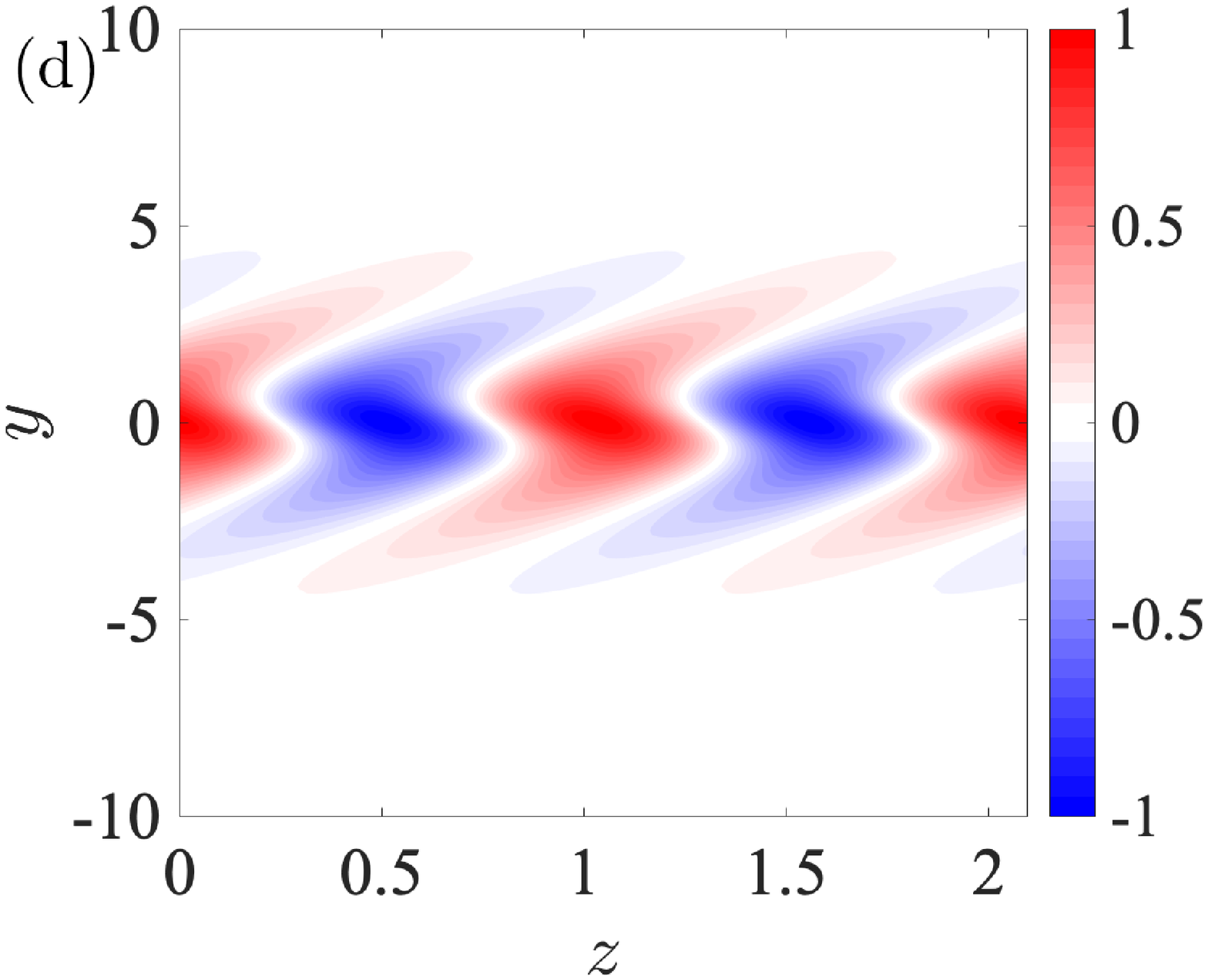}

   \caption{
   Eigenfunctions of the most unstable modes in Fig.~\ref{Fig_spectrum} at $Re=\infty,$ $N=1$, $Pe=1$, $f=0.5$, and $\tilde{f}=2$: (a) the mode shape $\hat{v}(y)$ and (b) velocity $v$ in the physical space $(x,y)$ at $z=0$ for $(k_{\rm{x}},k_{\rm{z}})=(0.25,0)$, and $\sigma=0.0712$, (c) the mode shape $\hat{v}(y)$ and (d) velocity $v$ in the physical space $(y,z)$ at $x=0$ for $(k_{\rm{x}},k_{\rm{z}})=(0,6)$, and $\sigma=0.250$. 
   In (a) and (c), black, blue, and red lines denote the absolute value, real part, and imaginary part of the mode shape $\hat{v}$, respectively.
   }
              \label{Fig_eigenfunctions}%
    \end{figure*}
In this section, we investigate examples of numerical stability results to see how the full Coriolis acceleration modifies the instabilities of the horizontal shear flow.
Figure \ref{Fig_spectrum} shows spectra of the eigenvalue $\sigma$ for various sets of $(k_{\rm{x}},k_{\rm{z}})$ at $Re=\infty,$ $N=1$, and $Pe=1$ for Coriolis components $f=0.5$ and $\tilde{f}=2$ (i.e., $2\Omega\simeq2.062$ and $\theta\simeq76^{\circ}$). 
In the eigenvalue spectra, the real part of the complex growth rate $\sigma_{r}$ determines the stability while the imaginary part $\sigma_{i}$ denotes the frequency of the mode.  
In Fig.~\ref{Fig_spectrum}a for $(k_{\rm{x}},k_{\rm{z}})=(0.25,0)$, we see different types of eigenvalues: widely distributed stable eigenvalues, clusters of neutral eigenvalues, and one unstable eigenvalue. 
The neutral or stable modes are continuous modes, which are not exponentially decreasing with $|y|\rightarrow\infty$, or gravito-inertial waves that can possess critical points \citep[]{Astoul2020}.
However, this paper will only investigate unstable modes that are the most likely to cause turbulence with horizontal shear. 

The eigenfunction of the most unstable mode in Fig.~\ref{Fig_spectrum}a is plotted in Fig.~\ref{Fig_eigenfunctions} in panels a and b.
The mode shape $\hat{v}$ has maxima around $y\simeq\pm2.96$, and the mode displayed in the physical space $(x,y)$ shows that the velocity $v(x,y)$ at $z=0$ has an inclined structure in the direction opposite to the shear.
This feature is a characteristic of the inflectional instability mode as previously studied for stratified fluids in the traditional $f$-plane \citep[]{Arobone2012,PPM2020}.

Figure~\ref{Fig_spectrum}b shows an eigenvalue spectrum that has nearly-neutral clustered eigenvalues, four neutral modes with frequency $|\sigma_{i}|>2$, a stable eigenvalue, and one unstable eigenvalue. 
We note that $f=0.5$ belongs to the inertial instability regime for $k_{\rm{z}}>0$ in the traditional approximation, and the instability is not from the inflectional point of the base flow since $k_{\rm{x}}=0$. 
For the unstable eigenvalue, we plot the corresponding mode in panels c and d of Fig.~\ref{Fig_eigenfunctions}.
We see that the mode shape $\hat{v}$ is oscillatory in the region $|y|<10$, and this wavelike behavior centered around $y=0$ reminds us of the inertial instability mode \citep[]{PPM2020}. 
We also observe clearly in Fig.~\ref{Fig_eigenfunctions}d a characteristic alternating pattern of $v$ when plotted in the physical space $(y,z)$.

In Fig.~\ref{Fig_spectrum}c, we plot the eigenvalue spectra for $(k_{\rm{x}},k_{\rm{z}})=(0.25,6)$. They have weakly unstable or stable clustered eigenvalues due to the critical point where the Doppler-shifted frequency becomes zero (i.e., $s=0$), separate neutral and stable eigenvalues, and one unstable eigenvalue. 
The mode arisen by the critical point will be beyond the scope of this study, and it will be covered by the follow-up work \citep[]{Astoul2020}.
At the wavenumbers $k_{\rm{z}}=6$ and $k_{\rm{x}}=0.25$, the inflectional instability is not present and we only have the inertial instability with the mode shape (not shown) similar to that in Fig.~\ref{Fig_eigenfunctions}c. 
  
  \begin{figure*}
   \centering
   \includegraphics[height=5.1cm]{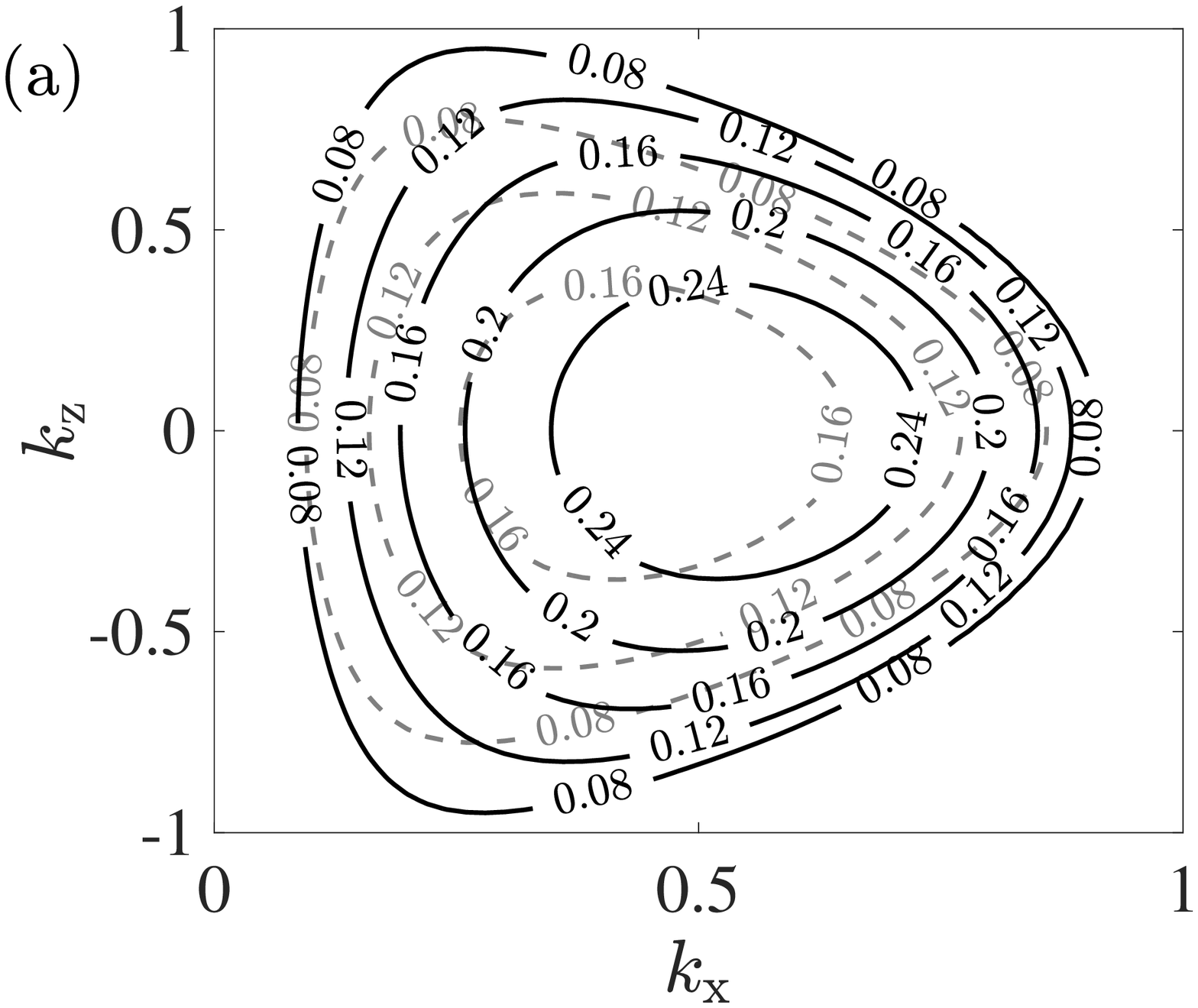}
   \includegraphics[height=5.1cm]{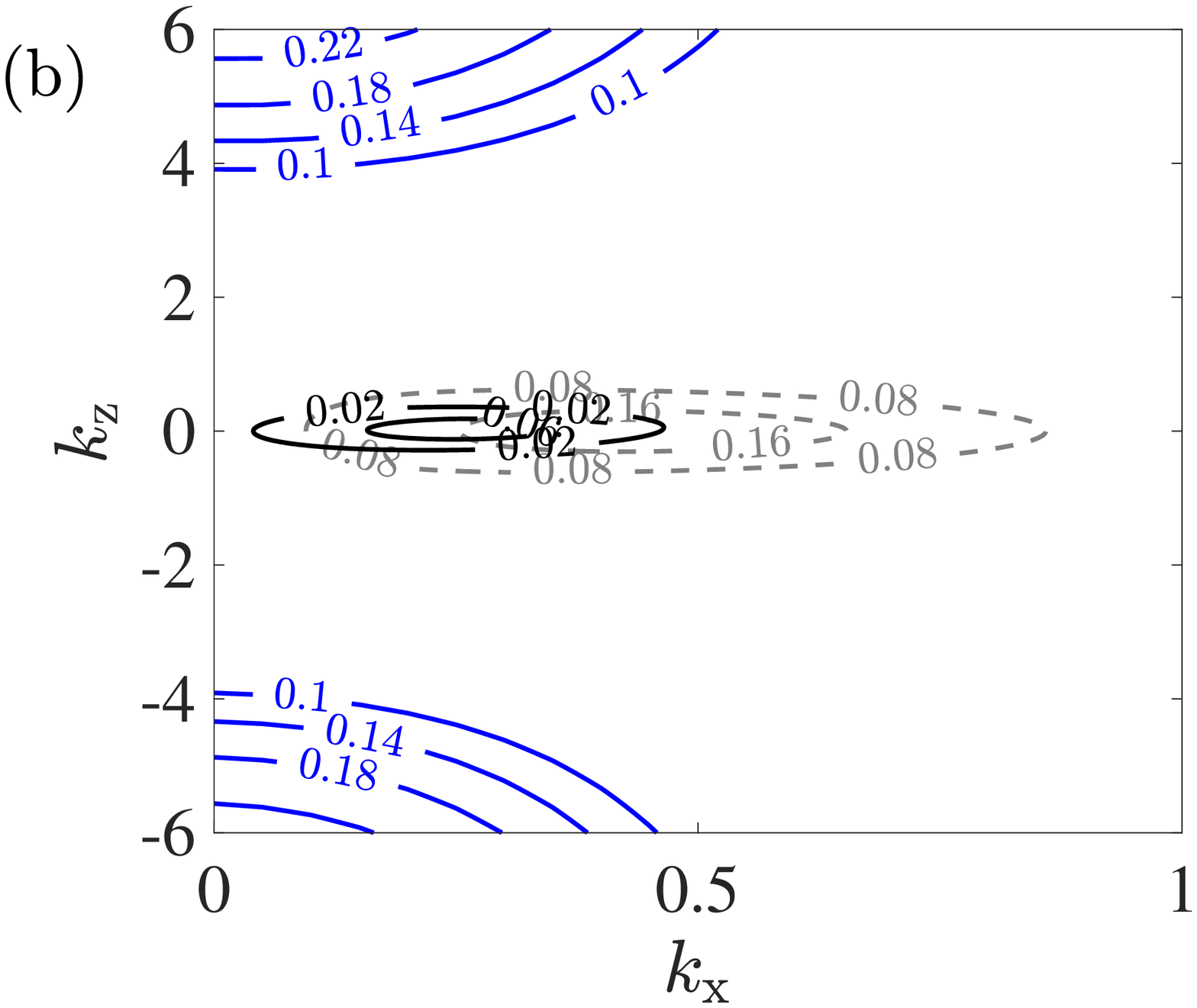}
   \includegraphics[height=5.1cm]{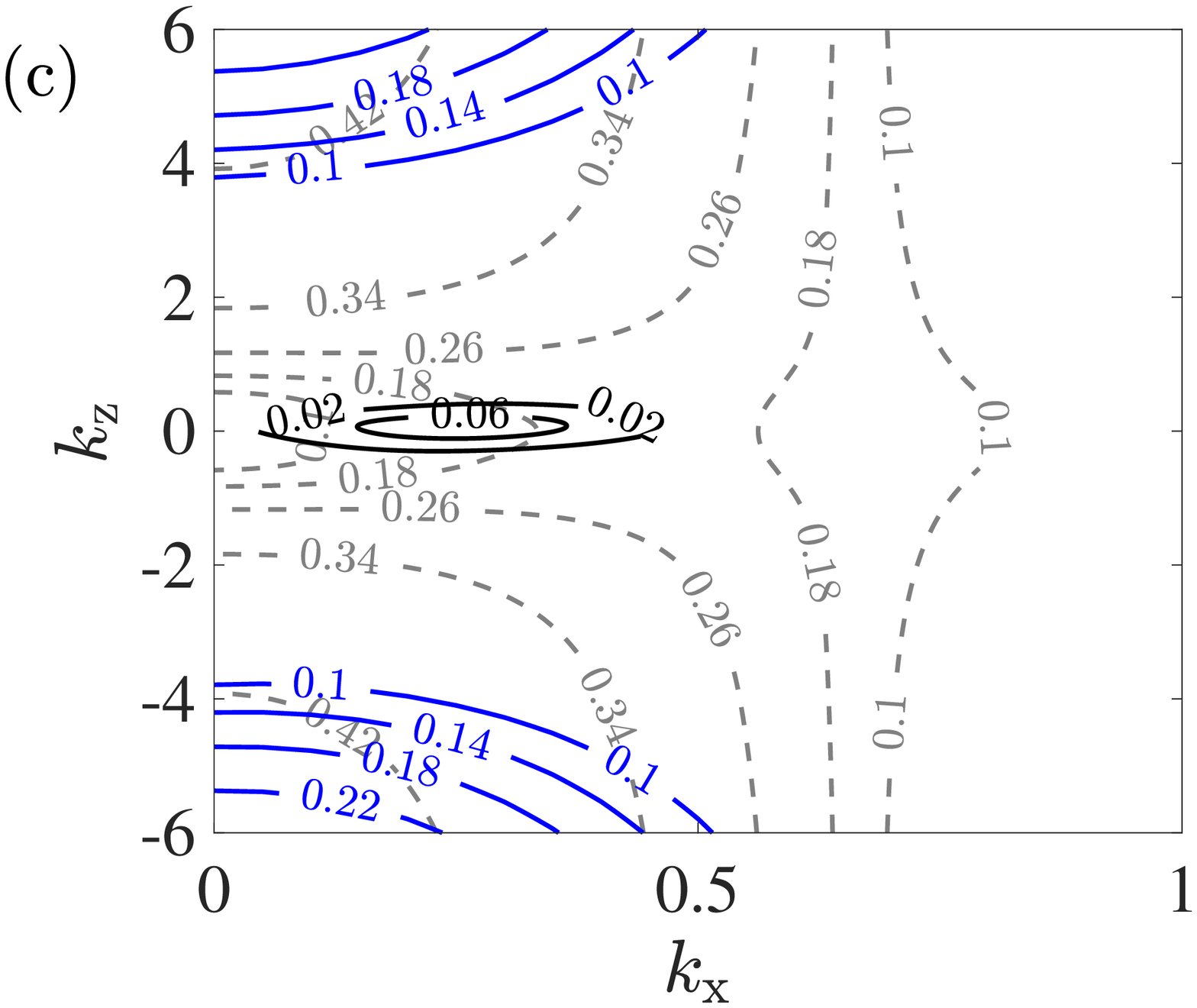}
   \caption{
   Contours of the maximum growth rate $\max(\sigma_{r})$ in the parameter space of $(k_{\rm{x}},k_{\rm{z}})$ (solid lines) for $Re=\infty$ and $N=1$ for (a) $(f,\tilde{f},Pe)=(0,2,\infty)$, (b) $(f,\tilde{f},Pe)=(0,2,1)$, and (c) $(f,\tilde{f},Pe)=(0.5,2,1)$. {For (a) and (b), this corresponds to $(\Omega,\theta)=(1,90^{\circ})$, and $(\Omega,\theta)=(1.031,76^{\circ}$) for (c)}. 
   Black and blue solid lines are contours of the maximum growth rate of the inflectional and inertial instabilities, respectively, and gray dashed lines in the background are contours of the maximum growth rate for the same parameters but in the traditional $f$-plane approximation where $\tilde{f}=0$.  
   }
              \label{Fig_contours_growth_preliminary}%
    \end{figure*}
Figure~\ref{Fig_contours_growth_preliminary} shows contours of the growth rate $\sigma_{r}$ for the most unstable mode in the parameter space $(k_{\rm{x}},k_{\rm{z}})$ for different sets of parameters $(f,\tilde{f},Pe)$ at $N=1$ and $Re=\infty$. 
We found that the most unstable modes have zero temporal frequency $\sigma_{i}$; therefore, we plot only the real part of the maximum growth rate $\sigma_{r}$. 
In Fig.~\ref{Fig_contours_growth_preliminary}a, we consider the non-diffusive limit $Pe=\infty$ in an inertially-stable regime at $f=0$ according to the criterion in the traditional approximation \citep[]{Arobone2012}.
If we consider $\tilde{f}=0$ (i.e., $2\Omega=0$), only the inflectional instability exists and its maximum growth rate $\sigma_{\max}=0.1897$ is attained at $k_{\rm{x}}=0.445$ and $k_{\rm{z}}=0$ \citep[see also,][]{Deloncle2007}.
As $\tilde{f}$ increases (i.e., on the equator for rotating stars with $2\Omega>0$), we see that the regime of the inflectional instability is widened in the parameter space $(k_{\rm{x}},k_{\rm{z}})$, and the higher maximum growth rate $\sigma_{\max}\simeq0.277$ is attained at $k_{\rm{z}}=0$ and at a higher value of $k_{\rm{x}}\simeq0.54$.

In Fig.~\ref{Fig_contours_growth_preliminary}b, we now consider a finite P\'eclet number $Pe=1$.
At $f=\tilde{f}=0$ (i.e., $2\Omega=0$), the instability is stabilized as the thermal diffusivity increases \citep[see also,][]{PPM2020} and the unstable regime in the parameter space $(k_{\rm{x}},k_{\rm{z}})$ is slightly shrunk compared to that at $Pe=\infty$ in Fig.~\ref{Fig_contours_growth_preliminary}a. 
The maximum growth rate of the inflectional instability at $Pe=1$ and $\tilde{f}=0$ still remains as $\sigma_{\max}=0.1897$ at $k_{\rm{x}}=0.445$.
As $\tilde{f}$ increases, we see that the inflectionally-unstable regime at $\tilde{f}=2$ is even more shrunk than the unstable regime at $\tilde{f}=0$. 
The maximum growth rate is decreased to $\sigma_{\max}\simeq0.0717$ and is found at $(k_{\rm{x}},k_{\rm{z}})\simeq(0.26,0.02)$. 
This implies that a slightly three-dimensional inflectional instability is now more unstable than the two-dimensional inflectional instability at finite $Pe$ and $\tilde{f}>0$. 
What is also very interesting in Fig.~\ref{Fig_contours_growth_preliminary}b is that we observe other unstable regimes.
The growth rate contours for large $|k_{\rm{z}}|$ are reminiscent of those for the inertial instability, which has a maximum growth rate as $|k_{\rm{z}}|\rightarrow\infty$ at $k_{\rm{x}}=0$. 
It is remarkable to observe this inertial instability at $f=0$ (i.e. in the inertially stable regime in the traditional approximation) {when} the horizontal Coriolis component $\tilde{f}$ becomes positive  {and the fluid is thermally diffusive}.

We now consider in Fig.~\ref{Fig_contours_growth_preliminary}c the case $f=0.5$, which belongs to the inertially unstable regime in the traditional $f$-plane. 
We see a clear difference between growth-rate contours at $\tilde{f}=0$ (i.e., on the pole with $2\Omega=0.5$) and those at $\tilde{f}=2$ (i.e., $2\Omega\simeq2.062$ and $\theta\simeq76^{\circ}$). 
While the growth rate smoothly increases as $|k_{\rm{z}}|$ increases in the traditional case at $\tilde{f}=0$, the regime of the inflectional instability is well separated from the regime of the inertial instability for $\tilde{f}=2$.
The inflectional instability at $\tilde{f}=2$ has a maximum growth rate $\sigma_{\max}\simeq0.0725$ around $(k_{\rm{x}},k_{\rm{z}})\simeq(0.26,0.07)$ while the inertial instability has a maximum growth rate as $|k_{\rm{z}}|\rightarrow\infty$ at $k_{\rm{x}}=0$.

%
   \begin{figure}
   \centering
   \includegraphics[height=5.5cm]{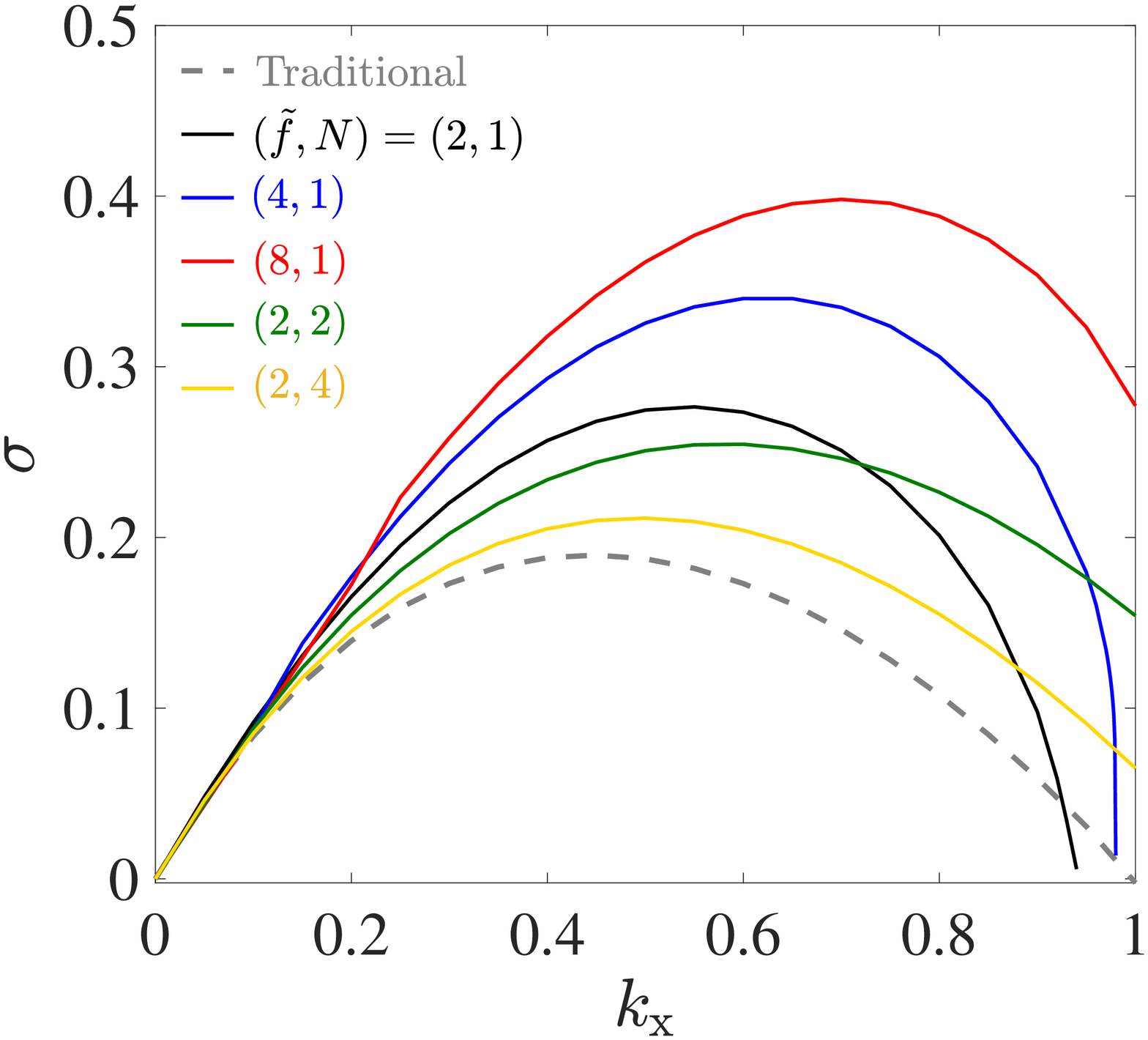}
      \caption{Inviscid growth rate of the inflectional instability for various parameter sets of $\tilde{f}$ and $N$ at $f=0$ {(i.e. $\theta=90^{\circ}$)}, $Pe=\infty$, and $k_{\rm{z}}=0$. 
              }
         \label{Fig_growth_inflectional}
   \end{figure}
In the traditional $f$-plane approximation, the inflectional instability reaches its maximum growth rate $\sigma_{\max}\simeq0.1897$ in the inviscid limit $Re=\infty$ at $k_{\rm{x}}\simeq0.445$ and $k_{\rm{z}}=0$ (also shown by the gray dashed line in Fig.~\ref{Fig_growth_inflectional}).
In this case, the maximum growth rate $\sigma_{\max}$ is independent from $f$ and $N$ \citep[]{PPM2020}.
But without the traditional approximation, the growth rate of the inflectional instability now becomes dependent of $N$ and $\tilde{f}$ as shown in Fig.~\ref{Fig_growth_inflectional}  at $Pe=\infty$ and $k_{\rm{z}}=0$. 
We verified numerically that the most unstable mode is reached for a finite $k_{\rm{x}}$ at $k_{\rm{z}}=0$ in the parameter space of $(k_{\rm{x}},k_{\rm{z}})$ in the inviscid and non-diffusive limits (i.e., $Re=Pe=\infty$). 
In these limits, we have the following 2nd-order ODE for $\hat{v}$
\begin{equation}
\label{eq:2ODE_v_kz0}
\begin{aligned}
	&\frac{\mathrm{d}^{2}\hat{v}}{\mathrm{d}y^{2}}+\left(\frac{2sk_{\rm{x}} U'}{s^{2}-N^{2}-\tilde{f}^{2}}-\frac{2sk_{\rm{x}} U'}{s^{2}-N^{2}}\right)\frac{\mathrm{d}\hat{v}}{\mathrm{d} y}\\
	&-\left(k_{\rm{x}}^{2}+\frac{k_{\rm{x}} U''}{s}+\frac{2k_{\rm{x}}^{2}U^{'2}}{s^{2}-N^{2}-\tilde{f}^{2}}-\frac{2k_{\rm{x}}^{2}U^{'2}}{s^{2}-N^{2}}\right)\hat{v}=0.
\end{aligned}
\end{equation}
We clearly see that Eq.~(\ref{eq:2ODE_v_kz0}) is still independent from $f$ but it depends on $N$ if $\tilde{f}>0$. 
For a fixed $N$, we see in Fig.~\ref{Fig_growth_inflectional} that the maximum growth rate increases with $\tilde{f}$, while it decreases with $N$ at a fixed $\tilde{f}$.
This implies that the stratification stabilizes the inflectional instability, as similarly observed in the traditional approximation \citep[]{PPM2020}.
While the wavenumber range of the inflectional instability is $0<k_{\rm{x}}<1$ at $\tilde{f}=0$, it is noticeable that the instability can sustain for $k_{\rm{x}}>1$ as $\tilde{f}$ increases.

%
   \begin{figure}
   \centering
   \includegraphics[height=5.5cm]{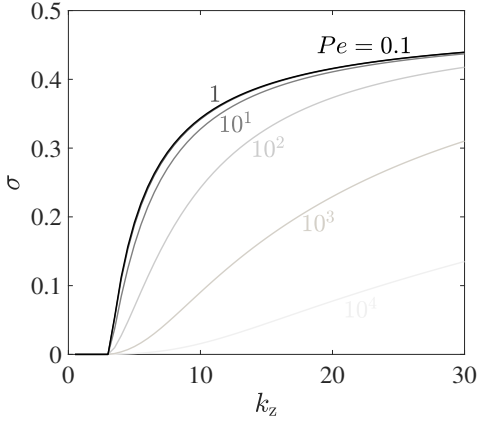}
      \caption{Inviscid growth rate of the inertial instability for various P\'eclet numbers at $N=1$ {and $k_{\rm{x}}=0$ for $f=0$ and $\tilde{f}=2$, i.e. $(\Omega,\theta)=(1,90^{\circ})$}. 
              }
         \label{Fig_growth_inertial_preliminary}
   \end{figure}
Figure \ref{Fig_growth_inertial_preliminary} shows growth-rate curves for various values of $Pe$ at $N=1$, $f=0$, $k_{\rm{x}}=0$, and $\tilde{f}=2$ (i.e., on the equator with $2\Omega=2$). 
Although $f=0$ implies that it is inertially stable rin the traditional approximation, it is still unstable for finite $Pe$ at $\tilde{f}=2$ and the growth rate increase as $Pe$ increases. 
At a fixed $Pe$, the growth rate increases with $k_{\rm{z}}$ and the maximum growth rate is reached as $k_{\rm{z}}\rightarrow\infty$.
Furthermore, the growth-rate curves converge to a single curve as $Pe\rightarrow0$.

Preliminary results presented in this section tell us that non-traditional effects can significantly modify the properties of the inflectional and inertial instabilities.
For instance, the unstable regime of the inertial instability is changed as $\tilde{f}$ increases. 
The maximum growth rate of the inflectional instability increases with $\tilde{f}$ and depends on $N$. 
Similarly to the traditional case at $\tilde{f}=0$, thermal diffusion destabilizes the inertial instability when $\tilde{f}>0$. 
In the following section, we discuss more thoroughly how the inertial instability depends on physical parameters such as $f$, $\tilde{f}$, $N$, or $Pe$ using the WKBJ approximation by taking the asymptotic limit $k_{\rm{z}}\rightarrow\infty$ for the inviscid case at $Re=\infty$.
We derive analytical expressions of the dispersion relations for the inertial instability in the non-diffusive ($Pe=\infty$) and highly-diffusive ($Pe\rightarrow0$) cases to understand how the inertial instability is modified as the horizontal Coriolis parameter $\tilde{f}$ increases.

\section{Asymptotic description of the inertial instability}
\label{sect:WKBJ}
Our previous work in \citet{PPM2020} with the traditional approximation ($\tilde{f}=0$) was successful to perform detailed analyses on the inertial instability employing the WKBJ approximation in the inviscid limit $Re=\infty$. 
For instance, asymptotic dispersion relations were explicitly proposed for large $k_{\rm{z}}$ in two limits: $Pe\rightarrow0$ and $Pe\rightarrow\infty$. 
From the dispersion relations, we found that the unstable regime of the inertial instability is $0<f<1$ and the maximum growth rate $\sigma_{\max}=\sqrt{f(1-f)}$ is reached as $k_{\rm{z}}\rightarrow\infty$. 
{In the traditional approximation, $\sigma_{\max}$ is related to the epicyclic frequency $\omega_{\rm ep}$, which satisfies $\sigma_{\max}^{2}=-\omega^{2}_{\rm ep}$ and leads to the Solberg-H\o iland criterion for stability when $\omega^{2}_{\rm ep}+N^{2}>0$ \citep[][]{Solberg1936,Hoiland1941}. 
Besides, $\sigma_{\max}$ is analogous to the growth rate suggested in the limit $Pr\rightarrow0$ for the Goldreich-Schubert-Fricke (GSF) instability \citep[][]{Goldreich1967,Fricke1968} induced by the {horizontal/vertical} shear \citep[for more details, we refer the reader to][]{Knobloch1982,Maeder2013,Barker2019}.}
Besides, the maximum growth rate $\sigma_{\max}$ is attained independently of the stratification and thermal diffusivity (i.e., $N$ and $Pe$) while the first-order term of the growth rate strongly depends on them. 
However, this argument is not applicable without the traditional approximation as demonstrated by numerical results in the previous section. 
Thus, it is imperative to investigate the inertial instability in the non-traditional case to see how the properties of the instability are changed as the horizontal Coriolis parameter $\tilde{f}$ increases. 

In the following subsections, we perform the WKBJ analysis for large $k_{\rm{z}}$ at $k_{\rm{x}}=0$ in the two limits of the P\'eclet number $Pe$: the non-diffusive case as $Pe\rightarrow\infty$ and the higly-diffusive case as $Pe\rightarrow0$.
This analysis is an extension of the work by \citet{PPM2020} in the traditional $f$-plane approximation.  
For the two cases, we describe below the properties of the inertial instability such as dispersion relations, the maximum growth rate, or regimes of the instability in the parameter space.

\subsection{WKBJ formulation in the non-diffusive limit $Pe\rightarrow\infty$} 
In this subsection, the thermally non-diffusive case with $Pe\rightarrow\infty$ is considered. 
We verified numerically that the inviscid maximum growth rate of the inertial instability is found as $|k_{\rm{z}}|\rightarrow\infty$ at $k_{\rm{x}}=0$, and it has a zero temporal frequency (i.e., $\sigma_{i}=0$).  
For simplicity, we focus on this most unstable case at $k_{\rm{x}}=0$ and consider only $k_{\rm{z}}>0$ due to the symmetry $\sigma(0,k_{\rm{z}})=\sigma(0,-k_{\rm{z}})$ at $k_{\rm{x}}=0$.
The 2nd-order ODE for $\hat{v}$ in Eq.~(\ref{eq:2ODE_v}) at $k_{\rm{x}}=0$ becomes
\begin{equation}
\label{eq:WKBJ_2ODE_v}
\frac{\mathrm{d}^{2}\hat{v}}{\mathrm{d}y^{2}}+\frac{2\mathrm{i}k_{\rm{z}}\tilde{f}f}{\sigma^{2}+N^{2}+\tilde{f}^{2}}\frac{\mathrm{d}\hat{v}}{\mathrm{d}y}+k_{\rm{z}}^{2}\frac{\Gamma}{\sigma^{2}+N^{2}+\tilde{f}^{2}}\hat{v}=0.
\end{equation}
We note that the term $2\mathrm{i}k_{\rm{z}}\tilde{f}f/(\sigma^{2}+N^{2}+\tilde{f}^{2})$ multiplied by the first derivative $\mathrm{d}\hat{v}/\mathrm{d}y$ has the order of $k_{\rm{z}}$ and is comparable with all the other terms when the WKBJ approximation is applied for large $k_{\rm{z}}$. 
For convenience, we introduce a new variable $\hat{V}$ prior to the WKBJ analysis
\begin{equation}
\hat{V}(y)=\hat{v}(y)\exp\left(\frac{\mathrm{i}k_{\rm{z}}\tilde{f}f}{\sigma^{2}+N^{2}+\tilde{f}^{2}}y\right).
\end{equation}
Then Eq. (\ref{eq:WKBJ_2ODE_v}) becomes
\begin{equation}
\label{eq:WKBJ_2ODE_V}
\frac{\mathrm{d}^{2}\hat{V}}{\mathrm{d}y^{2}}+\frac{k_{\rm{z}}^{2}}{\sigma^{2}+N^{2}+\tilde{f}^{2}}\left[\Gamma+\frac{\tilde{f}^{2}f^{2}}{\sigma^{2}+N^{2}+\tilde{f}^{2}}\right]\hat{V}=0,
\end{equation}
which is now in the form of the Poincar\'e equation.
Applying to Eq.~(\ref{eq:WKBJ_2ODE_V}) the WKBJ approximation of $\hat{V}$ for large $k_{\rm{z}}$:
\begin{equation}
\label{eq:WKBJ}
\hat{V}(y)\sim\exp\left[\frac{1}{\delta}\sum_{l=0}^{\infty}\delta^{l}S_{l}(y)\right],
\end{equation}
we get
\begin{equation}
	\delta=\frac{1}{k_{\rm{z}}},~~
	S_{0}^{'2}=-\frac{\tilde{\Gamma}}{{\sigma^{2}+N^{2}+\tilde{f}^{2}}},~~
	S'_{1}=-\frac{S''_{0}}{2S'_{0}},
\end{equation}
where 
\begin{equation}
\label{eq:tilde_Gamma}
	\tilde{\Gamma}\equiv\Gamma+\frac{\tilde{f}^{2}f^{2}}{\sigma^{2}+N^{2}+\tilde{f}^{2}}.
\end{equation}
Since $\sigma^{2}+N^{2}+\tilde{f}^{2}>0$, the exponential behavior of $\hat{V}$ determined by the sign of $S_{0}$ depends on the sign of $\tilde{\Gamma}$.
On the one hand, the solution is evanescent if $\tilde{\Gamma}<0$:
\begin{equation}
\begin{aligned}
\label{eq:WKBJ_2ODE_exponential}
	\hat{V}(y)=&(-\tilde{\Gamma})^{-\frac{1}{4}}\\
	&\left[A_{1}\exp\left(\tilde{k}\int_{y}\sqrt{-\tilde{\Gamma}}\mathrm{d}\upsilon\right)+A_{2}\exp\left(-\tilde{k}\int_{y}\sqrt{-\tilde{\Gamma}}\mathrm{d}\upsilon\right)\right],
\end{aligned}
\end{equation}
where $A_{1}$ and $A_{2}$ are constants, and $\tilde{k}=k_{\rm{z}}/\sqrt{\sigma^{2}+N^{2}+\tilde{f}^{2}}$. 
On the other hand, the solution is wavelike if $\tilde{\Gamma}>0$:
\begin{equation}
\begin{aligned}
\label{eq:WKBJ_2ODE_wavelike}
	\hat{V}(y)=\tilde{\Gamma}^{-\frac{1}{4}}\left[B_{1}\exp\left(\mathrm{i}\tilde{k}\int_{y}\sqrt{\tilde{\Gamma}}\mathrm{d}\upsilon\right)+B_{2}\exp\left(-\mathrm{i}\tilde{k}\int_{y}\sqrt{\tilde{\Gamma}}\mathrm{d}\upsilon\right)\right],\end{aligned}
\end{equation}
where $B_{1}$ and $B_{2}$ are constants. 

%
   \begin{figure}
   \centering
   \includegraphics[height=5.5cm]{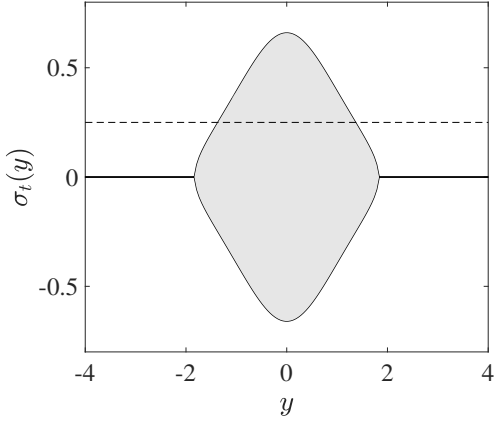}
      \caption{Function $\sigma_{t}(y)$ (solid lines) at {$N=1$ for $(f,\tilde{f})=(0.5,2)$ (i.e. $\Omega=1.031$ and $\theta=76^{\circ}$)}. Grey and white areas denote the regimes where $\tilde{\Gamma}$ is positive and negative, respectively. The dashed line represents an example of the growth rate $\sigma=0.250$. 
              }
         \label{Fig_sigma_t}
   \end{figure}
The wavelike or evanescent solution behavior changes at turning points $y_{t}$ where $\tilde{\Gamma}(y_{t})=0$. 
To find $y_{t}$, it is useful to define the function $\sigma_{t}(y)$ such that
\begin{equation}
\label{eq:def_turning_growthrate}
\begin{aligned}
	2\sigma_{t}^{2}(y)&=f(U'-f)-\left(N^{2}+\tilde{f}^{2}\right)\\
	&+\sqrt{\left(f(U'-f)+N^{2}+\tilde{f}^{2}\right)^{2}+4\tilde{f}^{2}f^{2}},
\end{aligned}
\end{equation}
\citep[see also,][]{Park2012JFM,Park2013JFM,Park2017JFM}.
By finding $\sigma_{t}(y)=\sigma$ at a given $\sigma$, we can find turning points where $\tilde{\Gamma}(y)=0$. 
An example of $\sigma_{t}(y)$ is shown in Fig.~\ref{Fig_sigma_t} for $f=0.5$, $\tilde{f}=2$, and $N=1$. 
The gray area denotes where the WKBJ solution is wavelike (i.e., $\tilde{\Gamma}>0$) while the white area denotes where the solution is evanescent (i.e., $\tilde{\Gamma}<0$).
At a given instability growth rate such that $|\sigma|<\sigma_{\rm t, max}$, there exist two turning points, one for $y>0$ and the other for $y<0$. 
For instance, if the growth rate is $\sigma=0.250$, there exist two turning points: one at $y_{t+}=1.365$ and the other at $y_{t-}=-1.365$. 
In this case, we can construct an eigenfunction such that the solution $\hat{V}$ decays exponentially as $|y|\rightarrow\infty$ while it is wavelike between the two turning points $y_{t\pm}$. 
If the growth rate is greater than the maximum of $\sigma_{t}$ (i.e., $|\sigma|>|\sigma_{\rm t,max}|$), the solution $\hat{V}$ is evanescent everywhere with no turning point and we cannot construct an eigenfunction $\hat{V}$ satisfying the decaying boundary conditions as $|y|\rightarrow\infty$. 
Therefore, to construct the eigenfunction, the growth rate has to lie in the range $0<|\sigma|<\sigma_{t,\max}$ where 
\begin{equation}
\label{eq:sigma_t_max}
2\sigma_{\rm t,\max}^{2}=f(1-f)-N^{2}-\tilde{f}^{2}+\sqrt{[f(1-f)+N^{2}+\tilde{f}^{2}]^{2}+4\tilde{f}^{2}f^{2}}.
\end{equation}

By performing a turning point analysis in the growth rate range $0<\sigma_{r}<\sigma_{\rm t,\max}$, we can further derive an asymptotic dispersion relation as done in \citet{PPM2020}. 
We consider first the evanescent WKBJ solution outside the turning point $y>y_{t+}$:
\begin{equation}
\label{eq:WKBJ_solution_ytplus}
	\hat{V}(y)=A_{\infty}(-\tilde{\Gamma})^{-\frac{1}{4}}\exp\left(-\tilde{k}\int_{y_{t+}}^{y}\sqrt{-\tilde{\Gamma}(\upsilon)}\mathrm{d}\upsilon\right),
\end{equation}
where $A_{\infty}$ is a constant. 
As $y$ approaches the turning point $y_{t+}$, the WKBJ solution (\ref{eq:WKBJ_solution_ytplus}) is not valid and a local solution is needed to find the WKBJ solution below the turning point $y<y_{t+}$. 
We use a new scaled coordinate $\tilde{y}=(y-y_{t+})/\epsilon$ where $\epsilon$ is a small parameter defined as $\epsilon=\left[\tilde{k}^{2}(-\tilde{\Gamma}'_{t+})\right]^{-1/3}$, $\tilde{\Gamma}'_{t+}$ is the derivative of $\tilde{\Gamma}$ at $y_{t+}$, and we assume at leading order that $\tilde{\Gamma}(y)\sim\tilde{\Gamma}'_{t+}\epsilon\tilde{y}$ around the turning point $y_{t+}$. 
The following local equation is obtained
\begin{equation}
\label{eq:local_equation_ytplus}
\frac{\mathrm{d}^{2}\hat{V}}{\mathrm{d}\tilde{y}^{2}}-\tilde{y}\hat{V}=O(\epsilon).
\end{equation}
Solutions of the local equation (\ref{eq:local_equation_ytplus}) are the Airy functions: $\hat{V}(\tilde{y})=a_{1}\mathrm{Ai}(\tilde{y})+b_{1}\mathrm{Bi}(\tilde{y})$, where $a_{1}$ and $b_{1}$ are constants \citep[]{Abramowitz}. 
From the asymptotic behavior of the Airy functions as $\tilde{y}\rightarrow\pm\infty$ and of the WKBJ solution as $y\rightarrow y_{t+}$, we obtain the matched WKBJ solution in the region $y_{t-}<y<y_{t+}$:
\begin{equation}
\begin{aligned}
\label{eq:WKBJ_2ODE_leftytplus}
	\hat{V}(y)=\tilde{\Gamma}^{-\frac{1}{4}}&\left[C_{+}\exp\left(\mathrm{i}\tilde{k}\int_{y}^{y_{t+}}\sqrt{\tilde{\Gamma}}\mathrm{d}\upsilon\right)+C_{-}\exp\left(-\mathrm{i}\tilde{k}\int^{y_{t+}}_{y}\sqrt{\tilde{\Gamma}}\mathrm{d}\upsilon\right)\right],
\end{aligned}
\end{equation}
where constants $C_{\pm}$ satisfy
\begin{equation}
\label{eq:WKBJ_amplitudes}
	C_{+}=\exp\left(-\mathrm{i}\frac{\pi}{4}\right)A_{\infty}~~\mathrm{and}~~
	C_{-}=\exp\left(\mathrm{i}\frac{\pi}{4}\right)A_{\infty}.
\end{equation}
Similarly, we consider the WKBJ solution below the lower turning point $y<y_{t-}$ decaying exponentially as $y\rightarrow-\infty$:
\begin{equation}
\label{eq:WKBJ_solution_ytminus}
	\hat{V}(y)=A_{-\infty}(-\tilde{\Gamma})^{-\frac{1}{4}}\exp\left(-\tilde{k}\int^{y_{t-}}_{y}\sqrt{-\tilde{\Gamma}(\upsilon)}\mathrm{d}\upsilon\right).
\end{equation}
After the local analysis around $y_{t-}$, we find the matched wavelike solution in the range $y_{t-}<y<y_{t+}$:
\begin{equation}
\begin{aligned}
\label{eq:WKBJ_2ODE_rightytminus}
	\hat{V}(y)=\tilde{\Gamma}^{-\frac{1}{4}}&\left[B_{+}\exp\left(\mathrm{i}\tilde{k}\int_{y_{t-}}^{y}\sqrt{\tilde{\Gamma}}\mathrm{d}\upsilon\right)+B_{-}\exp\left(-\mathrm{i}\tilde{k}\int_{y_{t-}}^{y}\sqrt{\tilde{\Gamma}}\mathrm{d}\upsilon\right)\right],
\end{aligned}
\end{equation}
where constants $B_{\pm}$ satisfy
\begin{equation}
\label{eq:WKBJ_amplitudes_B}
	B_{+}=\exp\left(-\mathrm{i}\frac{\pi}{4}\right)A_{-\infty}~~\mathrm{and}~~
	B_{-}=\exp\left(\mathrm{i}\frac{\pi}{4}\right)A_{-\infty}.
\end{equation}
Finally, matching the wavelike solutions (\ref{eq:WKBJ_2ODE_leftytplus}) and (\ref{eq:WKBJ_2ODE_rightytminus}) between two turning points leads to the following dispersion relation in the quantized form
\begin{equation}
\label{eq:WKBJ_dispersion_quantized}
	\tilde{k}\int_{y_{t-}}^{y_{t+}}\sqrt{\Gamma+\frac{\tilde{f}^{2}f^{2}}{\sigma^{2}+N^{2}+\tilde{f}^{2}}}\mathrm{d}y=\left(m-\frac{1}{2}\right)\pi,
\end{equation}
where $m$ is the positive integer denoting the mode number. 
This quantized dispersion relation becomes equivalent to that in the traditional approximation as $\tilde{f}$ becomes zero \citep[]{PPM2020}.

\begin{figure*}
   \centering
   \includegraphics[height=5.2cm]{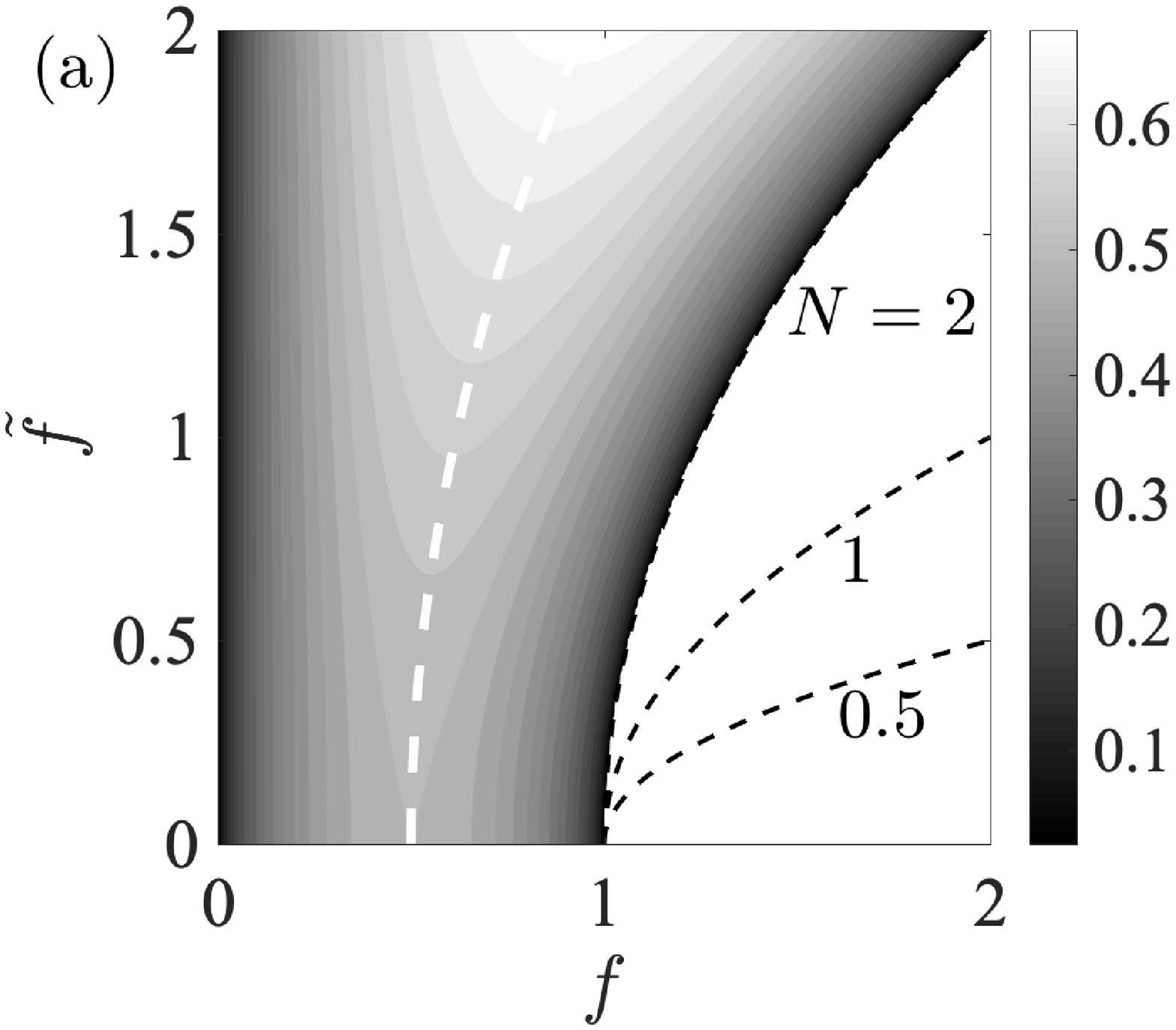}
   \includegraphics[height=5.2cm]{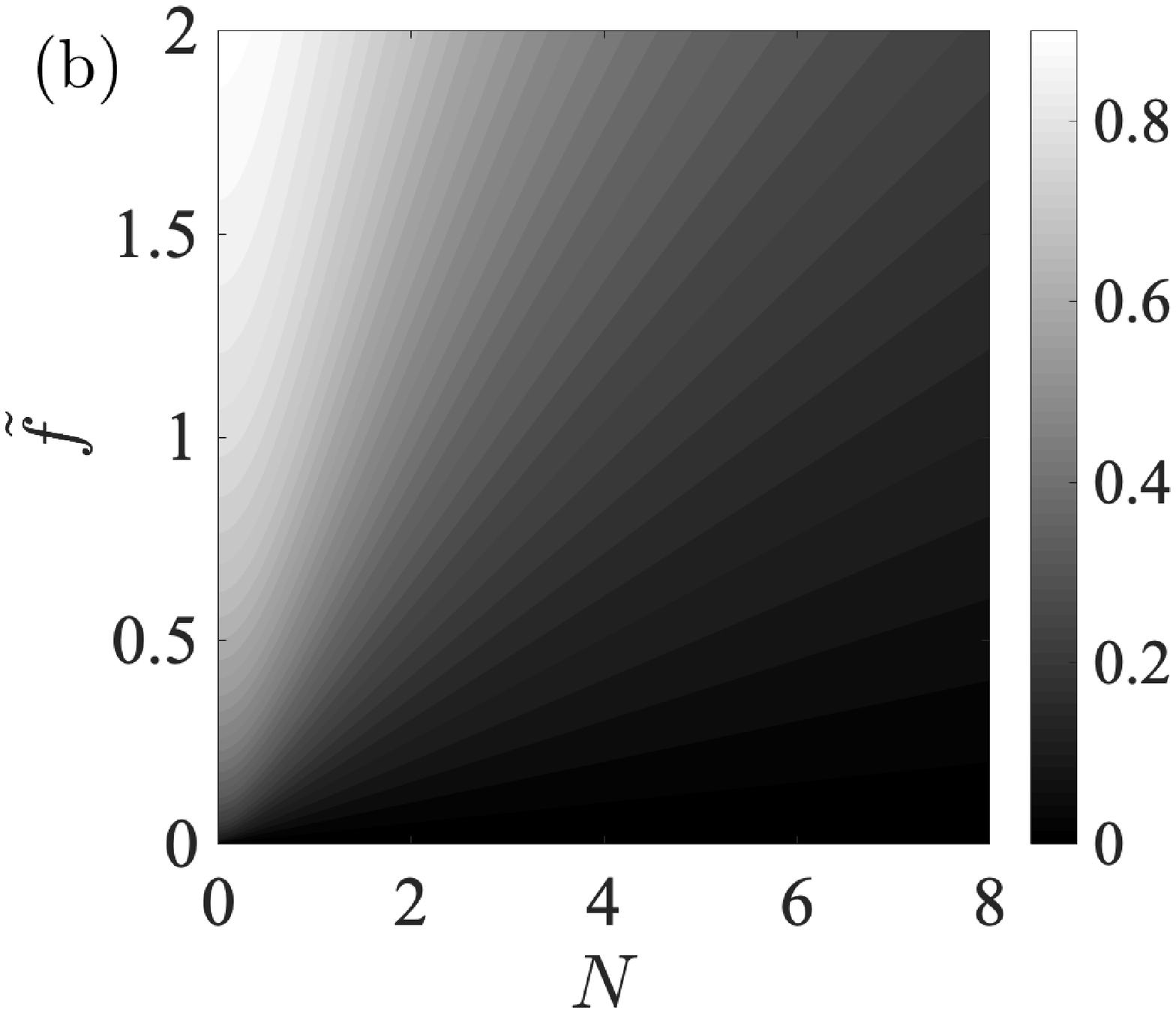}
   \includegraphics[height=5.2cm]{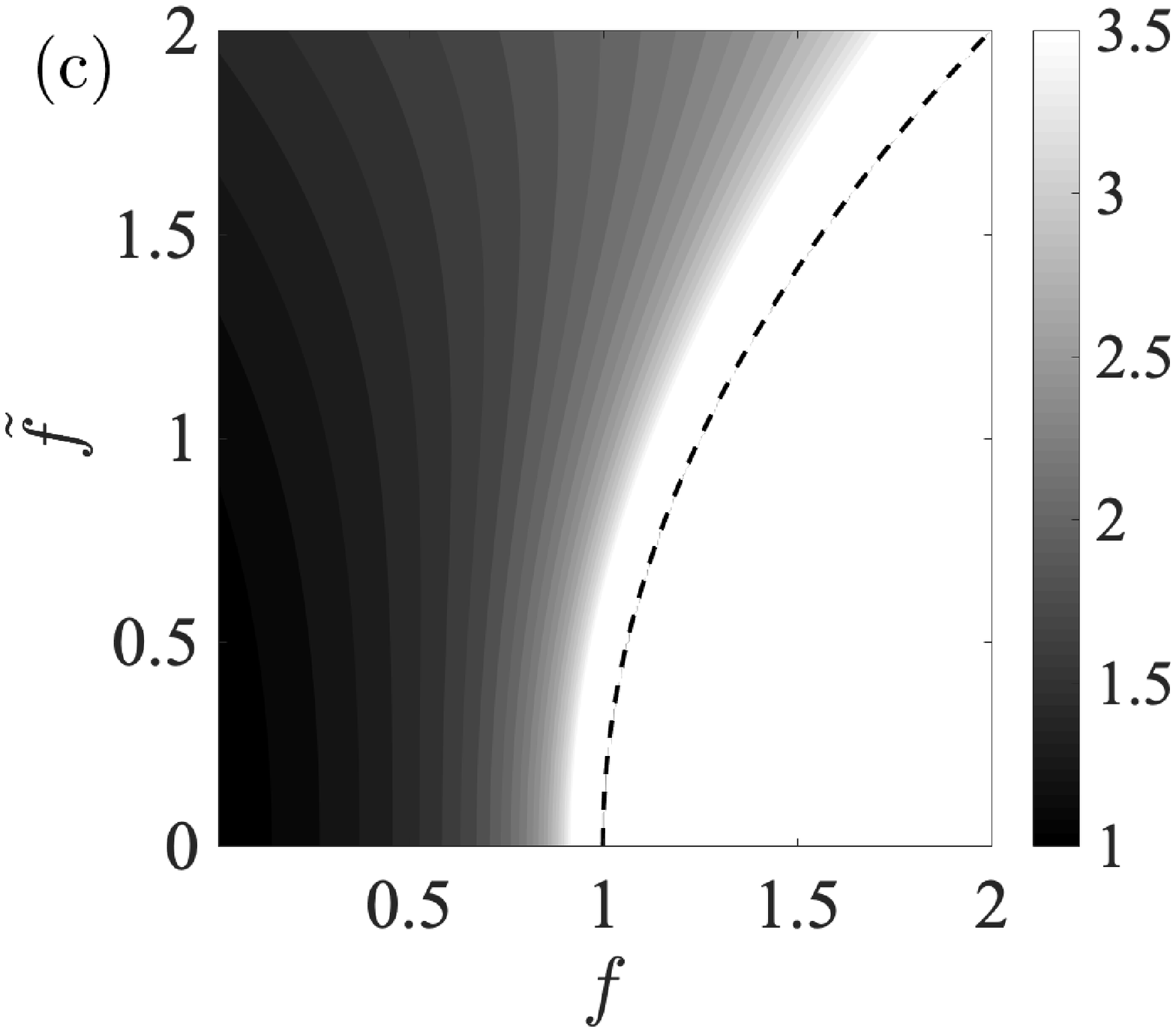}
   \caption{
   (a) Contours of the maximum growth rate $\sigma_{0}$ from Eq.~(\ref{eq:WKBJ_2ODE_dispersion_1st_leading}) in the parameter space $(f,\tilde{f})$ for $N=2$. Black dashed lines represent the upper limit of the instability $f=1+\tilde{f}^{2}/N^{2}$ from Eq.~(\ref{eq:range_inertial}) for different $N$, and the white dashed line represents $f_{\max}$ from (\ref{eq:f_max_WKBJ_2ODE}) where the maximum growth rate is attained. (b) Contours of $\sigma_{0}$ from Eq.~(\ref{eq:WKBJ_2ODE_dispersion_1st_leading}) in the parameter space $(N,\tilde{f})$ at $f=1$. (c) Contours of the first-order term $\sigma_{1}$ in the parameter space $(f,\tilde{f})$ for $N=2$ and the first branch at $m=1$.
   }
              \label{Fig_contour_sigma}%
    \end{figure*}
The right-hand side of Eq.~(\ref{eq:WKBJ_dispersion_quantized}) is always fixed at a finite $m$ so that the integral on the left-hand side of Eq.~(\ref{eq:WKBJ_dispersion_quantized}) should go to zero as $\tilde{k}\rightarrow\infty$.
This implies that the two turning points converge to zero as $\tilde{k}\rightarrow\infty$.
Using this property, we can find a more explicit dispersion relation in the Taylor expansion form for the growth rate $\sigma$ as
\begin{equation}
\label{eq:WKBJ_2ODE_dispersion_1st}
	\sigma=\sigma_{0}-\frac{\sigma_{1}}{k_{\rm{z}}}+O\left(\frac{1}{k_{\rm{z}}^{2}}\right),
\end{equation}
where
\begin{equation}
\label{eq:WKBJ_2ODE_dispersion_1st_leading}
\begin{aligned}
	2\sigma^{2}_{0}=&f(1-f)-N^{2}-\tilde{f}^{2}+\sqrt{\left[f(1-f)+N^{2}+\tilde{f}^{2}\right]^{2}+4\tilde{f}^{2}f^{2}}, 
\end{aligned}
\end{equation}
and
\begin{equation}
\label{eq:WKBJ_2ODE_dispersion_1st_1st}
\begin{aligned}
	\sigma_{1}=\frac{\left(m-\frac{1}{2}\right)\sqrt{f}\sqrt{\sigma_{0}^{2}+N^{2}+\tilde{f}^{2}}}{\sigma_{0}\left[1+f^{2}\tilde{f}^{2}/\left(\sigma_{0}^{2}+N^{2}+\tilde{f}^{2}\right)^{2}\right]}.
\end{aligned}
\end{equation}
Since $\sigma_{1}$ is positive, the first mode with $m=1$ is the most unstable mode.
There are also higher-order modes with $m>1$ as investigated in \citet{PPM2020}, but their growth rate is smaller than that of the first mode.
Thus, we hereafter consider only the $m=1$ mode for asymptotic results.  
We note that $\sigma=\sigma_{0}$ is the maximum growth rate $\sigma_{\max}$ as $k_{\rm{z}}\rightarrow\infty$, and $\sigma_{0}$ equals to $\sigma_{t,\max}$ of Eq.~(\ref{eq:sigma_t_max}).
While the maximum growth rate in the traditional $f$-plane approximation is $\sigma_{\max}=\sqrt{f(1-f)}$, which depends only on $f$, we see that the maximum growth rate $\sigma_{0}$ now depends on $f$, $\tilde{f}$, and $N$ altogether. 
For instance, at a fixed $N$, the maximum growth rate $\sigma_{0}$ increases with $\tilde{f}$ as shown in Fig.~\ref{Fig_contour_sigma}(a).
Compared to the inertially-unstable regime $0<f<1$ in the traditional approximation, we take $\sigma_{\max}$ real and positive and find the following range of the inertial instability for fixed $N$ and $\tilde{f}$:
\begin{equation}
\label{eq:range_inertial}
	0<f<1+\frac{\tilde{f}^{2}}{N^{2}}.
\end{equation}
If we use the relations $f=2\Omega\cos\theta$ and $\tilde{f}=2\Omega\sin\theta$, we have the corresponding range
\begin{equation}
\label{eq:range_inertial_new}
	0<2\Omega\cos\theta<1+\frac{4\Omega^{2}\sin^{2}\theta}{N^{2}},
\end{equation}
which is equivalent to
\begin{equation}
\label{eq:range_inertial_newnew}
	\tan^{-1}\left(\frac{N}{2}\right)<\theta<90^{\circ},
\end{equation}
if $2\Omega>0$.

Moreover, at the vertical Coriolis parameter $f=f_{\max}$
\begin{equation}
\label{eq:f_max_WKBJ_2ODE}
f_{\max}=\frac{1}{4}-N^{2}+\sqrt{\left(\frac{1}{4}-N^{2}\right)^{2}+N^{2}+\tilde{f}^{2}},
\end{equation}
we can find the maximum growth rate  at fixed $\tilde{f}$ and $N$ as
\begin{equation}
\max\left(\sigma_{\max}\right)\Large|_{\tilde{f},N}=\sigma_{0}(f_{\max}).
\end{equation}

In Fig.~\ref{Fig_contour_sigma}b, we see that the maximum growth rate $\sigma_{0}$ at fixed $f$ and $\tilde{f}$ decreases as $N$ increases, which highlights the stabilizing role of the stratification. 
We also display in Fig.~\ref{Fig_contour_sigma}c the dependence of the first-order term $\sigma_{1}$ on $f$ and $\tilde{f}$ at $N=2$. 
In this case, $\sigma_{1}$ increases with $f$ at a fixed $\tilde{f}$ and goes to infinity as $f$ reaches its upper limit $f=1+\tilde{f}^{2}/N^{2}$ since $\sigma_{0}$ is in the denominator of $\sigma_{1}$ and $\sigma_{0}\rightarrow0$ as $f\rightarrow1+\tilde{f}^{2}/N^{2}$. 

   \begin{figure}
   \centering
   \includegraphics[height=6cm]{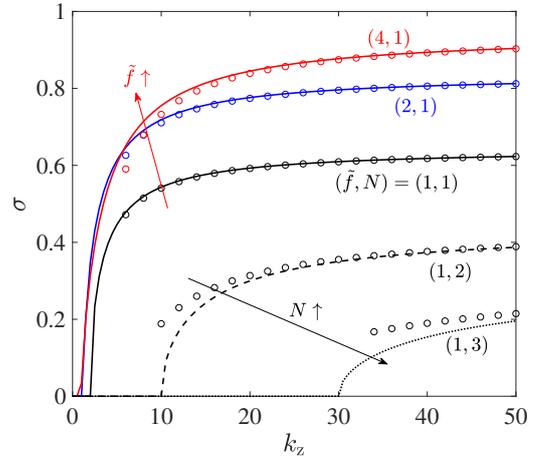}
      \caption{Numerical results (lines) and asymptotic results (circles) from (\ref{eq:WKBJ_2ODE_dispersion_1st}) for various values of $\tilde{f}$ and $N$ at $f=1$, $k_{\rm{x}}=0$, $Re=\infty$, and $Pe=\infty$. 
              }
         \label{Fig_numerical_WKBJ}
   \end{figure}
Figure \ref{Fig_numerical_WKBJ} shows the growth rate $\sigma$ versus the vertical wavenumber $k_{\rm{z}}$ for different values of $\tilde{f}$ and $N$ at $f=1$ and $k_{\rm{x}}=0$ in the inviscid and non-diffusive limits. 
The vertical Coriolis parameter $f=1$ belongs to the inertially stable regime under the traditional approximation, but we see that it becomes unstable if we consider the non-traditional effects with the horizontal component $\tilde{f}>0$, and the growth rate increases with $\tilde{f}$, as predicted from the WKBJ analysis. 
We notice that numerical stability results are in good agreement with asymptotic predictions from Eq.~(\ref{eq:WKBJ_2ODE_dispersion_1st}), especially for large $k_{\rm{z}}$. 
We also verify that the stratification stabilizes the inertial instability, and the onset of the inertial instability appears at a higher $k_{\rm{z}}$ as $N$ increases.

\subsection{The WKBJ formulation in the limit $Pe\rightarrow 0$}
In this subsection, we now turn our attention to the highly diffusive case in the limit of $Pe\rightarrow0$. 
This limit is relevant for stellar interiors.  
To analyze the inertial instability, we consider the case with $k_{\rm{x}}=0$ and use the 4th-order ODE (\ref{eq:4ODE_v}) for the WKBJ analysis. 
As performed in the previous subsection, we introduce for convenience the variable $\hat{W}$
\begin{equation}
\hat{W}(y)=\hat{v}(y)\exp\left[-\frac{\mathrm{i}k_{\rm{z}}\tilde{f}(U-2fy)}{2(\sigma^{2}+\tilde{f}^{2})}\right],
\end{equation}
to understand more clearly the exponential or wavelike behavior around turning points. 
The 4th-order ODE (\ref{eq:4ODE_v}) can be rewritten in terms of $\hat{W}$ as follows:
\begin{equation}
\label{eq:4ODE_W}
\begin{aligned}
	&\frac{1}{k_{\rm{z}}^{4}}\frac{\mathrm{d}^{4}\hat{W}}{\mathrm{d}y^{4}}+\frac{\mathrm{i}\tilde{f}(U'-2f)}{k_{\rm{z}}^{3}(\sigma^{2}+\tilde{f}^{2})}\frac{\mathrm{d}^{3}\hat{W}}{\mathrm{d}y^{3}}+\frac{1}{k_{\rm{z}}^{2}}\left(\frac{\Gamma}{\sigma^{2}+\tilde{f}^{2}}-1\right)\frac{\mathrm{d}^{2}\hat{W}}{\mathrm{d}y^{2}}\\
	&+\left[\frac{\mathrm{i}\tilde{f}(U'-2f)}{k_{\rm{z}}(\sigma^{2}+\tilde{f}^{2})}\left(\frac{\Gamma}{\sigma^{2}+\tilde{f}^{2}}+\frac{\tilde{f}^{2}(U'-2f)^{2}}{4(\sigma^{2}+\tilde{f}^{2})^{2}}\right)+\frac{U''}{k_{\rm{z}}^{2}(\sigma^{2}+\tilde{f}^{2})}\right.\\
	&\left.\times\left(2f+\frac{3(U'-2f)}{2(\sigma^{2}+\tilde{f}^{2})}\right)\right]\frac{\mathrm{d}\hat{W}}{\mathrm{d}y}
	+\left\{\left[\frac{-\Gamma}{\sigma^{2}+\tilde{f}^{2}}-\frac{\tilde{f}^{2}(U'-2f)^{2}}{4(\sigma^{2}+\tilde{f}^{2})^{2}}\right]\right.\\
	&\times\left[1+\frac{\tilde{f}^{2}(U'-2f)^{2}}{4(\sigma^{2}+\tilde{f}^{2})^{2}}\right]+\frac{1}{k_{\rm{z}}}\left[\frac{\mathrm{i}\tilde{f}U''}{2(\sigma^{2}+\tilde{f}^{2})}\left(1+\frac{\Gamma}{\sigma^{2}+\tilde{f}^{2}}\right)\right.\\
	&\left.\left.+\frac{\mathrm{i}\tilde{f}U''(U'-2f)}{(\sigma^{2}+\tilde{f}^{2})^{2}}\left(f+\frac{3\tilde{f}^{2}(U'-2f)}{4(\sigma^{2}+\tilde{f}^{2})}\right)\right]\right\}\hat{W}=O\left(\frac{1}{k_{\rm{z}}^{2}}\right),
\end{aligned}
\end{equation}
where the right-hand side of order $O(k_{\rm{z}}^{-2})$ will be neglected in the WKBJ analysis for large $k_{\rm{z}}$.
We apply the WKBJ approximation
\begin{equation}
	\hat{W}(y)\sim\exp\left[\frac{1}{\delta}\sum_{l=0}^{\infty}\delta^{l}S_{l}(y)\right],
\end{equation}
to the 4th-order ODE (\ref{eq:4ODE_W}), and we find $\delta=k_{\rm{z}}^{-1}$ and four solutions for $S_0$ where
\begin{equation}
\label{eq:WKBJ_leading_order_4ODE}
\begin{aligned}
	&S_{0}^{'(1,2)}=\pm\sqrt{\frac{-\Gamma}{\sigma^{2}+\tilde{f}^{2}}-\frac{\tilde{f}^{2}(U'-2f)^{2}}{4(\sigma^{2}+\tilde{f})^{2}}},\\
	&S_{0}^{'(3,4)}=-\frac{\mathrm{i}\tilde{f}(U'-2f)}{2(\sigma^{2}+\tilde{f}^{2})}\pm1.
\end{aligned}
\end{equation}
The WKBJ solution with $S_{0}^{(3,4)}$ is
\begin{equation}
\label{eq:WKBJ_solution_S034}
\begin{aligned}
	\hat{W}(y)&=A_{3}\exp\left[\frac{\mathrm{i}k_{\rm{z}}\tilde{f}(2fy-U)}{2(\sigma^{2}+\tilde{f}^{2})}+k_{\rm{z}} y+O(1)\right]\\
	&+A_{4}\exp\left[\frac{\mathrm{i}k_{\rm{z}}\tilde{f}(2fy-U)}{2(\sigma^{2}+\tilde{f}^{2})}-k_{\rm{z}} y+O(1)\right],
\end{aligned}
\end{equation}
where $A_{3}$ and $A_{4}$ are constants. 
This solution implies that $\hat{v}(y)=A_{3}\exp(k_{\rm{z}} y)+A_{4}\exp(-k_{\rm{z}} y)$ thus $\hat{v}$ simply increases or decreases exponentially as $|y|\rightarrow\infty$ without turning points. 
Therefore, an eigenfunction satisfying the decaying boundary conditions as $|y|\rightarrow\infty$ cannot be constructed.
Thus, we impose $A_{3}=A_{4}=0$. 
The WKBJ solution $\hat{W}$ with $S_{0}^{(1,2)}$ can be either evanescent or wavelike depending on the sign of ${\widetilde{\Gamma}}$ where
\begin{equation}
	{\widetilde{\Gamma}}=\Gamma+\frac{\tilde{f}^{2}(U'-2f)^{2}}{4(\sigma^{2}+\tilde{f}^{2})}.
\end{equation}
On the one hand, the WKBJ solution is evanescent if ${\widetilde{\Gamma}}<0$:
\begin{equation}
\label{eq:WKBJ_solution_4ODE_exponential}
\begin{aligned}
	\hat{W}&=C_{1}\exp\left[{\widetilde{k}}\int_{y}\sqrt{-{\widetilde{\Gamma}(\upsilon)}}\mathrm{d}\upsilon+O(1)\right]\\
	&+C_{2}\exp\left[-{\widetilde{k}}\int_{y}\sqrt{-{\widetilde{\Gamma}(\upsilon)}}\mathrm{d}\upsilon+O(1)\right],
\end{aligned}
\end{equation}
where $C_{1}$ and $C_{2}$ are constants, and $\widetilde{k}=k_{\rm{z}}/\sqrt{\sigma^{2}+\tilde{f}^{2}}$. 
On the other hand, the WKBJ solution is wavelike if $\widetilde{\Gamma}>0$:
\begin{equation}
\label{eq:WKBJ_solution_4ODE_wavelike}
\begin{aligned}
	\hat{W}&=D_{1}\exp\left[\mathrm{i}\widetilde{k}\int_{y}\sqrt{\widetilde{\Gamma}(\upsilon)}\mathrm{d}\upsilon+O(1)\right]\\
	&+D_{2}\exp\left[-\mathrm{i}\widetilde{k}\int_{y}\sqrt{\widetilde{\Gamma}(\upsilon)}\mathrm{d}\upsilon+O(1)\right],
\end{aligned}
\end{equation}
where $D_{1}$ and $D_{2}$ are constants.  
In these expressions, we consider only the leading order term $S_{0}$ and neglect higher-order terms.

As conducted in the previous subsection, to derive asymptotic dispersion relations, we need to perform analyses around turning points $\widetilde{y}_{t}$ where $\widetilde{\Gamma}(\widetilde{y}_{{t}})=0$. 
To construct an eigenfunction, we first note that $\widetilde{\Gamma}$ is negative as $y\rightarrow\infty$ since ${\widetilde{\Gamma}(y\rightarrow\infty)}=-\sigma^{2}-f^{2}\sigma^{2}/(\sigma^{2}+\tilde{f}^{2})$.
We thus have the exponentially decaying WKBJ solution for $y>\widetilde{y}_{t+}$ as follows:
\begin{equation}
\label{eq:WKBJ_4ODE_y_large_ytplus}
\hat{W}=C_{+}\exp\left[-\widetilde{k}\int_{\widetilde{y}_{t+}}^{y}\sqrt{-\widetilde{\Gamma}(\upsilon)}\mathrm{d}\upsilon\right].
\end{equation}
Around the turning point $\widetilde{y}_{t+}$ where ${\widetilde{\Gamma}}(\widetilde{y}_{t+})=0$, we expand ${\widetilde{\Gamma}}\sim{\widetilde{\Gamma}}_{t+}^{'}\widetilde{\epsilon}\widetilde{{y}}$, where $\widetilde{\Gamma}_{t+}^{'}$ is the derivative of $\widetilde{\Gamma}$ at the turning point, $\widetilde{{y}}=(y-\widetilde{y}_{t+})/\widetilde{\epsilon}$, and $\widetilde{\epsilon}$ is a small parameter satisfying $\widetilde{\epsilon}=\left[(-{\widetilde{\Gamma}}^{'}_{t+})\widetilde{k}^{2}\right]^{-1/3}$.
Using this expansion in Eq.~(\ref{eq:4ODE_W}), we obtain the local equation around $\widetilde{y}_{t+}$:
\begin{equation}
\label{eq:local_4ODE}
\frac{\mathrm{d}^{2}\hat{W}}{\mathrm{d}\widetilde{{y}}^{2}}-\widetilde{{y}}\hat{W}=0,
\end{equation}
whose solutions are the Airy functions $\hat{W}=c_{1}\mathrm{Ai}({\widetilde{y}})+d_{1}\mathrm{Bi}(\widetilde{{y}})$, where $c_{1}$ and $d_{1}$ are constants. 
Matching the asymptotic behavior of the Airy functions as ${\widetilde{y}}\rightarrow\pm\infty$ and of the WKBJ solution (\ref{eq:WKBJ_4ODE_y_large_ytplus}) as $y\rightarrow \widetilde{y}_{t+}$, we have the following WKBJ solution in the range $\widetilde{y}_{t-}<y<\widetilde{y}_{t+}$ between the two turning points:
\begin{equation}
\label{eq:WKBJ_4ODE_y_small_ytplus_large_ytminus}
\begin{aligned}
	\hat{W}&=C_{+}\exp\left[\mathrm{i}\widetilde{k}\int^{\widetilde{y}_{t+}}_{y}\sqrt{{\widetilde{\Gamma}(\upsilon)}}\mathrm{d}\upsilon-\mathrm{i}\frac{\pi}{4}\right]\\
	&+C_{+}\exp\left[-\mathrm{i}\widetilde{k}\int^{\widetilde{y}_{t+}}_{y}\sqrt{{\widetilde{\Gamma}(\upsilon)}}\mathrm{d}\upsilon+\mathrm{i}\frac{\pi}{4}\right].
\end{aligned}
\end{equation}
For $y\rightarrow-\infty$, we consider the decaying WKBJ solution for $y<\widetilde{y}_{t-}$ as
\begin{equation}
\label{eq:WKBJ_4ODE_y_small_ytminus}
\hat{W}=C_{-}\exp\left[-\widetilde{k}\int^{\widetilde{y}_{t-}}_{y}\sqrt{-\widetilde{\Gamma}(\upsilon)}\mathrm{d}\upsilon\right].
\end{equation}
From the local solution behavior around the turning point $\widetilde{y}_{t-}$, we can match the two WKBJ solutions (\ref{eq:WKBJ_4ODE_y_small_ytplus_large_ytminus}) and (\ref{eq:WKBJ_4ODE_y_small_ytminus}).
It leads to the following dispersion relation in a quantized form:
\begin{equation}
\label{eq:WKBJ_4ODE_dispersion_quantization}
\widetilde{k}\int_{\tilde{y}_{t-}}^{\tilde{y}_{t+}}\sqrt{\Gamma+\frac{\tilde{f}^{2}(U'-2f)^{2}}{4(\sigma^{2}+\tilde{f}^{2})}}dy=\left(m_{0}-\frac{1}{2}\right)\pi,
\end{equation}
where $m_{0}$ is the positive integer for the mode number. 

%
   \begin{figure}
   \centering
   \includegraphics[height=6cm]{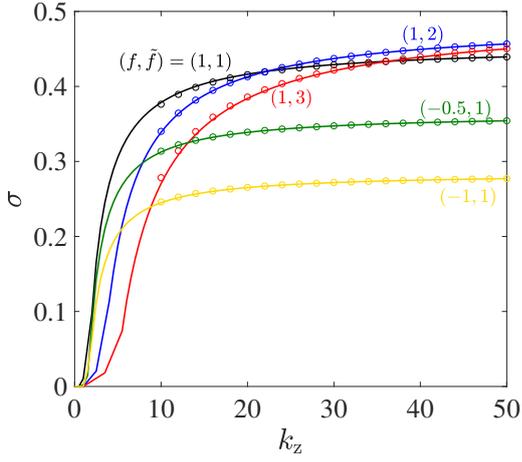}
      \caption{Numerical results (lines) of the growth rate $\sigma$ as a function of $k_{\rm{z}}$ for various parameter sets of $(f,\tilde{f})${: (1,1) (black, $\Omega=0.707$ and $\theta=45^{\circ}$), (1,2) (blue, $\Omega=1.118$ and $\theta=63.4^{\circ}$), (1,3) (red, $\Omega=1.581$ and $\theta=71.6^{\circ}$), (-0.5,1) (green, $\Omega=0.559$ and $\theta=116.6^{\circ}$), (-1,1) (yellow, $\Omega=0.707$ and $\theta=135^{\circ}$)}, at $Pe=0.1$, $k_{\rm{x}}=0$, $N=1$, and $Re=\infty$, and asymptotic results (circles) from Eq.~(\ref{eq:WKBJ_4ODE_dispersion_1st}) in the limit $Pe\rightarrow0$. 
              }
         \label{Fig_numerical_WKBJ_Pe0}
   \end{figure}
Furthermore, we apply the Taylor expansion for $\sigma$
\begin{equation}
\label{eq:WKBJ_4ODE_dispersion_1st}
\sigma=\sigma_{0,0}-\frac{\sigma_{1,0}}{k_{\rm{z}}^{1} }+O\left(\frac{1}{k_{\rm{z}}^{2}}\right),
\end{equation}
to the quantized dispersion relation (\ref{eq:WKBJ_4ODE_dispersion_quantization}), and we get 
\begin{equation}
\label{eq:WKBJ_4ODE_sigma00}
\sigma_{0,0}=\sqrt{\frac{f(1-f)-\tilde{f}^{2}+\sqrt{\left[f(1-f)-\tilde{f}^{2}\right]^{2}+\tilde{f}^{2}}}{2}},
\end{equation}
and
\begin{equation}
\sigma_{1,0}=\frac{\left(m_{0}-1/2\right)\sqrt{f\sigma^{2}_{0,0}+\tilde{f}^{2}/2}}{\sigma_{0,0}\left[1+\tilde{f}^{2}(f-1/2)^{2}/(\sigma_{0,0}^{2}+\tilde{f}^{2})^{2}\right]}.
\end{equation}
Since $\sigma_{1,0}$ is positive, we consider hereafter the first mode with $m_{0}=1$ for the asymptotic results of the most unstable mode.  
In Fig.~\ref{Fig_numerical_WKBJ_Pe0}, we display the growth rate $\sigma$ as a function of the vertical wavenumber $k_{\rm{z}}$ for various values of $f$ and $\tilde{f}$ at small $Pe=0.1$ for $k_{\rm{x}}=0$, $N=1$, and $Re=\infty$, and we compare the numerical results with the asymptotic predictions from Eq.~(\ref{eq:WKBJ_4ODE_dispersion_1st}). 
We see that they are in very good agreement, especially as $k_{\rm{z}}\rightarrow\infty$. 

%
   \begin{figure}
   \centering
   \includegraphics[height=5.8cm]{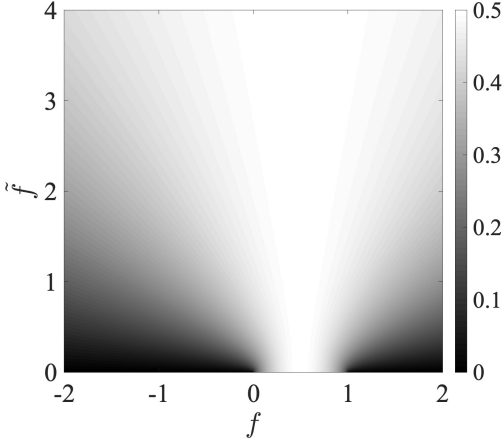}
      \caption{Contours of $\sigma_{0,0}$ in the parameter space of $(f,\tilde{f})$. 
              }
         \label{Fig_contours_sig00}
   \end{figure}
%
Figure \ref{Fig_contours_sig00} shows how the maximum growth rate $\sigma_{0,0}$ depends on the Coriolis parameters $f$ and $\tilde{f}$.
It is very remarkable that for $\tilde{f}>0$, $\sigma_{0,0}$ is positive for any value of $f$ including negative ($f<0$) and large ($f>1$) values, which are the two ranges outside the inertially unstable regime in the traditional approximation. 
The maximum value of $\sigma_{0,0}$ at a fixed $\tilde{f}$ is always attained at $f=f_{\max}=0.5$ with the corresponding value $\sigma_{\max,0}=0.5$. 

   \begin{figure}
   \centering
   \includegraphics[height=6cm]{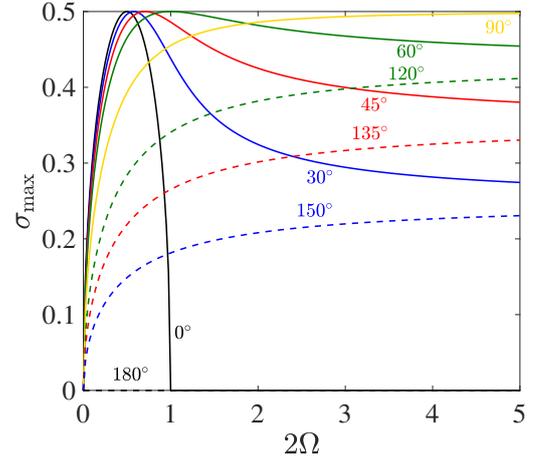}
      \caption{Maximum growth rate $\sigma_{\max}=\sigma_{0,0}$ of Eq.~(\ref{eq:WKBJ_4ODE_sigma00}) as a function of the absolute Coriolis parameter $2\Omega$ at various colatitudes $\theta$.
      Solid lines and dashed lines denote the results in the northern and southern hemispheres, respectively.
              }
         \label{Fig_sigma_max_2Omega}
   \end{figure}
Using the expressions of the Coriolis parameters $f=2\Omega\cos\theta$ and $\tilde{f}=2\Omega\sin\theta$, we can see how the maximum growth rate $\sigma_{0,0}$ depends on the rotation rate $\Omega$ and colatitude $\theta$.
Figure \ref{Fig_sigma_max_2Omega} shows the maximum growth rate $\sigma_{0,0}$ plotted against the absolute Coriolis parameter $2\Omega$ for various colatitudes.
At the northern pole ($\theta=0^\circ$), the case considered in the traditional approximation, the inertial instability only exists in the range $0<2\Omega<1$.
On the other hand, in the northern hemisphere $0^{\circ}<\theta\leq90^{\circ}$, we see that the inertial instability exists in any values of $2\Omega>0$, and the growth rate reaches its maximum $\sigma_{\max}=0.5$ at $2\Omega_{\max}=0.5/\cos\theta$.
After the peak, $\sigma_{\max}$ decreases with $2\Omega$ and asymptotes to a constant value as $2\Omega\rightarrow\infty$.
By considering the limit $2\Omega\rightarrow\infty$, we can find the asymptote of $\sigma_{\max}$ as follows:
\begin{equation}
\label{eq:sigma_00_aymptote}
\sigma_{\max}(2\Omega\rightarrow\infty)=\frac{\sin\theta}{2}.
\end{equation}
In the southern hemisphere $90^{\circ}<\theta<180^{\circ}$, the growth rate always increases with $2\Omega$ and it reaches the maximum $\sin\theta/2$ as $2\Omega\rightarrow\infty$.
At the southern pole ($\theta=180^{\circ}$), we find no inertial instability.

As a summary, it is remarkable that the inertial instability for the high-diffusivity case ($Pe\rightarrow0$) exists in the ranges
\begin{equation}
\label{eq:inertial_range_Pe0_pole}
0<2\Omega<1~~\mathrm{if}~~\theta=0^{\circ},
\end{equation} 
or
\begin{equation}
\label{eq:inertial_range_Pe0}
2\Omega>0~~\mathrm{if}~~0^{\circ}<\theta<180^{\circ}.
\end{equation} 
This colatitude range $0^{\circ}\leq\theta<180^{\circ}$ is much wider than the range of Eq.~(\ref{eq:range_inertial_newnew}), which belongs to the northern hemisphere in the non-diffusive case ($Pe=\infty$).
Since the high-thermal-diffusivity regime is relevant for stellar interiors, the inertial instability due to the horizontal shear can be an important source of turbulence in stars at any colatitude except at the southern pole.  
{This unstable regime in the limit $Pe\rightarrow0$ is equivalent to the regime of the GSF instability that occurs when the rotation profile varies along the rotation axis for small $Pe$.} \footnote{{The configuration for the GSF instability is relevant to our case for the inertial instability where the rotation varies along the latitudinal $y$-direction due to the local horizontal shear we consider.}}

\section{Detailed parametric investigation}
\label{sect:parametric}
The detailed WKBJ analysis performed in the previous section provided explicit expressions for the dispersion relation of the inertial instability in two cases: one with no thermal diffusion and the other with high thermal diffusivity, without the traditional approximation.
However, we do not fully understand how the inertial and inflectional instabilities are modified in other regimes such as finite $Pe$, finite $Re$, etc. 
In this section, we present more broad and detailed numerical results to understand the parametric behavior of the inflectional and inertial instabilities on $Pe$, $Re$, $N$, $f$, and $\tilde{f}$.  

\subsection{Maximum growth rate of the inflectional instability: effects of $N$, $\tilde{f}$ and $Pe$}
%
   \begin{figure}
   \centering
   \includegraphics[height=6cm]{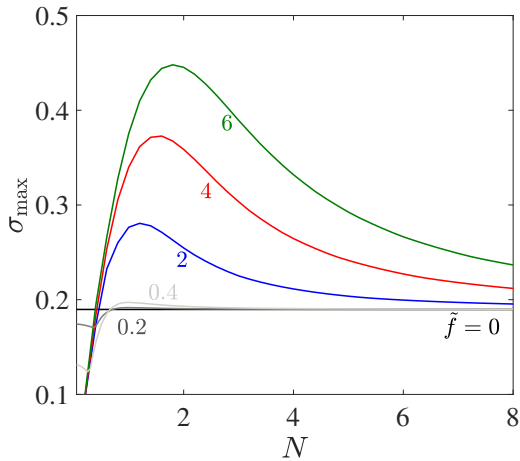}
      \caption{Maximum growth rate $\sigma_{\max}$ of the inflectional instability as a function of the Brunt-V\"ais\"al\"a frequency $N$ for different values of $\tilde{f}$ at $Re=Pe=\infty$. 
              }
         \label{Fig_inflectional_max}
   \end{figure}
%
In the traditional approximation (i.e., $\tilde{f}=0$), the inflectional instability always has the maximum growth rate $\sigma_{\max}\simeq0.1897$ in the inviscid limit $Re=\infty$ at $k_{\rm{x}}\simeq0.445$ and $k_{\rm{z}}=0$, independently from the Coriolis parameter $f$, the Brunt-V\"ais\"al\"a frequency $N$, and the P\'eclet number $Pe$ \citep[]{Deloncle2007,Arobone2012,PPM2020}. 
But as shown in Figs.~\ref{Fig_contours_growth_preliminary} and \ref{Fig_growth_inflectional}, the maximum growth rate of the inflectional instability depends on the values of $N$ and $Pe$ if $\tilde{f}>0$. 
For instance, we plot in Fig.~\ref{Fig_inflectional_max} the maximum growth rate $\sigma_{\max}$ of the inflectional instability over the parameter space $(k_{\rm{x}},k_{\rm{z}})$ at $f=0$ (i.e., on the equator) in the inviscid and non-diffusive limits (i.e., $Re=\infty$ and $Pe=\infty$).
We see how the maximum growth rate $\sigma_{\max}$ is modified by the stratification for various values of $\tilde{f}$.
We found numerically that the maximum growth rate $\sigma_{\max}$ of the inflectional instability is attained for a finite $k_{\rm{x}}$ at $k_{\rm{z}}=0$ for the non-diffusive case $Pe=\infty$, and that it is independent of $f$ as explained by Eq.~(\ref{eq:2ODE_v_kz0}).
While the maximum growth rate $\sigma_{\max}=0.1897$ is invariant for $\tilde{f}=0$, we found that $\sigma_{\max}$ at a fixed $\tilde{f}>0$ is reduced for weakly stratified fluids with $N\ll0.7$ while $\sigma_{\max}$ surpasses 0.1897 for stratified fluids with $N\gtrsim0.7$. 
The maximum of $\sigma_{\max}$ over $N$, namely $\max(\sigma_{\max})$, is thus greater than 0.1897 and it increases with $\tilde{f}$.

   \begin{figure}
   \centering
   \includegraphics[height=5.5cm]{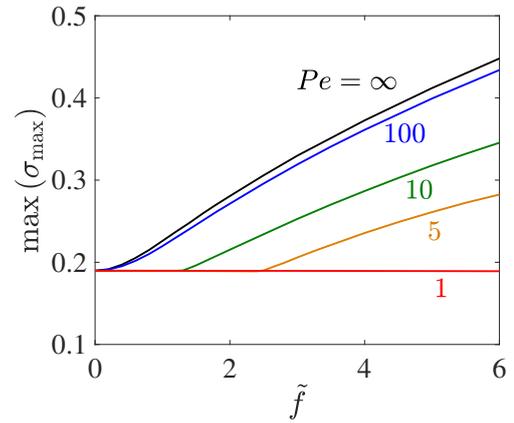}
      \caption{Maximum of $\sigma_{\max}$ of the inflectional instability over $N$ as a function of $\tilde{f}$ for different values of $Pe$ (solid lines) at $k_{\rm{z}}=0$ and $f=0$. 
              }
         \label{Fig_sigmaxmax}
   \end{figure}
Figure \ref{Fig_sigmaxmax} shows the maximum growth rate of the inflectional instability over $N$ as a function of $\tilde{f}$ for various values of $Pe$ at $k_{\rm{z}}=0$ and $f=0$.
We see that at a finite $Pe$, $\max\left(\sigma_{\max}\right)$ remains constant around $0.1897$ for small $\tilde{f}$ and then increases with $\tilde{f}$. 
At $Pe=1$, we see that the maximum remains constant in the range $0\leq\tilde{f}\leq6$.
The onset of the growth-rate increase is delayed as $Pe$ decreases; therefore, we can say that the inflectional instability is stabilized as the thermal diffusivity increases (i.e., as $Pe$ decreases).
This stabilization of the inflectional instability was also reported in the traditional approximation \citep[see e.g.,][]{PPM2020}, thus we can say that the inflectional instability is always stabilized by the thermal diffusivity for $\tilde{f}\geq0$.
And we can expect that the inflectional instability will not sustain in stars due to their high thermal diffusivity.
As a summary, we display in Table \ref{table:1} the parametric dependence for the inflectional instability in the non-traditional case \citep[for other parameters $f$ and $N$ in the traditional case, refer to][]{Deloncle2007,Arobone2012,PPM2020}. 
\begin{table}[!t]
\caption{Summary table for the inflectional-instability growth rate by its variation with parameters $\tilde{f}$, $Pe$, and $Re$}
\label{table:1}
\centering
\begin{tabular}{ c  c    c  c}
\hline
\hline
Inflectional instability & $\tilde{f}$ & $Pe\downarrow$ $(\kappa_{0}\uparrow)$  & $Re\downarrow$ $(\nu_{0}\uparrow)$\\
\hline
Growth rate $\sigma_{r}$ & $\uparrow$ & $\downarrow$ & $\downarrow$\\
\hline
\end{tabular}
\end{table}

\subsection{Maximum growth rate of the inertial instability at finite $Pe$}
%
   \begin{figure}
   \centering
   \includegraphics[height=6cm]{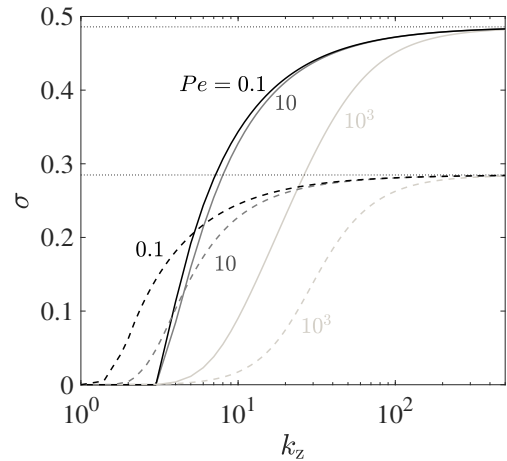}
      \caption{Growth rate $\sigma_{r}$ of the inertial instability as a function of $k_{\rm{z}}$ for values of $Pe$ at $N=1$ {and $k_{\rm{x}}=0$ for $(f,\tilde{f})=(0,2)$ (i.e. $\Omega=1$ and $\theta=90^{\circ}$; solid lines) and for $(f,\tilde{f})=(-1,1)$ (i.e. $\Omega=0.707$ and $\theta=135^{\circ}$; dashed lines)}. 
      Dotted lines indicate asymptotes $\sigma_{0,0}$ of Eq.~(\ref{eq:WKBJ_4ODE_sigma00}) predicted from the WKBJ analysis in the limit $Pe\rightarrow0$.   
              }
         \label{Fig_sigmax_finite_Pe}
   \end{figure}
Figure \ref{Fig_sigmax_finite_Pe} shows plots of the growth rate as a function of $k_{\rm{z}}$ for different values of $Pe$ at $N=1$ and $k_{\rm{x}}=0$.
Two parameter sets are considered: $(f,\tilde{f})=(0,2)$ (i.e. $\Omega=1$ and $\theta=90^{\circ}$) and $(f,\tilde{f})=(-1,1)$ (i.e. $2\Omega=\sqrt{2}$ and $\theta=135^{\circ}$). 
These Coriolis parameters belong to the inertially stable regime in the non-diffusive limit $Pe\rightarrow\infty$ according to the criterion (\ref{eq:range_inertial}).
However, we see that it becomes unstable as $Pe$ becomes finite, and the growth rate increases as $Pe$ decreases.
For a given parameter set of $(f,\tilde{f})$, we see that all growth-rate curves for any $Pe$ reach the same maximum value $\sigma_{0,0}$ given by Eq.~(\ref{eq:WKBJ_4ODE_sigma00}) derived in the limit $Pe\rightarrow0$ as $k_{\rm{z}}\rightarrow\infty$.
We also verified numerically that this asymptotic behavior is the same for other values of $N$ at finite $Pe$. 
Therefore, the largest growth rate of the inertial instability is always $\sigma_{0,0}$, and we have analogously the same relation between the maximum growth rate of the inertial instability and the Coriolis parameters $(f,\tilde{f})$ at finite $Pe$.

In Table \ref{table:2}, we summarize the variation of the growth rate of the inertial instability with parameters $\tilde{f}$ and $Pe$, and the unstable regimes in the two limits $Pe\rightarrow\infty$ and $Pe\rightarrow0$.
The destabilization of the inertial instability by $Pe$ was already reported previously \citep[see e.g.,][]{PPM2020}, and we confirm in this paper that this destabilization holds beyond the traditional approximation.

\begin{table*}[!t]
\caption{Summary table for the variation of the inertial-instability growth rate and the inertially unstable regime}
\label{table:2}
\centering
\begin{tabular}{c  c  c  c c c}
\hline
\hline
Type & $\tilde{f}$ & $Pe\downarrow$ $(\kappa_{0}\uparrow)$  & $Re\downarrow$ $(\nu_{0}\uparrow)$ & Unstable regime at $Pe=\infty$ & Unstable regime as $Pe\rightarrow0$\\
\hline
\multirow{2}{*}{Inertial instability} & \multirow{2}{*}{$\uparrow$} & \multirow{2}{*}{$\uparrow$} & \multirow{2}{*}{$\downarrow$} & \multirow{2}{*}{$\tan^{-1}\left(\frac{N}{2}\right)<\theta<90^{\circ}~~~\mathrm{if}~~~\Omega>0$ } 
& $0<2\Omega<1$~~~if~~~$\theta=0^{\circ}$\\
& & & & &
~~~~~~~~~~~~~~~~~~~~~$\Omega>0$~~~if~~~$0^{\circ}<\theta<180^{\circ}$\\
\hline
\end{tabular}
\end{table*}
\subsection{Effect of the viscosity at finite $Re$}
\begin{figure*}
   \centering
   \includegraphics[height=5.1cm]{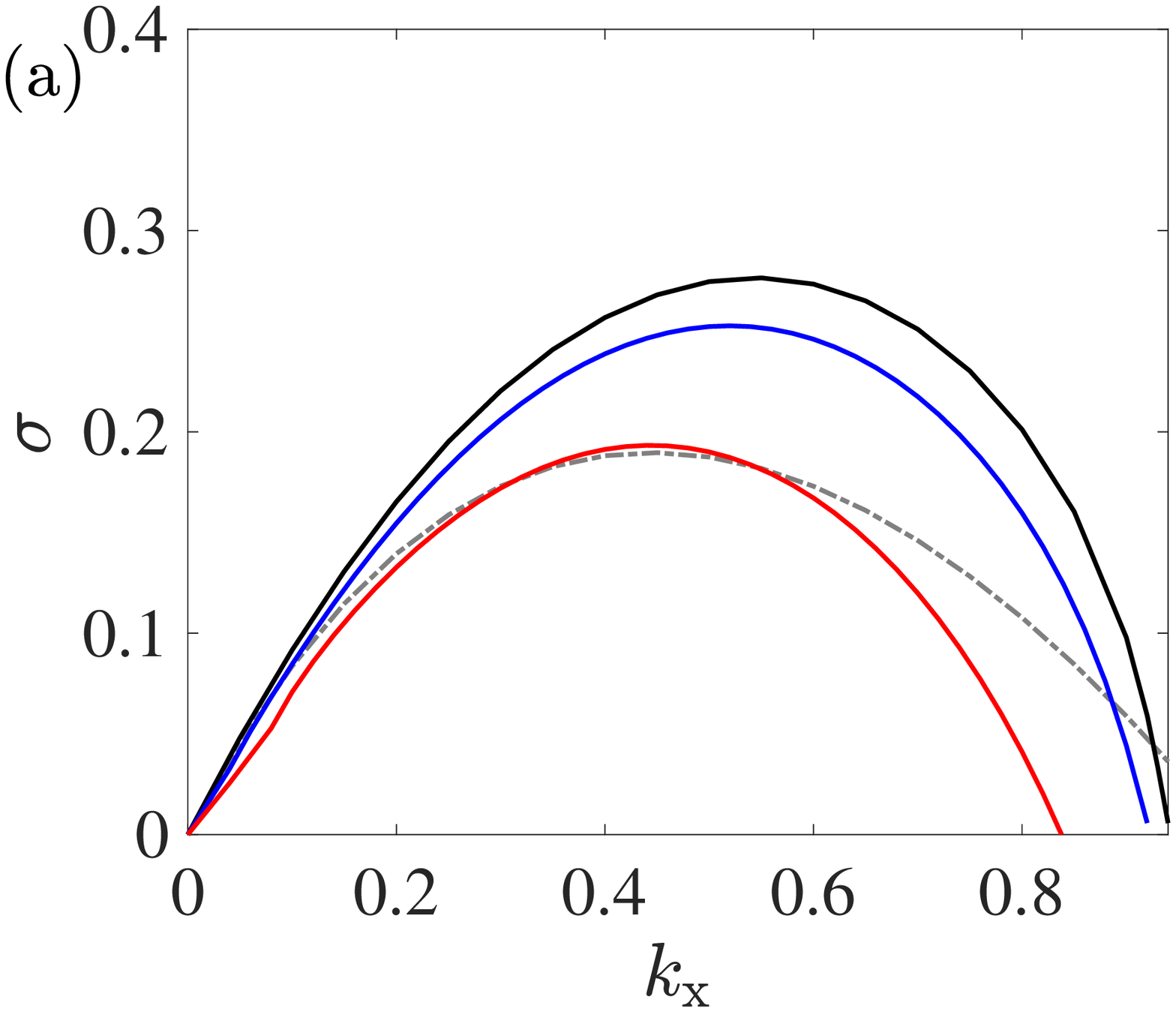}
   \includegraphics[height=5.1cm]{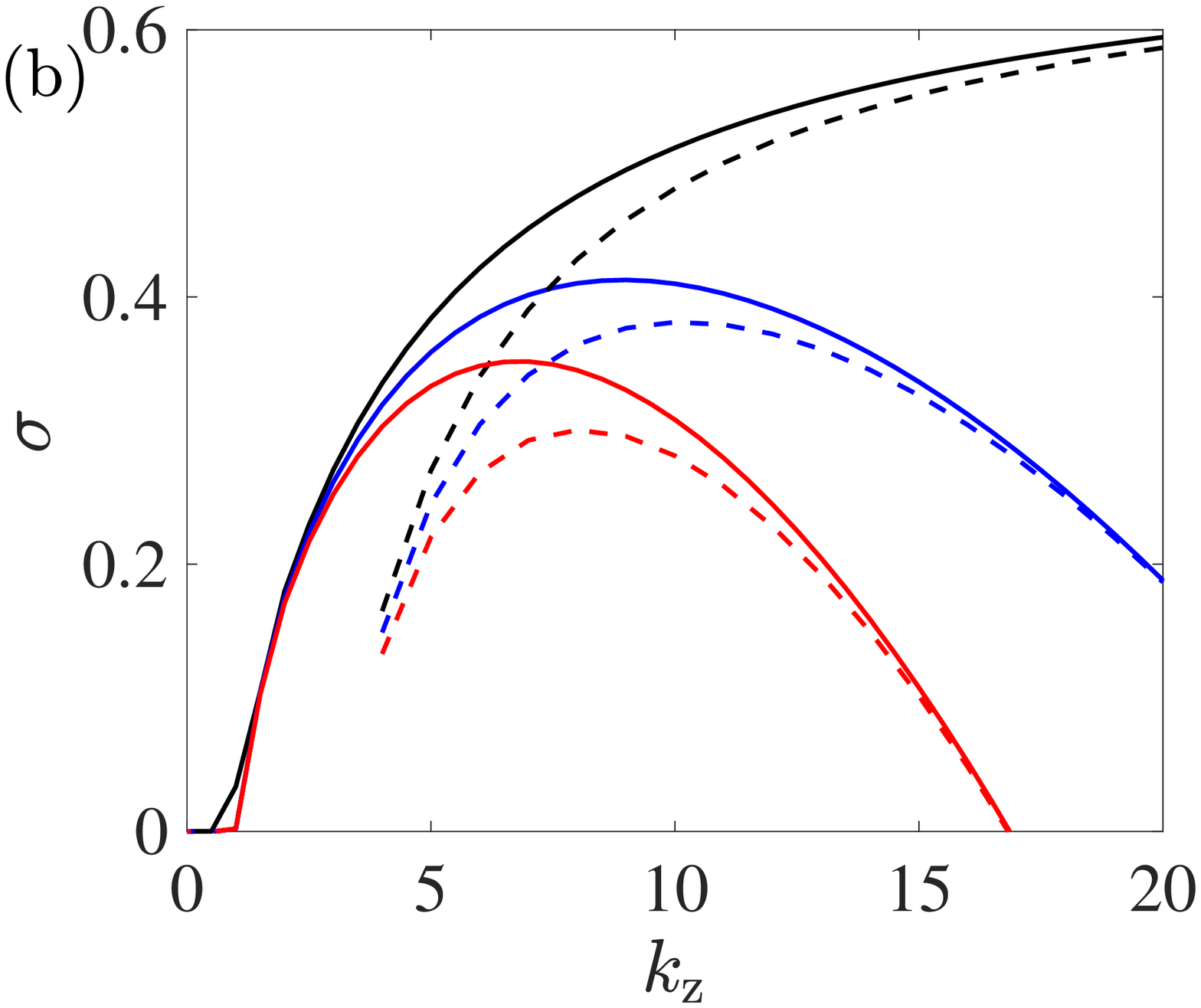}
   \includegraphics[height=5.1cm]{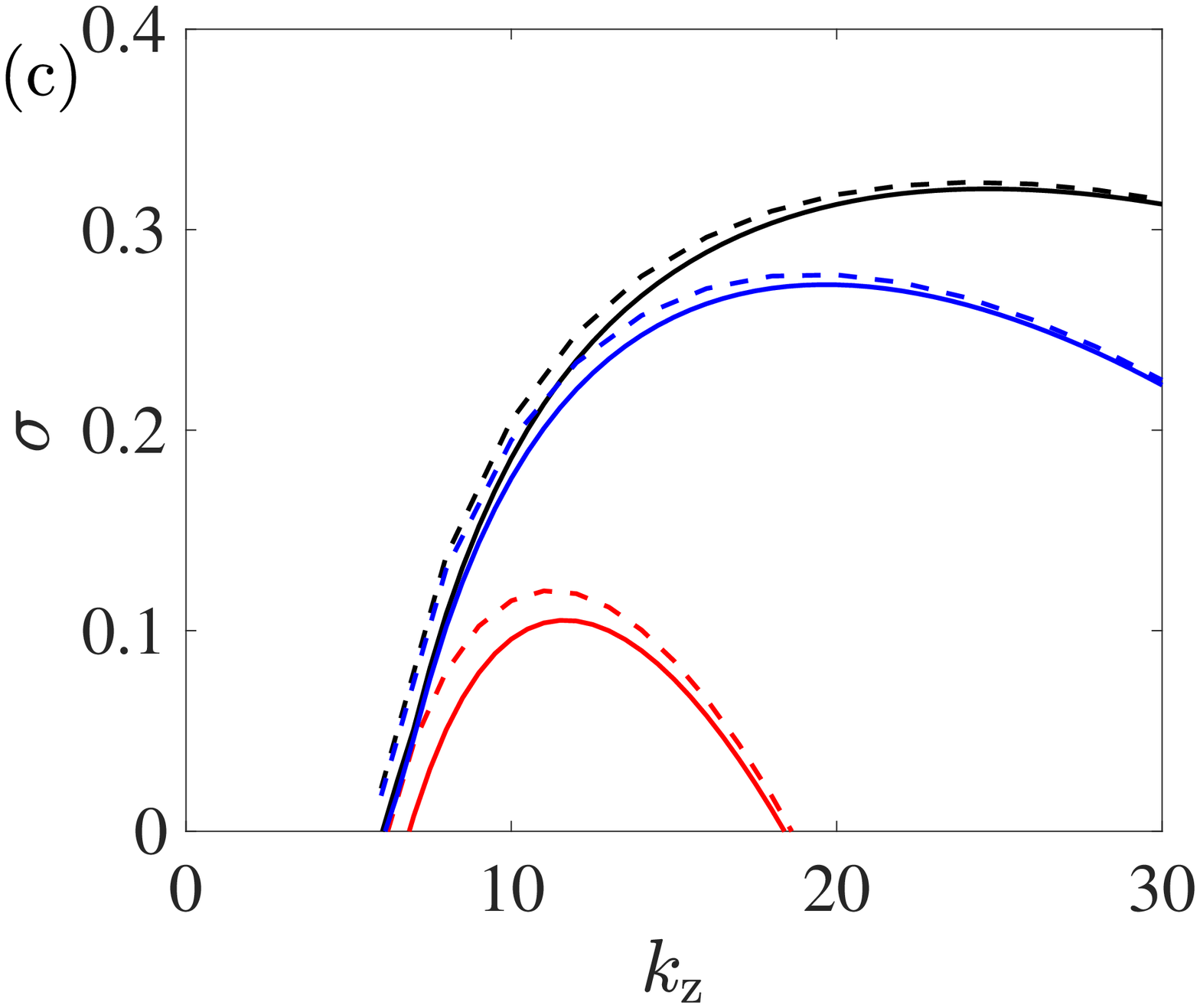}
   \caption{
   (a) Growth rate $\sigma$ of the inflectional instability as a function of the streamwise wavenumber $k_{\rm{x}}$ for $k_{\rm{z}}=0$, $N=1$, {$Pr=1$, $\tilde{f}=2$, $f=0$ (i.e., $\Omega=1$ and $\theta=90^{\circ}$)} for $Re=\infty$ (black), $Re=50$ (blue), and $Re=10$ (red). 
   The gray dash-dot line denotes the inviscid growth rate of the inflectional instability under the traditional approximation (i.e., $\tilde{f}=0$). 
   (b,c) Growth rate $\sigma$ of the inertial instability as a function of the vertical wavenumber $k_{\rm{z}}$ for $k_{\rm{x}}=0$, $N=1$, $\tilde{f}=4$, and $f=0.5$ {(i.e., $\Omega=2.02$ and $\theta=82.9^{\circ}$)} for (b) $Pr=1$ and $Re=\infty$ (black), $Re=1000$ (blue), and $Re=500$ (red), and (c) $Pr=10^{-6}$ and $Re=10^{4}$ (black), $Re=5000$ (blue), and $Re=1000$ (red). 
   Solid lines are numerical results, and dashed lines in (b) and (c) are asymptotic predictions from Eqs.~(\ref{eq:asymptotic_viscous_growth_Pr1}) and (\ref{eq:asymptotic_viscous_growth_Pr0}), respectively.
      }
              \label{Fig_viscous_growth}%
\end{figure*}
Now, we investigate how the viscosity modifies the inflectional and inertial instabilities. 
Figure \ref{Fig_viscous_growth} shows the growth rate of the inflectional and inertial instabilities for various parameters at different Reynolds numbers $Re$. 
For viscous cases, we fix the Prandtl number $Pr$ defined as $Pr=Pe/Re$.
We clearly see that the viscosity has a stabilizing effect on both instabilities. 
For the inflectional instability in Fig.~\ref{Fig_viscous_growth}a, the inviscid growth rate curve is increased for $\tilde{f}=2$ (black solid line) compared to that of the traditional case at $\tilde{f}=0$ (gray dashed line), while the viscous growth rate at $\tilde{f}=2$ decreases as $Re$ decreases. 
It is remarkable that the inflectional instability persists at low Reynolds number $Re=10$, and that the curve at $\tilde{f}=2$ has the same order of magnitude as the inviscid growth rate at $\tilde{f}=0$. 
Therefore, we can conclude the inflectional instability is stabilized by the viscosity while it is destabilized as $\tilde{f}$ increases.
This trend is summarized in Table \ref{table:1}.

Figures \ref{Fig_viscous_growth}b and c show the growth rate of the inertial instability at $k_{\rm{x}}=0$, $N=1$, $f=0.5$, and $\tilde{f}=4$ for two Prandtl numbers: $Pr=1$ and $Pr=10^{-6}$.
For both cases, the maximum growth rate is now attained at finite $k_{\rm{z}}$ due to a stronger stabilization by the viscosity at large wavenumber $k_{\rm{z}}$. 
For large Reynolds numbers and a unity Prandtl number ($Pr=1$), we can apply the multiple-scale analysis to perturbation equations and propose the viscous growth rate expressed in terms of the inviscid growth rate as
\begin{equation}
\label{eq:asymptotic_viscous_growth_general}
\sigma_{\rm{viscous}}=\sigma_{\rm{inviscid}}-\frac{k^{2}}{Re}+O\left(\frac{1}{Re^{2}}\right).
\end{equation}
This expression of the growth rate has already been proposed in other works \citep[see e.g.,][]{Arobone2012,Yim2016}. 
Using the asymptotic expression of the inviscid growth rate obtained from the WKBJ analysis in Sect.~\ref{sect:WKBJ}, we can express explicitly the viscous growth rate of the inertial instability at $k_{\rm{x}}=0$ as
\begin{equation}
\label{eq:asymptotic_viscous_growth_Pr1}
\sigma \big|_{Pr=1}=\sigma_{0}-\frac{\sigma_{1}}{k_{\rm{z}}}-\frac{k_{\rm{z}}^{2}}{Re}.
\end{equation} 
We see in Fig.~\ref{Fig_viscous_growth}b that this expression for the viscous growth rate (\ref{eq:asymptotic_viscous_growth_Pr1}) is in good agreement with numerical results at finite Reynolds numbers and $Pr=1$. 

   \begin{figure*}
   \centering
   \includegraphics[height=5cm]{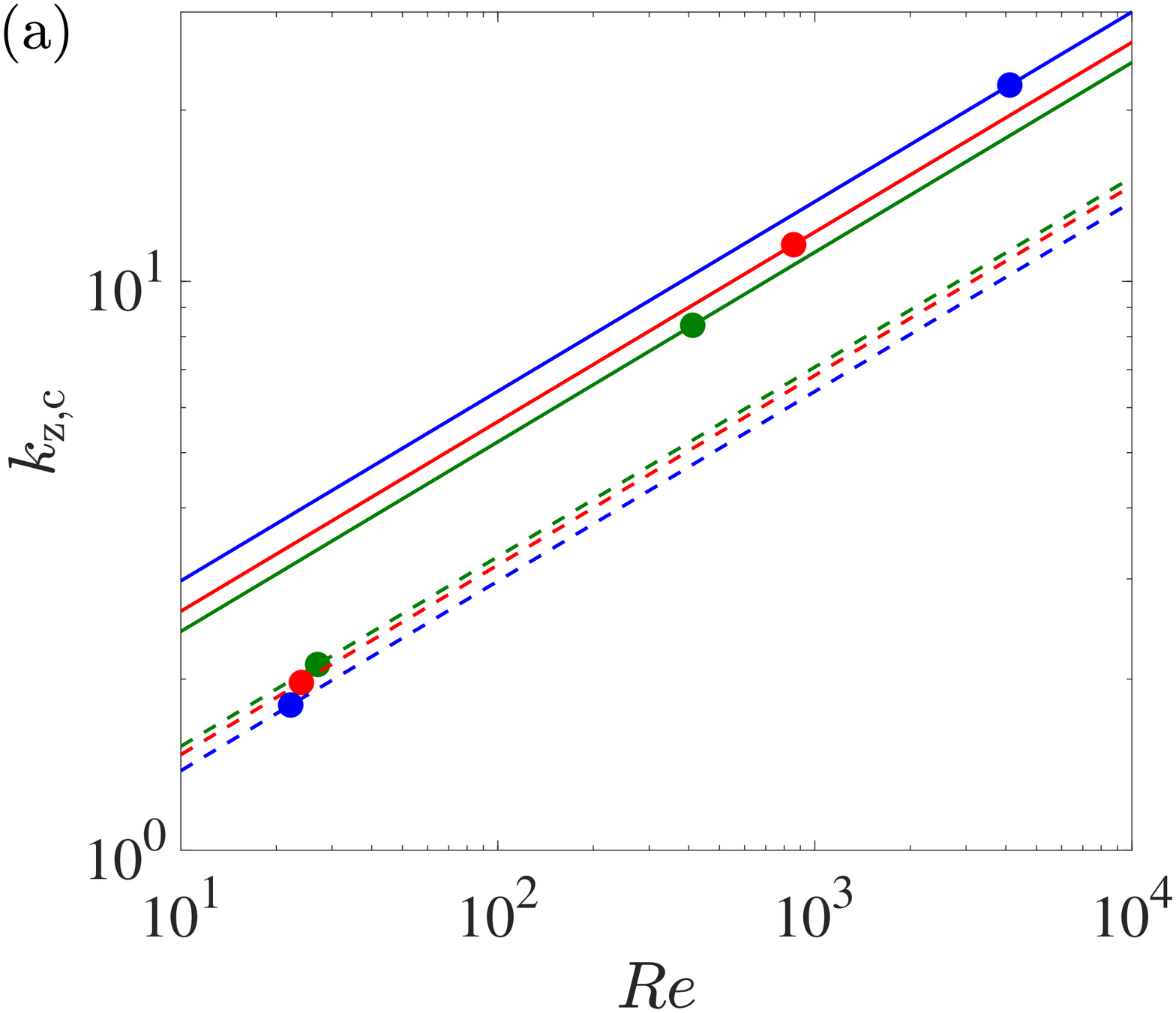}
   \includegraphics[height=5.1cm]{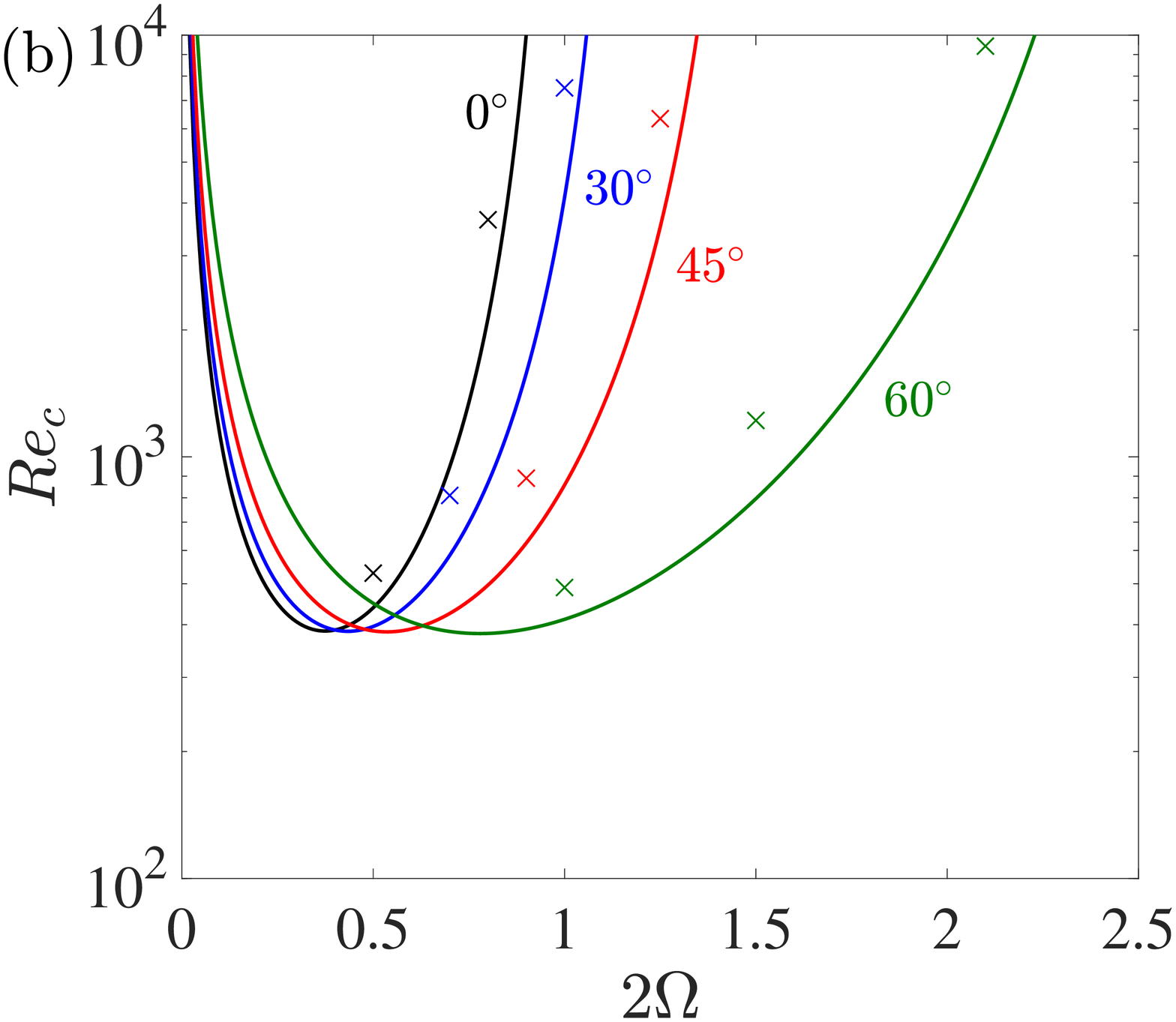}
   \includegraphics[height=5.1cm]{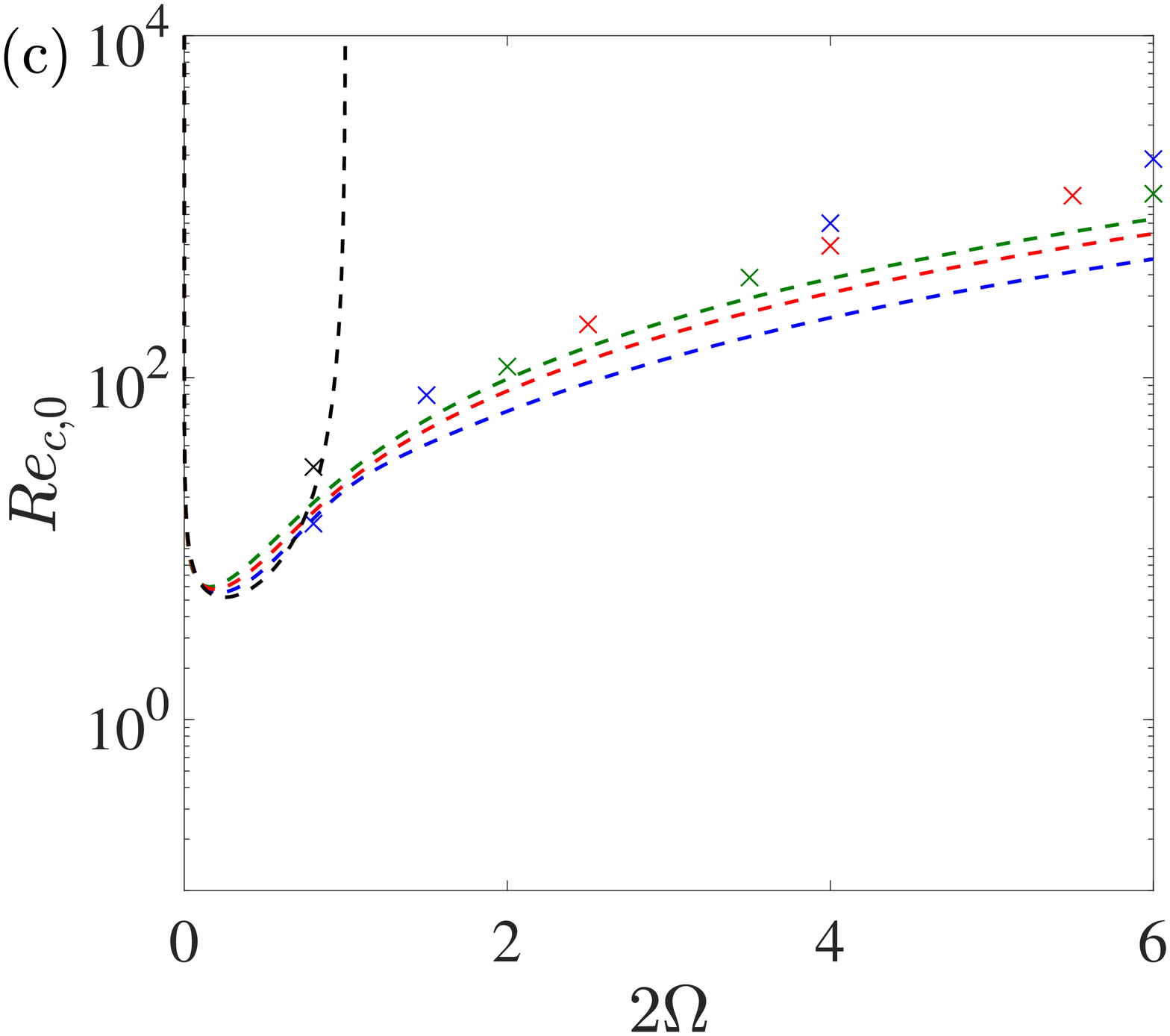}
      \caption{
      (a) Optimal wavenumbers $k_{\rm{z},c}$ from (\ref{eq:k_c}) (solid lines) and $k_{\rm{z},c0}$ from (\ref{eq:k_c0}) (dashed lines) as a function of $Re$ at $2\Omega=1$, $N=4$, and $k_{\rm{x}}=0$ for colatitudes $\theta$: $30^{\circ}$ (blue), $45^{\circ}$ (red), and $60^{\circ}$ (green).
      Filled circles denote the critical Reynolds numbers. 
      (b,c) Critical Reynolds number $Re_{c}$ as a function of the absolute Coriolis parameter $2\Omega$ for $N=4$, and (b) $Pr=1$, (c) $Pr=10^{-6}$ at different colatitudes $\theta$: $0^{\circ}$ (black), $30^{\circ}$ (blue), $45^{\circ}$ (red), and $60^{\circ}$ (green) from numerical results (crosses) and asymptotic results of $Re_{c}$ (solid lines in b) from (\ref{eq:Re_c}) and $Re_{c,0}$ (dashed lines in c) from (\ref{eq:Re_c0}). 
              }
         \label{Fig_critical_Re}
   \end{figure*}
From (\ref{eq:asymptotic_viscous_growth_Pr1}), we can further derive the optimal vertical wavenumber $k_{\rm{z},c}$, at which the maximum growth rate is attained, by solving the equation $\partial\sigma/\partial k_{\rm{z}}=0$:
\begin{equation}
\label{eq:k_c}
k_{\rm{z},c}=\left(\frac{\sigma_{1}Re}{2}\right)^{\frac{1}{3}}.
\end{equation}
Applying the optimal wavenumber $k_{\rm{z},c}$ to the viscous growth rate (\ref{eq:asymptotic_viscous_growth_Pr1}), we find the critical Reynolds number $Re_{c}$ above which the inertial instability occurs.
\begin{equation}
\label{eq:Re_c}
\begin{aligned}
&Re_{c}=\frac{27\sigma_{1}^{2}}{4\sigma_{0}^{3}}=\frac{27\left(m-\frac{1}{2}\right)^{2}f\left(\sigma_{0}^{2}+N^{2}+\tilde{f}^{2}\right)}{4\sigma^{5}_{0}\left[1+f^{2}\tilde{f}^{2}/\left(\sigma_{0}^{2}+N^{2}+\tilde{f}^{2}\right)^{2}\right]^{2}}.
\end{aligned}
\end{equation}
In Fig.~\ref{Fig_critical_Re}a, we display the optimal wavenumber $k_{\rm{z},c}$ (\ref{eq:k_c}) as a function of the Reynolds number $Re$ and the corresponding critical $Re_{c}$ at $2\Omega=1$, $N=4$, and $k_{\rm{x}}=0$ at various colatitudes $\theta$. 
In Fig.~\ref{Fig_critical_Re}b, we plot the critical Reynolds number $Re_{c}$ (\ref{eq:Re_c}) as a function of $2\Omega$ for different colatitudes. 
Asymptotic predictions from Eq.~(\ref{eq:Re_c}) are of the same order of magnitude as numerical results for $Re_{c}$, and the small differences come from the small values of $k_{\rm{z},c}$ predicted from the WKBJ analysis, which works well for large $k_{\rm{z}}$. 
However, the interest of the asymptotic expression of Eq.~(\ref{eq:Re_c}) is that it can provide information about the critical Reynolds number $Re_{c}$ in wide parameter spaces without exhaustive numerical computations. 

For very small Prandtl numbers, which is the relevant regime for the radiative zone of stars \citep[]{Lignieres1999}, we can similarly propose the viscous growth rate using the asymptotic growth rate (\ref{eq:WKBJ_4ODE_dispersion_1st}) in the limit $Pe=0$:
\begin{equation}
\label{eq:asymptotic_viscous_growth_Pr0}
\sigma \big|_{Pr\rightarrow 0}=\sigma_{0,0}-\frac{\sigma_{1,0}}{k_{\rm{z}}}-\frac{k_{\rm{z}}^{2}}{Re}.
\end{equation}  
In Fig.~\ref{Fig_viscous_growth}c, we see that this viscous growth rate agrees very well with numerical results at finite Reynolds numbers and $Pr=10^{-6}$. 
Moreover, we can further derive the optimal wavenumber $k_{\rm{z},c0}$ and the critical Reynolds number $Re_{c,0}$ in the limit $Pr\rightarrow0$ as follows:
\begin{equation}
\label{eq:k_c0}
k_{\rm{z},c0}=\left(\frac{\sigma_{1,0}Re}{2}\right)^{\frac{1}{3}},
\end{equation}
\begin{equation}
\label{eq:Re_c0}
Re_{c,0}=\frac{27\sigma_{1,0}^{2}}{4\sigma_{0,0}^{3}}=\frac{27\left(m_{0}-\frac{1}{2}\right)^{2}\left(f\sigma_{0,0}^{2}+\tilde{f}^{2}/2\right)}{4\sigma_{0,0}^{5}\left[1+\tilde{f}^{2}(f-1/2)^{2}/(\sigma_{0,0}^{2}+\tilde{f}^{2})^{2}\right]^{2}}.
\end{equation}
Figure \ref{Fig_critical_Re}a and c show the asymptotic predictions of $k_{\rm{z},c0}$ and $Re_{c,0}$ and comparisons with numerical results at $Pr=10^{-6}$.
The critical Reynolds number is roughly within the same order of magnitude as numerical results, but the difference is large since the predicted critical wavenumber $k_{\rm{z},c0}$ is very small, of order $O(1)$, which lies much below the validity range of the WKBJ approximation. 
However, we can see that the critical Reynolds number $Re_{c,0}$ increases with $2\Omega$ except at the northern pole $\theta=0^{\circ}$ where it is unstable in the range $0<2\Omega<1$.   
We also found both numerically and theoretically that $Re_{c,0}$ does not change significantly with colatitude $\theta$ at a given $\Omega$.

\section{Effective horizontal turbulent viscosity}
\label{sect:turbulent_viscosity}
\subsection{Turbulent viscosity induced by the inertial instability}
As small-amplitude perturbations grow due to the instability mechanism, they first follow the linear instability growth, then reach a saturated equilibrium state and undergo a transition to turbulence as the perturbation amplitude increases and nonlinear effects become crucial. 
The nonlinear saturation of instabilities has been investigated in different astrophysical contexts and has been used to predict the order of magnitude of angular momentum transport, mixing, or dynamo actions \citep[]{Spruit2002,Denissenkov2007,Zahn2007,Fuller2019}. 

In this section, we use our results on horizontal shear instabilities to derive an effective turbulent viscosity{. 
We consider perturbations that are developed by the horizontal shear instabilities and reach a saturated state due to nonlinear interactions, which we model here through a turbulent viscosity. 
We adopt the approach described by \citet{Spruit2002,Fuller2019} that allows the saturated state when }
the turbulent damping rate $\gamma_{\rm{turb}}$ equals the growth rate of the instability $\sigma$. 
{This} leads to the following expression of the local effective turbulent viscosity in the horizontal direction $\nu_{\rm{h},\rm{h}}$:
\begin{equation}
	\nu_{\rm{h},\rm{h}}=\frac{\sigma}{\bar{k}_{\rm{y}}^{2}},
\end{equation}
where $\bar{k}_{\rm{y}}$ is the horizontal wavenumber in the $y$-direction required for instability. 
We use here the notation $\nu_{\rm{h},\rm{h}}$ first introduced by \citet{Mathis2018} that corresponds to the turbulent transport triggered in the horizontal direction by the instability of a latitudinal horizontal shear. 
{We thus make the choice to {consider} the horizontal wavenumber $\bar{k}_{\rm{y}}$ in latitudinal $y$-direction to characterize the horizontal effective turbulent viscosity.}

We first consider the case of the effective turbulent viscosity $\nu_{\rm{h},\rm{h}}$ induced by the inertial instability, whose dispersion relation has been derived using the WKBJ approximation in Sect.~\ref{sect:WKBJ}. 
Since the horizontal shear flow $U(y)$ is inhomogeneous in the $y$-direction, the associated wavenumber ${k}_{\rm{y}}$ in the $y$-direction is also a function of $y$. 
Moreover, the local wavenumber ${k}_{\rm{y}}(y)$ only exists when the solution is wavelike in the regime between the two turning points. 
For the nondiffusive case at $Pe=\infty$, we have the WKBJ expression for $\hat{v}$ as
\begin{equation}
	\hat{v}(y)\sim\exp\left[\mathrm{i}\tilde{k}\int_{y_{t-}}^{y}\left(\sqrt{\Gamma+\frac{f^{2}\tilde{f}^{2}}{\sigma^{2}+N^{2}+\tilde{f}^{2}}}-\frac{f\tilde{f}}{\sqrt{\sigma^{2}+N^{2}+\tilde{f}^{2}}}\right)\mathrm{d}y\right].
\end{equation} 
Since the exponent changes with $y$, we define the horizontal wavenumber $\bar{k}_{\rm{y}}$ as the average of the absolute part of the exponent over the two turning points as follows:
\begin{equation}
\label{eq:k_y_Pe_inf}
	\bar{k}_{\rm{y}}\int_{y_{t-}}^{y_{t+}}\mathrm{d}y=\tilde{k}\int_{y_{t-}}^{y_{t+}}\left|\sqrt{\Gamma+\frac{f^{2}\tilde{f}^{2}}{\sigma^{2}+N^{2}+\tilde{f}^{2}}}-\frac{f\tilde{f}}{\sqrt{\sigma^{2}+N^{2}+\tilde{f}^{2}}}\right|\mathrm{d}y.
\end{equation}
Similarly, we compute $\nu_{\rm{h},\rm{h}}$ using the asymptotic dispersion relation (\ref{eq:WKBJ_4ODE_dispersion_1st}) in the highly diffusive limit $Pe\rightarrow0$:
\begin{equation}
\label{eq:k_y_Pe_0}
	\bar{k}_{0,\rm{y}}\int_{\tilde{y}_{t-}}^{\tilde{y}_{t+}}\mathrm{d}y=\widetilde{k}\int_{\widetilde{y}_{t-}}^{\widetilde{y}_{t+}}\left|\sqrt{{\widetilde{\Gamma}}}+\frac{\tilde{f}(U'-2f)}{2\sqrt{\sigma^{2}+\tilde{f}^{2}}}\right|\mathrm{d}y.
\end{equation}

   \begin{figure}
   \centering
   \includegraphics[height=5cm]{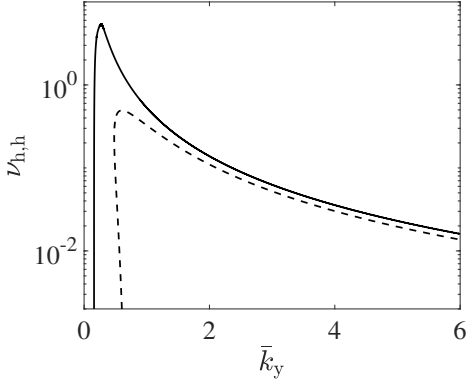}
      \caption{Effective turbulent viscosity $\nu_{\rm{h}}$ as a function of the horizontal wavenumber $\bar{k}_{\rm{y}}$ for $f=0.5$, $\tilde{f}=1$ {(i.e., $\Omega=0.559$ and $\theta=63.4^{\circ}$)}, and $k_{\rm{x}}=0$ for $Pe=\infty$ and $N=1$ (solid line), and as $Pe\rightarrow0$ (dashed line).              }
         \label{Fig_nuh_ky}
   \end{figure}
Figure \ref{Fig_nuh_ky} shows an example of $\nu_{\rm{h},\rm{h}}=\sigma/\bar{k}^{2}_{\rm{y}}$ as a function of $\bar{k}_{\rm{y}}$ in the two limits $Pe=\infty$ (solid line) and $Pe\rightarrow0$ (dashed line).
The asymptotic growth rates $\sigma$ from Eq.~(\ref{eq:WKBJ_2ODE_dispersion_1st}) for $Pe=\infty$ and from Eq.~({\ref{eq:WKBJ_4ODE_dispersion_1st}}) for $Pe\rightarrow0$ derived with the WKBJ analyses are used to compute $\nu_{\rm{h},\rm{h}}$. 
We see that the effective turbulent viscosity reaches its maximum at a finite $\bar{k}_{\rm{y}}$ and decreases monotonically with $\bar{k}_{\rm{y}}$. 
The monotonic decrease of $\nu_{\rm{h},\rm{h}}$ is due to the fact that the growth rate reaches asymptotically the maximum growth rate $\sigma_{\max}$ as $k_{\rm{z}}\rightarrow\infty$ while the horizontal wavenumber $\bar{k}_{\rm{y}}$, proportional to $k_{\rm{z}}$, goes to infinity.

\begin{figure*}
   \centering
   \includegraphics[height=4.7cm]{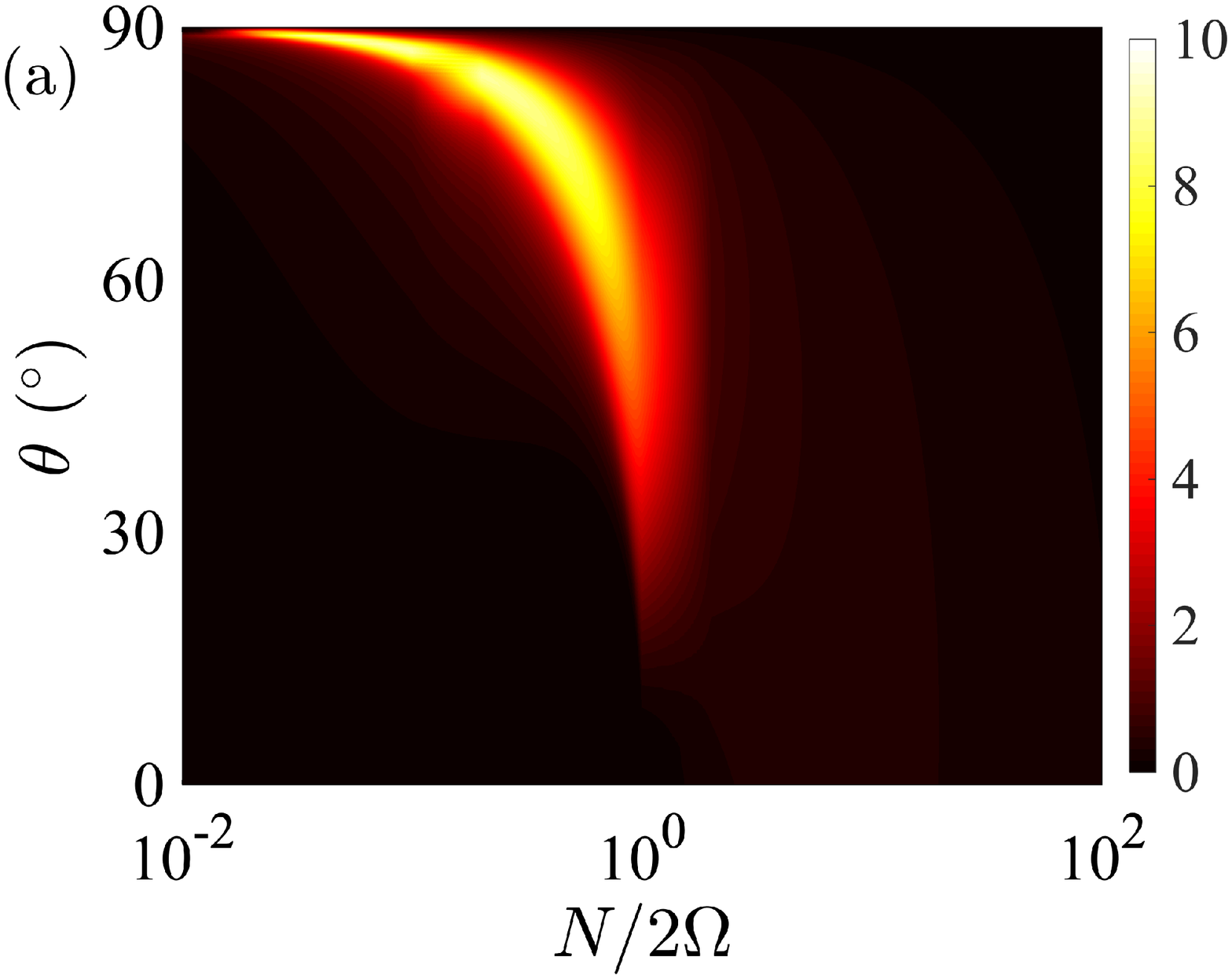}
   \includegraphics[height=4.7cm]{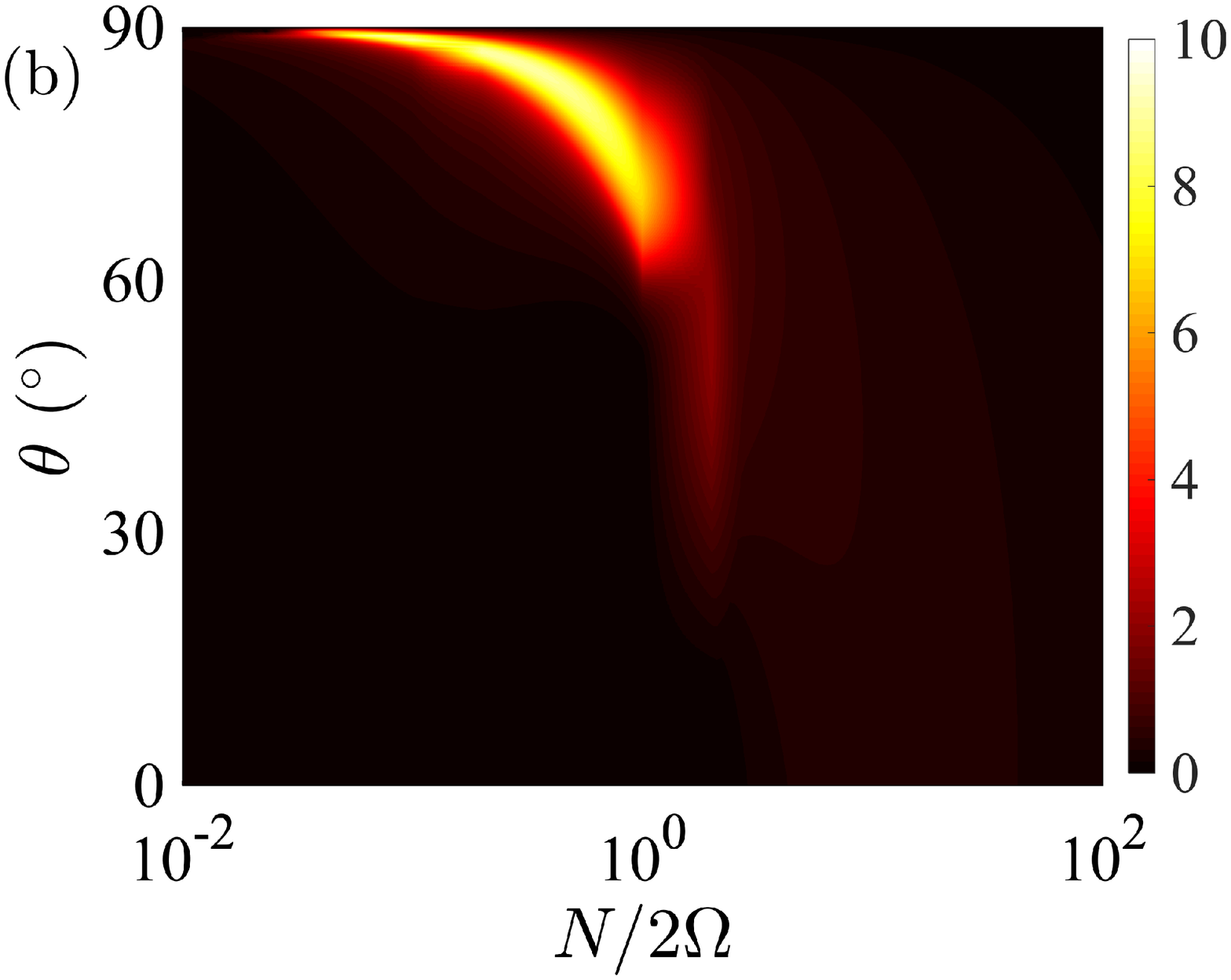}
   \includegraphics[height=4.7cm]{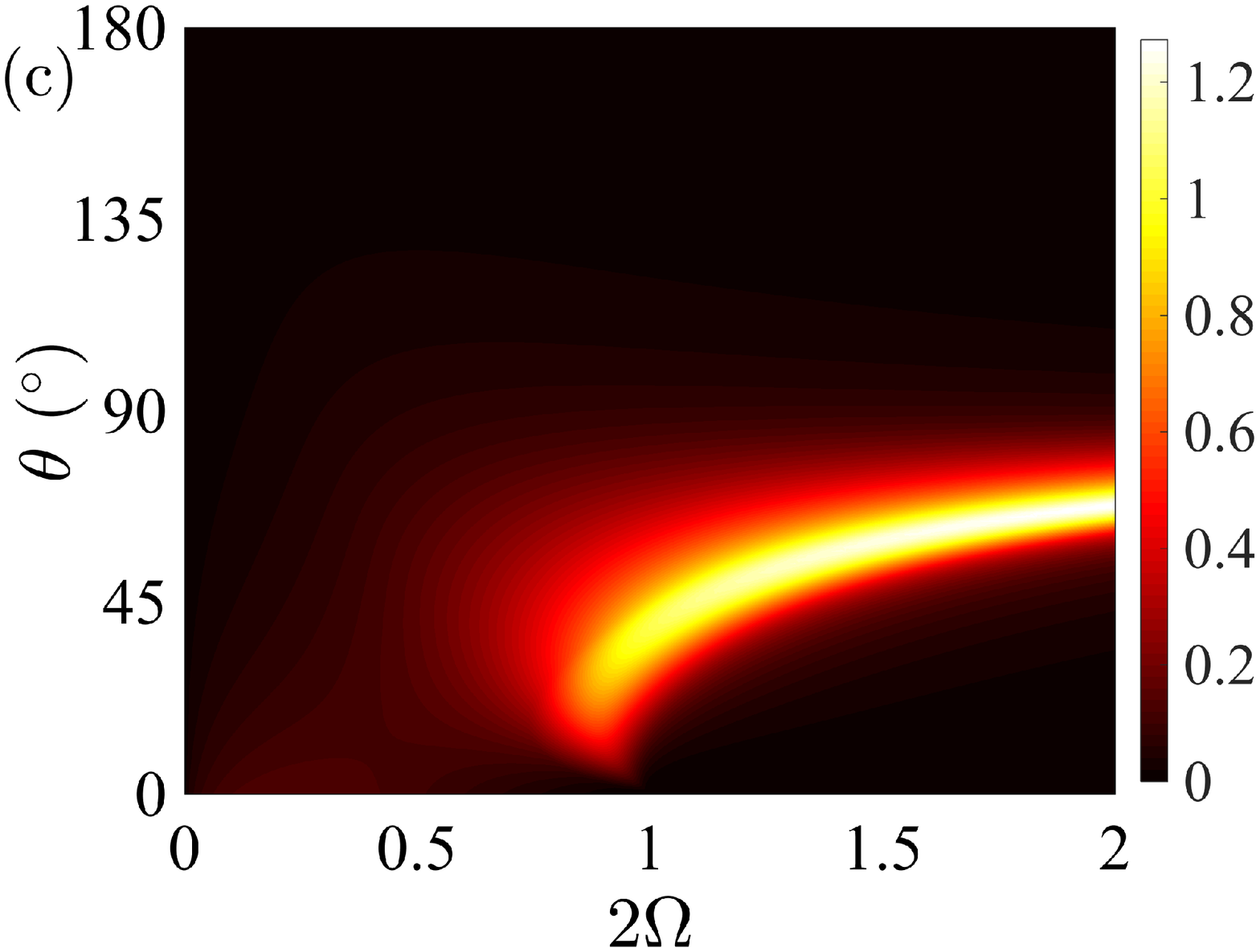}
   \caption{(a,b) Contours of the maximum of the effective turbulent viscosity $\max(\nu_{\rm{h},\rm{h}})$ in the parameter space $(N/2\Omega,\theta)$ for $Pe=\infty$ and (a) $N=1$, (b) $N=2$. 
   (c) Contours of the maximum of the effective turbulent viscosity $\max(\nu_{\rm{h},\rm{h}})$ in the parameter space $(2\Omega,\theta)$ as $Pe\rightarrow0$. 
   }
              \label{Fig_f_fs_nuh}%
\end{figure*}
Similarly to \citet{Fuller2019}, where the minimum of the wavenumber $\bar{k}_{\rm{y}}$ is considered to get $\nu_{\rm{h},\rm{h}}$, we consider the maximum peak of $\nu_{\rm{h},\rm{h}}$ as the representative value of the effective horizontal turbulent viscosity for given values of the physical parameters such as $f$ and $\tilde{f}$.  
In Fig.~\ref{Fig_f_fs_nuh}a and b, we plot contours of $\max(\nu_{\rm{h},\rm{h}})$ in the parameter space of $(N/2\Omega,\theta)$ for two different stratified fluids with $N=1$ and $N=2$ without thermal diffusion (i.e., $Pe=\infty$). 
For both cases, the turbulent viscosity $\max(\nu_{\rm{h},\rm{h}})$ reaches its maximum around the colatitude $\theta\simeq80^{\circ}$ in the northern hemisphere close to the equator around the value $N/2\Omega\sim 1$.
The effective turbulent viscosity in the southern hemisphere $\theta>90^{\circ}$ is zero since there is no instability for negative $f=2\Omega\cos\theta$. 
This implies that a strong turbulence and a large effective turbulent viscosity are expected near the equator in the northern hemisphere due to strong instability. 
In Fig.~\ref{Fig_f_fs_nuh}c, we plot $\max(\nu_{\rm{h},\rm{h}})$ in the parameter space of $(2\Omega,\theta)$ for high-diffusivity fluids in the limit $Pe\rightarrow0$ using Eq.~(\ref{eq:k_y_Pe_0}). 
In this case, the effective turbulent viscosity is not zero in the southern hemisphere due to the presence of the inertial instability for negative $f$.
The effective turbulent viscosity $\max(\nu_{\rm{h},\rm{h}})$ has a maximum around $\theta\simeq70^{\circ}$. 
This emphasizes that non-traditional effects with the positive horizontal Coriolis parameter $\tilde{f}>0$ play a crucial role on the effective turbulent viscosity, especially near the equator. 

\begin{figure*}
   \centering
   \includegraphics[height=4.7cm]{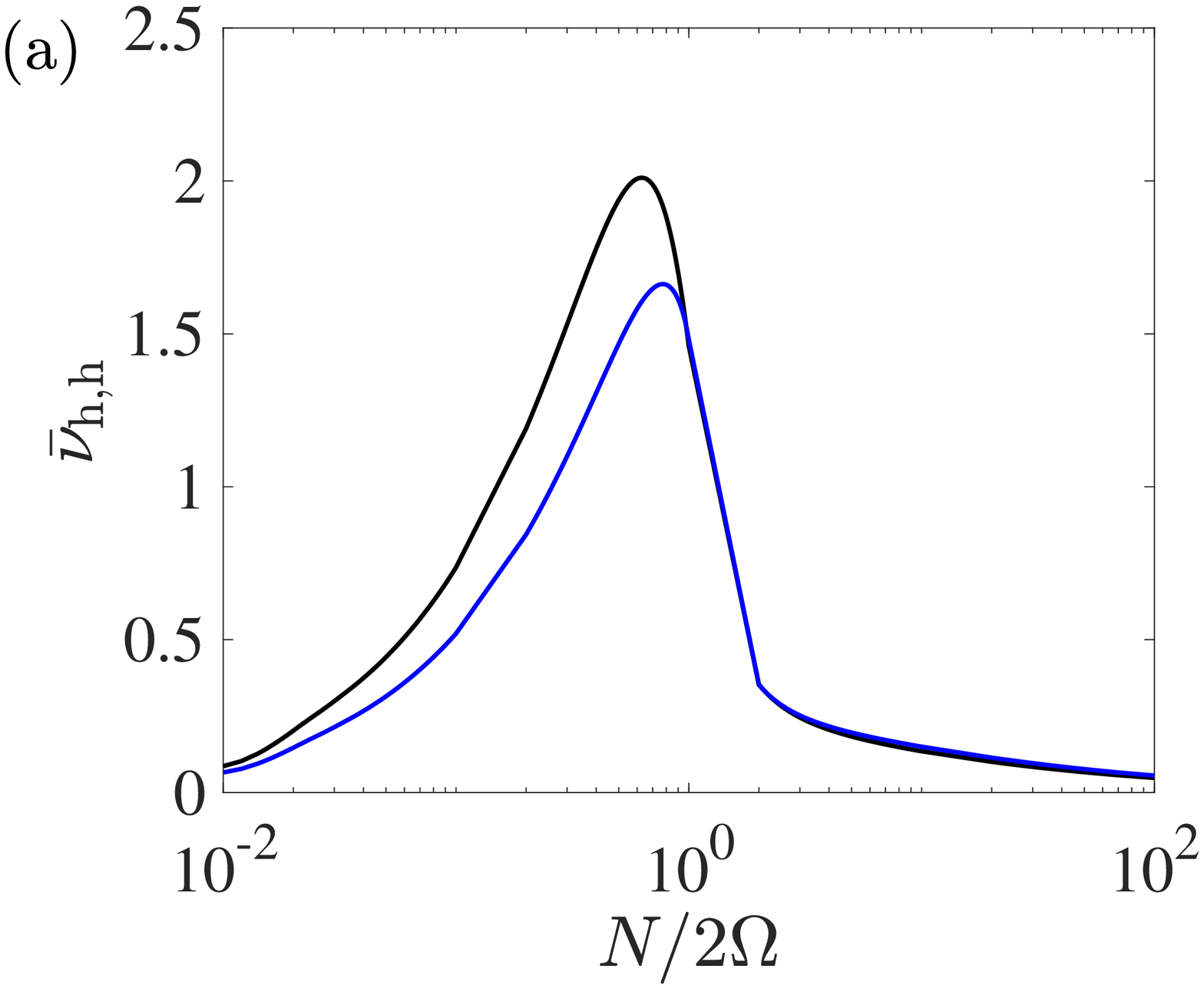}
   \includegraphics[height=4.7cm]{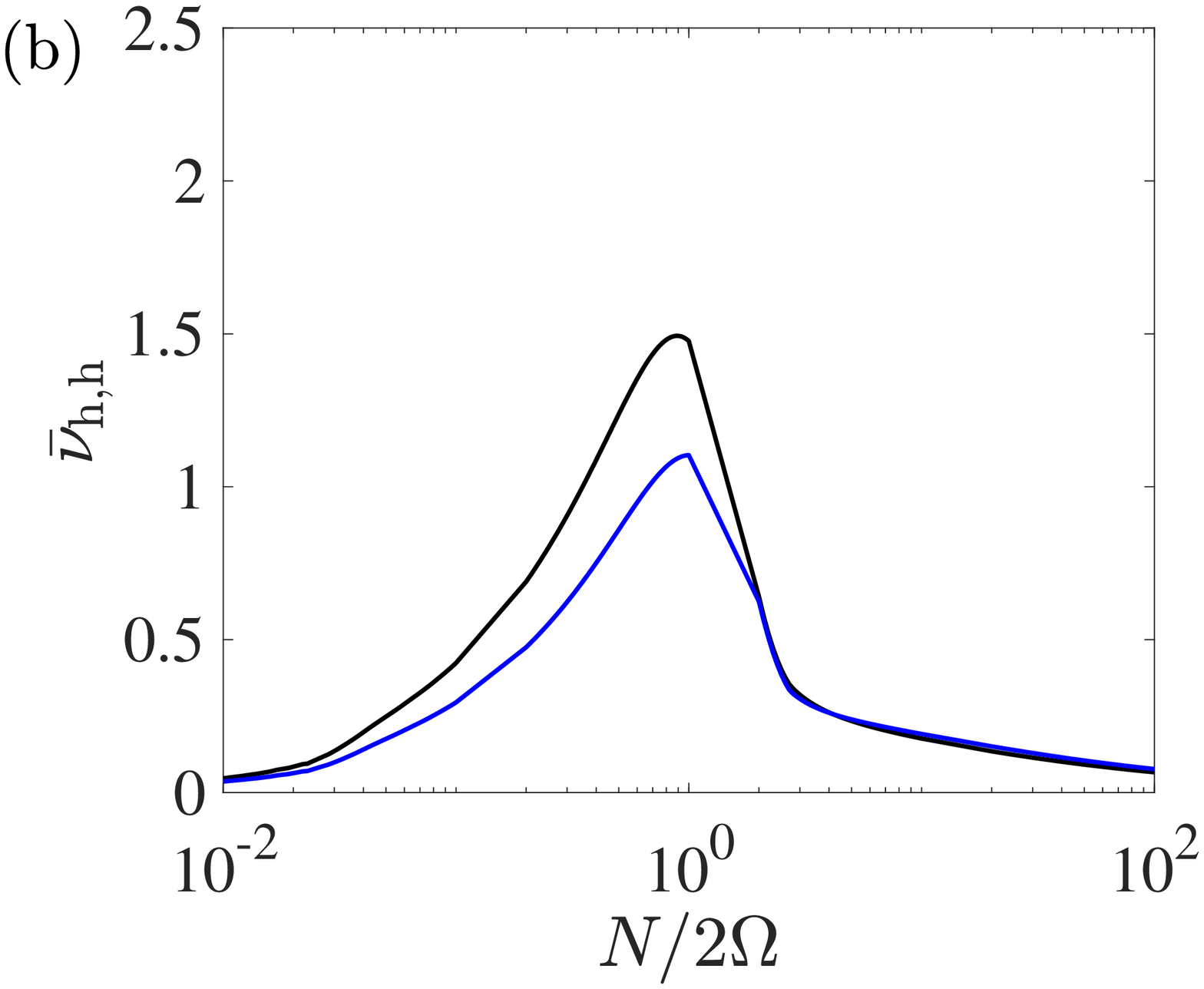}
   \includegraphics[height=4.7cm]{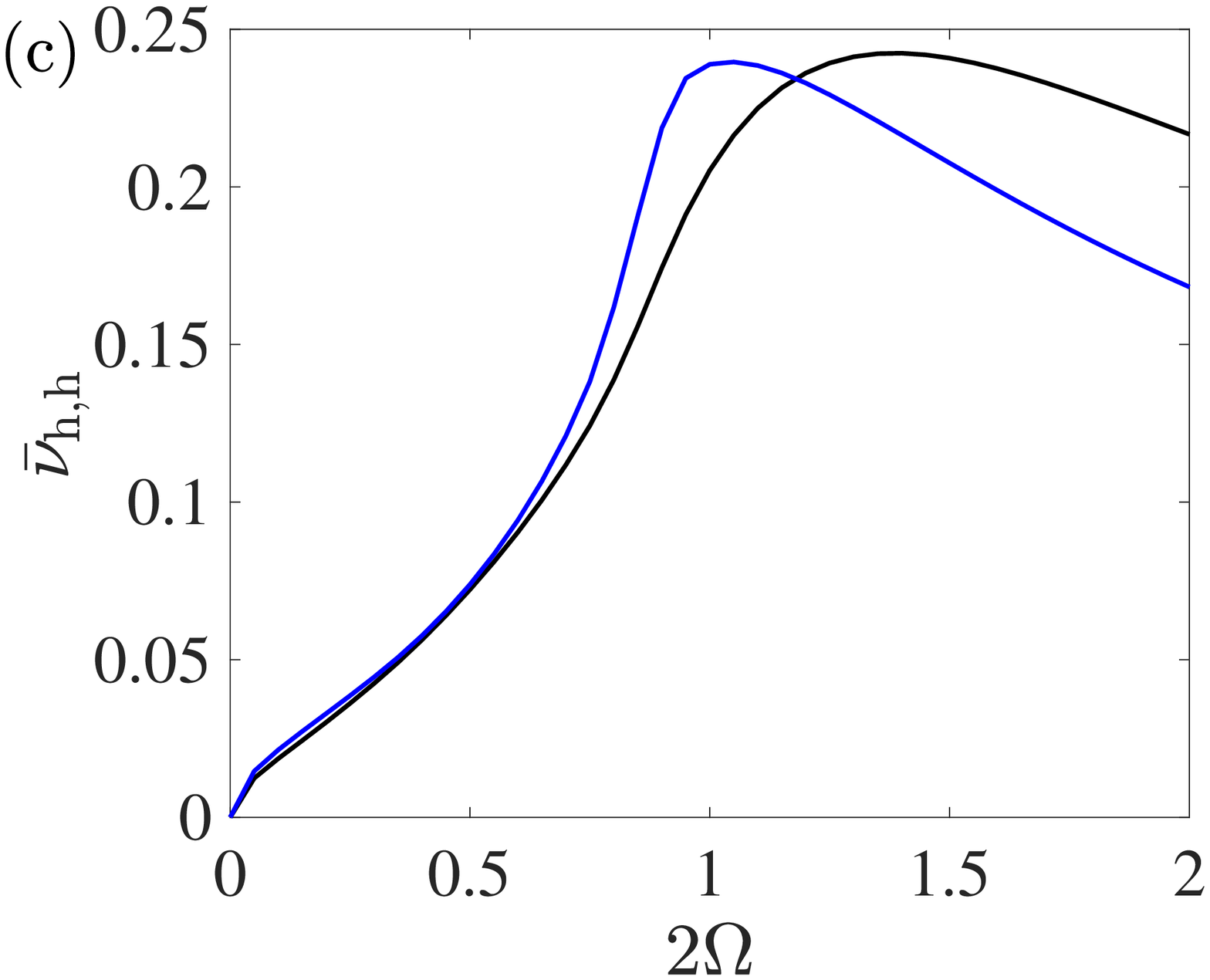}
   \caption{The latitudinally-averaged turbulent viscosities $\bar{\nu}_{\rm{h},\rm{h}}$ (black) and $\bar{\bar{\nu}}_{\rm{h},\rm{h}}$ (blue) computed from contours in Fig.~\ref{Fig_f_fs_nuh}.
   }
              \label{Fig_nu_avg}%
\end{figure*}
From $\nu_{\rm{h},\rm{h}}$, the latitudinally-averaged turbulent viscosity can be computed in two ways. 
On the one hand, we define the averaged turbulent viscosity $\bar{\nu}_{\rm{h},\rm{h}}$ following \citet{Zahn1992} as
\begin{equation}
\label{eq:averaged_turbulent_viscosity_sin3}
	\bar{\nu}_{\rm{h},\rm{h}}\int_{0}^{\pi}\sin^{3}\theta \mathrm{d}\theta=\int_{0}^{\pi}\max(\nu_{\mathrm{h},\rm{h}})\sin^{3}\theta  \mathrm{d}\theta.
\end{equation}
This average was used to compute the transport of angular momentum. 
On the other hand, we follow the definition by \citet{Mathis2018}:
\begin{equation}
\label{eq:averaged_turbulent_viscosity}
	\bar{\bar{\nu}}_{\rm{h},\rm{h}}\int_{0}^{\pi}\sin\theta  \mathrm{d}\theta=\int_{0}^{\pi}\max(\nu_{\mathrm{h},\rm{h}})\sin\theta  \mathrm{d}\theta,
\end{equation}
where $\bar{\bar{\nu}}_{\rm{h},\rm{h}}$ is the averaged turbulent viscosity obtained in the way the mean transport of chemicals is computed.  
In Fig.~\ref{Fig_nu_avg}, we plot these averaged viscosities.
For the non-diffusive case ($Pe=\infty$), $\bar{\nu}_{\rm{h},\rm{h}}$ is larger than $\bar{\bar{\nu}}_{\rm{h},\rm{h}}$ and they are reduced as $N$ increases.
The maxima of the two viscosities are reached around $N/2\Omega=1$ for both $N=1$ and $N=2$. 
For $Pe\rightarrow0$, the maximum is much lower than that for $Pe=\infty$, and it is reached around $2\Omega=1$.

\subsection{Turbulent viscosity induced by the inflectional instability}
   \begin{figure}
   \centering
   \includegraphics[height=5.5cm]{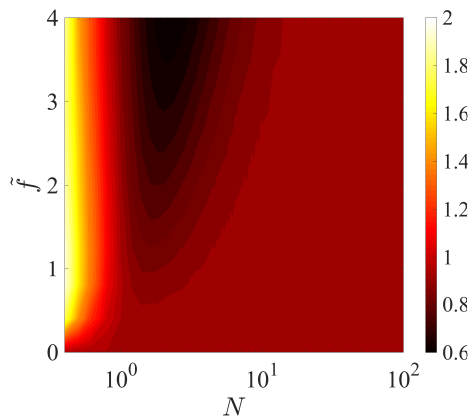}
      \caption{Contours of the maximum of the effective turbulent viscosity $\max(\nu_{\mathrm{h},\rm{h}})$ induced by the inflectional instability in the parameter space of $(N,\tilde{f})$ for $f=0$ (i.e., on the equator at $\theta=90^{\circ}$) and $Pe=\infty$.              }
         \label{Fig_nuh_inflectional}
   \end{figure}
The inflectional instability can also induce a turbulent transport in stellar radiation zones.
Therefore, we also compute the effective turbulent viscosity based on the growth rate $\sigma$ of the inflectional instability.
The issues with the inflectional instability are that we do not have an analytic expression for the growth rate $\sigma$, and it is difficult to define systematically the horizontal wavenumber $\bar{k}_{\rm{y}}$. 
One possible candidate for $\bar{k}_{\rm{y}}$ is to choose $\bar{k}_{\rm{y}}\simeq k_{\rm{x}}$ based on the solution's asymptotic behavior $\hat{v}(y)\sim\exp(-k_{\rm{x}}|y|)$ as $|y|\rightarrow\infty$. 
From numerical computations of the maximum growth rate of the inflectional instability in the parameter space $(k_{\rm{x}},k_{\rm{z}})$, we display in Fig.~\ref{Fig_nuh_inflectional} the effective turbulent viscosity $\max(\nu_{\rm{h},\rm{h}})=\sigma_{\max}/\bar{k}_{\rm{y}}^{2}$ in the parameter space of $(N,\tilde{f})$ at $Pe=\infty$ and $f=0$. 
It is important to note that the maximum growth rate $\sigma_{\max}$ of the inflectional instability is independent of $f$ at $Pe=\infty$. 
We see that at a fixed $N$, $\max(\nu_{\rm{h},\rm{h}})$ increases with $\tilde{f}$ for a weak stratification with $N<1$ while it decreases with $\tilde{f}$ for a strong stratification with $N>1$. 
The effective turbulent viscosity $\max(\nu_{\rm{h},\rm{h}})$ is generally of order of the unity $O(1)$ in the parameter space of $(N,\tilde{f})$ and is smaller than the effective turbulent viscosity $\max(\nu_{\rm{h},\rm{h}})$ induced by the inertial instability. 
We also verified numerically that the $\max(\nu_{\rm{h},\rm{h}})$ of the inflectional instability decreases as the thermal diffusivity becomes finite and small. 
Therefore, we can conjecture that the effective turbulent viscosity induced by the inertial instability will be larger than that induced by the inflectional instability. 

\section{Discussion with dimensional parameters}
\label{sect:discussion}
\subsection{Inertial instability with latitudinal differential rotation}
We discussed in the previous sections how the inertial instability is developed at a given latitude $\theta$ when the base shear of Eq.~(\ref{eq:base_shear}) is considered.
In our analysis, we investigated the growth rate of the inertial instability in the dimensionless form by taking the velocity scale $U_{0}$, length scale $L_{0}$, and time scale $t_{0}=L_{0}/U_{0}$.
{We note that the length scale $L_{0}$ is different from that of global large-scale shear flows in stellar interiors as we consider small-scale shear flows and associated perturbations on a local plane.}
An important thing to note is that the base flow with positive shear $S_{0}=U_{0}/L_{0}$ at $y=0$ is used in the normalization. 
This implies that the velocity in the azimuthal direction always decreases as the colatitude $\theta$ increases. 
The latitudinal differential rotation in stars is, however, different from this case. 
For instance, stars can have a conical differential rotation $\bar{\Omega}(\theta)$ in the simplest form as
\begin{equation}
\label{eq:differential_rotation}
\bar{\Omega}(\theta)=\Omega_{0}\left(1+\epsilon\sin^{2}\theta\right),
\end{equation}
where $\epsilon>0$ corresponds to the solar-like case, in which the equator rotates faster than the pole (e.g., $\epsilon\simeq0.3$ for the Sun), while $\epsilon<0$ corresponds to the anti-solar case, where the equator rotates slower than the pole \citep[]{Guenel2016}.
If instead of being constant, $\epsilon$ is proportional to $r^2$, one obtains the simplest form of cylindrical differential rotation \citep{BaruteauRieutord2013}.
At a constant radius, the latitudinal differential rotation follows the same law as in the conical case.

In the local frame rotating with $\Omega_{0}$ at the radius $r=R$ and the colatitude $\theta$, the relative azimuthal velocity $U_{\varphi}$ is
\begin{equation}
U_{\varphi}=R\epsilon \Omega_{0}\sin^{3}\theta,
\end{equation}
\citep[see also][]{Hypolite2018}.
By considering the increment in the latitudinal direction on the local frame:
\begin{equation}
\mathrm{d} y\simeq -R\mathrm{d} \theta,
\end{equation}
we obtain the base shear
\begin{equation}
S(\theta)=\frac{\mathrm{d}U_{\varphi}}{\mathrm{d} y}\simeq-\epsilon \Omega_{0}3\sin^{2}\theta\cos\theta.
\end{equation}
The latitudinal shear $S$ is negative (resp. positive) in the northern hemisphere and positive (resp. negative) in the southern hemisphere if $\epsilon>0$ (resp. $\epsilon<0$).
Moreover, the latitudinal shear $S$ is zero at the equator since the differential rotation $\bar{\Omega}$ of Eq.~(\ref{eq:differential_rotation}) is either at the maximum when $\epsilon>0$ or the minimum when $\epsilon<0$. 

   \begin{figure}
   \centering
   \includegraphics[height=5.5cm]{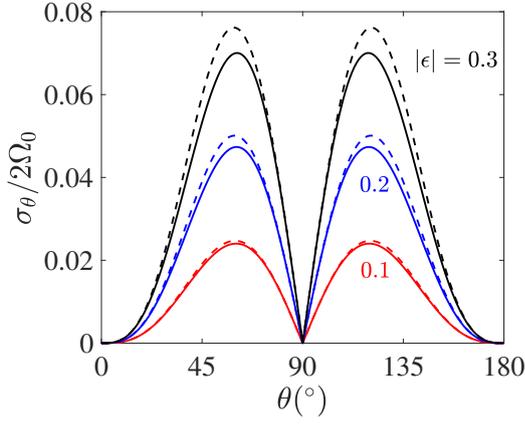}
      \caption{The growth rate $\sigma_{\theta}$ as a function of the colatitude $\theta$ for various values of $\epsilon$. 
      Solid lines denote $\sigma_{\theta}$ with $\epsilon>0$ and dashed lines denote $\sigma_{\theta}$ with $\epsilon<0$.}
         \label{Fig_growth_latitudinal}
   \end{figure}
The growth rate of Eq.~(\ref{eq:WKBJ_4ODE_dispersion_1st}) that we derived from the WKBJ analysis is local but the growth rate induced by the latitudinal differential rotation of Eq.~(\ref{eq:differential_rotation}) needs to be obtained by the global stability analysis \citep[see e.g.,][]{Guenel2016}.
However, we can approximately predict the growth rate ${\sigma}_{\theta}$ in the stellar radiation zones using the latitudinal shear $S$ and the maximum growth rate of Eq.~(\ref{eq:WKBJ_4ODE_sigma00}), which can be expressed in a dimensional form as follows:
\begin{equation}
\label{eq:growth_rate_dimensional}
{\sigma}_{\theta}=\sqrt{\frac{f_{\rm{v},0}(S-f_{\rm{v},0})-{f}_{\rm{h},0}^{2}+\sqrt{\left[f_{\rm{v},0}(S-f_{\rm{v},0})-{f}_{\rm{h},0}^{2}\right]^{2}+S^{2}{f}_{\rm{h},0}^{2}}}{2}}.
\end{equation}
Eq.~(\ref{eq:growth_rate_dimensional}) can further be expanded in terms of the stellar rotation $\Omega_{0}$, the colatitude $\theta$, and $\epsilon$ as
\begin{equation}
\begin{aligned}
2\left(\frac{{\sigma}_{\theta}}{2\Omega_{0}}\right)^{2}&=-1-\frac{3}{2}\epsilon\sin^{2}\theta\cos^{2}\theta\\
&+\sqrt{\left[1+\frac{3}{2}\epsilon\sin^{2}\theta\cos^{2}\theta\right]^{2}+\frac{9}{4}\epsilon^{2}\sin^{6}\theta\cos^{2}\theta}.
\end{aligned}
\end{equation}
Figure \ref{Fig_growth_latitudinal} shows the growth rate $\sigma_{\theta}$ versus the colatitude $\theta$ for various $\epsilon$.
The growth rate $\sigma_{\theta}$ increases with $|\epsilon|$ and reaches its maximum around $\theta\simeq60^{\circ}$ and $120^{\circ}$. 
It is zero at the poles and the equator. 
We see that, for the same $|\epsilon|$, the growth rate of the anti-solar case for $\epsilon<0$ is slightly larger than that of the solar-like case for $\epsilon>0$.

   \begin{figure}
   \centering
   \includegraphics[height=5.5cm]{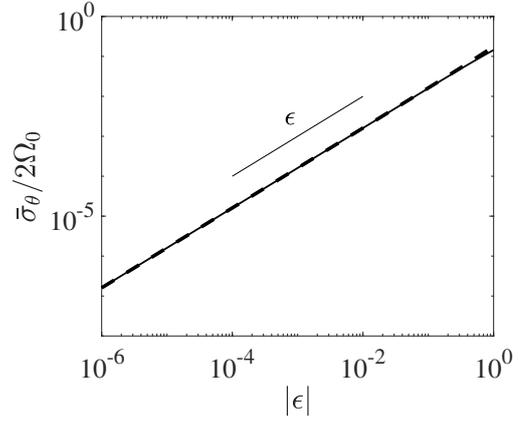}
      \caption{The latitudinally-averaged growth rate $\bar{\sigma}_{\theta}$ as a function of $|\epsilon|$ for the solar-like case ($\epsilon>0$, solid line) and the anti-solar case ($\epsilon<0$, dashed line).}
         \label{Fig_growth_latitudinal_avg}
   \end{figure}
Furthermore, the latitudinally-averaged\footnote{We here use the latitudinal average defined by \cite{Zahn1992} for quantities related to angular momentum. For instance, he defined the shellular rotation as ${\overline\Omega}\left(r\right)=\int_{0}^{\pi}\Omega\left(r,\theta\right)\sin^3\theta{\rm d}\theta/\int_{0}^{\pi}\sin^3\theta{\rm d}\theta$.} growth rate $\bar{\sigma}_{\theta}$ can be computed as 
\begin{equation}
\bar{\sigma}_{\theta}\int_{0}^{\pi}\sin^{3}\theta\mathrm{d}\theta=\int_{0}^{\pi}{\sigma}_{\theta}\sin^{3}\theta\mathrm{d}\theta.
\end{equation}
In Fig.~\ref{Fig_growth_latitudinal_avg}, we display $\bar{\sigma}_{\theta}$ as a function of $\epsilon$ for the solar-like and anti-solar cases. 
In the range $|\epsilon|<1$, $\bar{\sigma}_{\theta}$ almost does not depend on the sign of $\epsilon$.
The growth rate $\bar{\sigma}_{\theta}$ shows a linear relation with $|\epsilon|$ as
\begin{equation}
\label{eq:sigma_theta_epsilon}
\frac{\bar{\sigma}_{\theta}}{2\Omega_{0}}\simeq a_{\epsilon}|\epsilon|,
\end{equation}
where $a_{\epsilon}\simeq0.16$.
{We can derive a similar scaling law for the growth rate $\bar{\sigma}_{\theta}$ in the limit $Pe\rightarrow\infty$, more relevant to the solar tachocline case \citep[][]{Garaud2020}, if we use the growth rate (\ref{eq:WKBJ_2ODE_dispersion_1st_leading}) and a proper value of $N$.}

\subsection{Turbulent viscosities and characteristic time of turbulent transport}
Using the averaged growth rate $\bar{\sigma}_{\theta}$, we can define the turbulent viscosities induced by the horizontal shear due to the latitudinal differential rotation of Eq.~\eqref{eq:differential_rotation}.
Consistently with the notations from \citet{Mathis2018}, we define the turbulent viscosity in the latitudinal (horizontal) direction $\bar{\nu}_{\rm{h},\rm{h}}$ and the turbulent viscosity in the radial (vertical) direction $\bar{\nu}_{\rm{v},\rm{h}}$. 
Following the assumption proposed by \citet{Spruit2002,Fuller2019} that the shear instability grows and the momentum balances in nonlinear regime with the turbulent Reynolds stress, we find
\begin{equation}
\bar{\nu}_{\rm{h},\rm{h}}=\frac{\bar{\sigma}_{\theta}}{k_{\perp}^{2}}=\bar{\sigma}_{\theta}l_{\perp}^{2},
\end{equation}
\begin{equation}
\bar{\nu}_{\rm{v},\rm{h}}=\frac{\bar{\sigma}_{\theta}}{k_{\parallel}^{2}}=\bar{\sigma}_{\theta}l_{\parallel}^{2},
\end{equation}
where $\parallel$ denotes the direction parallel to the stratification, $\perp$ denotes the direction perpendicular to the stratification, $k_{\perp}$ and $k_{\parallel}$ are the characteristic wavenumbers in the horizontal and vertical directions, and $l_{\perp}$ and $l_{\parallel}$ are the length scales in the horizontal and vertical directions, respectively. 
{We note here that we consider small length scales of the unstable modes that trigger turbulent transport, the length scales different from those of global large-scale shear flows.}
The eddy viscosities are directly proportional to the horizontal shear as in the prescription derived by \cite{Mathis2004} (see their Eq. 18). Note that it would also be possible to define diffusion coefficients for chemicals $\bar{D}_{\rm h,h}$ and $\bar{D}_{\rm v,h}$; for this one should compute $\tilde{\sigma}_{\theta}=\int_{0}^{\pi}{\sigma}_{\theta}\sin\theta{\rm d}\theta/\int_{0}^{\pi}\sin\theta{\rm d}\theta$.\\   

While the turbulent viscosities depend on the direction with the scaling 
\begin{equation}
\frac{\bar{\nu}_{\rm{v},\rm{h}}}{\bar{\nu}_{\rm{h},\rm{h}}}=\frac{l_{\parallel}^{2}}{l_{\perp}^{2}},
\end{equation}
\citep[see also,][]{Mathis2018}, the dynamical time scale $\tau$ that characterizes the turbulence induced by the horizontal shear, and the transport of momentum and chemicals it triggers both in the vertical and latitudinal directions, is
\begin{equation}
\label{eq:tau}
\tau=\frac{l_{\perp}^{2}}{\bar{\nu}_{\rm{h},\rm{h}}}=\frac{l_{\parallel}^{2}}{\bar{\nu}_{\rm{v},\rm{h}}}=\frac{1}{\bar{\sigma}_{\theta}},
\end{equation}
which is simply the inverse of the growth rate $\bar{\sigma}_{\theta}$.

Using the relation of Eq.~(\ref{eq:sigma_theta_epsilon}), we can express the characteristic time $\tau$ in terms of $\Omega_{0}$ as
\begin{equation}
\label{eq:tau_scaling}
\tau=\frac{1}{2\Omega_{0}a_{\epsilon}|\epsilon|}\simeq\frac{\tau_{0}}{2|\epsilon|},
\end{equation}
where $\tau_{0}$ denotes the rotation period of the star at the pole. 
We note that the characteristic time scale $\tau$ is similar to that proposed by \citet{Mathis2018} in the form $\tau=1/S$ where $S(r,\theta)=r\sin\theta\partial_{r}\Omega$ characterizes the radial (vertical) shear. 

According to the scaling (\ref{eq:tau_scaling}), we estimate the transport time $\tau$ for the Sun at the level of the tachocline as $44.5~\rm{days}$ when we use $\Omega_{0}\simeq433~\rm{nHz}$ and $\epsilon=0.3$ \citep[][]{Garcia2007}.
For solar-like stars within the mass range $0.9-1.1~ {\rm M}_{\sun}$ (${\rm M}_{\sun}$ is the Solar mass), if we make the rough assumption that the latitudinal differential rotation with $\epsilon=0.3$ is maintained during the evolution\footnote{Using scaling laws computed in \cite{Brunetal2017} and grids of stellar models that take rotation into account using the STAREVOL code \citep{Amardetal2019}, Astoul et al. (2020, submitted) demonstrates that 0.15<$\vert \epsilon \vert$<0.4 that will not predict orders of magnitude for $\tau$}, we can roughly estimate that the transport time $\tau$ for the solar-like stars in the case of a slow initial rotation \citep[][]{Gallet2015} is about $13.3~\rm{days}$, $9.7~\rm{days}$, and $16.8~\rm{days}$ at the ages of $10~\rm{Myr}$, $100~\rm{Myr}$, and $1000~\rm{Myr}$, respectively.
{These time scales will be eventually used to compute the eddy viscosities in (\ref{eq:tau}) upon the choice of the length scales.}

For stars with a convective core, numerical simulations showed that the differential rotation in the core is mostly cylindrical \citep{Browning2004, Augustson2016}.
For a 2 $\rm M_\odot$ star, \citet{Browning2004} found a rotation contrast within the core of between 30 and 60\%. At the boundary between the convective core and the surrounding radiative zone, this is equivalent to $\epsilon\simeq0.3-0.6$.
Using this estimate as well as rotation rates typical of A-type stars computed using observed surface velocities \citep[see e.g.][]{ZorecRoyer2012} and MESA stellar structure and evolution models of a 2$M_{\odot}$ star with a Solar metallicity \citep{Paxtonetal2011} that provide us the stellar radius, the characteristic time $\tau$ ranges between 0.4 and 1 days during most of the main sequence and can go up to 1.5 d a the end of the main sequence (around 1 Gyr).

If the turbulence triggered by the horizontal shear acts to damp its source, the horizontal differential rotation, as proposed by \cite{Zahn1992} we thus predict a very efficient transport of angular momentum that can lead to the so-called shellular rotation where horizontal gradients of the angular velocity are weak. This also may be the origin of the observed very small radial extent of the Solar tachocline \citep{SpiegelZahn1992} and of an efficient mixing of chemicals in this region \citep[][]{Brunetal1999}. 

The behavior of the turbulence generated by horizontal shears has been recently explored in the non-rotating case by \cite{Copeetal2019} and \cite{Garaud2020}. They find when exploring the parameter space using Direct Numerical Simulations (DNS) a turbulent stratified regime where the turbulent transport can be modeled by an eddy-diffusivity. This opens an interesting path to verify our predictions when such DNS will take rotation into account.

\section{Conclusion}
\label{sect:conclusion}
In this paper, we studied horizontal shear flow instabilities in stably-stratified and thermally-diffusive fluids in a rotating plane where the full Coriolis acceleration with both vertical and horizontal components of the rotation vector is taken into account. 
For the canonical shear flow in the hyperbolic tangent profile, there exist two types of shear instabilities: the inflectional instability due to the presence of an inflection point and the inertial instability due to an inertial imbalance in the presence of the Coriolis acceleration. 
In the presence of positive horizontal Coriolis parameter $\tilde{f}>0$, we found that both the inflectional and inertial instabilities are strongly affected. 
For instance, the inflectional instability, whose maximum growth rate is known to be independent of $N$, $f$, and $Pe$ in the traditional approximation \citep[]{Deloncle2007,PPM2020}, has now a maximum growth rate that depends on $N$ and $Pe$. 
The horizontal Coriolis parameter $\tilde{f}$ destabilizes the inflectional instability for strong stratification while it stabilizes the instability at a small $N$. 
The thermal diffusivity at finite $Pe$ also plays a stabilizing role in the inflectional instability. 
On the other hand, the inertial instability is destabilized by the thermal diffusivity, and the unstable regime is widely extended. 
For instance, in the nondiffusive limit $Pe=\infty$, the inertially unstable regime is found to be $0<f<1+\tilde{f}^{2}/N^{2}$ (i.e., $\tan^{-1}(N/2)<\theta<90^{\circ}$). 
More strikingly, it is inertially unstable for any $f$ for high diffusivity fluids as $Pe\rightarrow0$ when $\tilde{f}>0$ (i.e., the inertial instability is active at any colatitude $0<\theta<180^{\circ}$ for the absolute Coriolis parameter $2\Omega>0$). 
These unstable regimes as well as the dispersion relations for the inertial instability are derived analytically using the WKBJ approximation for large vertical wavenumber $k_{\rm{z}}$, and this asymptotic analysis demonstrates a very good agreement with numerical results in the inviscid limit. 
Using the asymptotic expressions of the growth rate for the inertial instability, we also predicted the critical Reynolds number above which the growth rate becomes positive. 
Finally, we proposed prescriptions for the effective horizontal and vertical turbulent viscosities induced by the inertial and inflectional instabilities and found that the inertial instability plays an important role in the turbulent transport near the equator. 

Observational and numerical studies suggest that stellar radiative zones have a mostly uniform rotation, whereas stellar convective zones are differentially rotating, but neutrally stratified.
Therefore, one expects the present study to be relevant near the boundary between the convective and radiative zones.
Because of the small values predicted for the time that characterizes the turbulence triggered by horizontal shear flows in such regions, this turbulent transport can have a strong impact on the structure and the evolution of stars, for instance by interacting with overshooting flows and by extracting angular momentum and chemical elements from the core to the envelope.
In particular, the horizontal shear could be a crucial physical ingredient needed to explain the observed structure of the solar/stellar tachocline(s) \citep{SpiegelZahn1992,Brunetal1999}.
Other ingredients that are missing in the present work, such as magnetic fields \citep{GM1998,Strugareketal2011,AGW2013,Barnabeetal2017}, may also play an important role.

Our predictions concerning the turbulent viscosities need to be confirmed by fully turbulent numerical simulations.
In particular, global numerical simulations would allow us to validate the local approach used in this work, and the relevance of the latitudinally averaged quantities derived for stellar evolution codes.
Besides, the local results could be used to build subgrid models for large-eddy simulations and stellar evolution calculations to better capture small-scale transport processes.

\begin{acknowledgements}
The authors acknowledge support from the European Research Council through ERC grant SPIRE 647383 and from GOLF and PLATO CNES grants at the Department of Astrophysics at CEA Paris-Saclay.
{We thank the referee, Prof. A. J. Barker, for his constructive comments that allow us to improve the article. }
\end{acknowledgements}

\bibliographystyle{aa} 
\bibliography{aabib} 

\end{document}